\documentclass[useAMS,usenatbib]{mn2e}
\usepackage{txfonts} 
\usepackage{graphicx}

\graphicspath{{figures/}}

\renewcommand{\vec}[1]{\ensuremath{\bmath{#1}}}
\newcommand{\GJ}[1]{\ensuremath{#1_{\textrm{\tiny{}GJ}}}}    
\newcommand{\PC}[1]{\ensuremath{#1_\mathrm{pc}}}    
\newcommand{\RLC}{\ensuremath{R_\textrm{\tiny{}LC}}}
\newcommand{\jm}{\ensuremath{j_{\mathrm{m}}}}
\newcommand{\lambdaD}{\ensuremath{\lambda_\textrm{\tiny{}D}}}    
\newcommand{\lambdaDGJ}{\ensuremath{\lambda_{\textrm{\tiny{}D, GJ}}}}    
\newcommand{\omegaPGJ}{\ensuremath{\omega_{\textrm{\tiny{}p, GJ}}}}    
\newcommand{\xb}{\ensuremath{x_{\textrm{\tiny{}B}}}}
\newcommand{\Vm}{\ensuremath{\Phi_m}}
\newcommand{\Ss}{\S\S}
\newcommand{\PapI}{\citet{Timokhin2010::TDC_MNRAS_I}}

\title[Pair Cascades over Pulsar Polar Caps.]%
{Current Flow and Pair Creation at Low Altitude in Rotation Powered
  Pulsars' Force-Free Magnetospheres:\\ 
  Space-Charge Limited Flow}

\author[A.~N.~Timokhin and J.~Arons]%
{
  A.~N.~Timokhin$^{1,2,3}$\thanks{E-mail: andrey.timokhin@nasa.gov} and J.~Arons$^{2,4}$\\
  $^1$Astrophysics Science Division, NASA Goddard Space Flight Center,
  Greenbelt, MD 20771, USA\\
  $^2$Theoretical Astrophysics Center, University of California at Berkeley,
  Berkeley, CA 94720, USA\\
  $^3$Moscow State University, Sternberg Astronomical Institute,
  Universitetskij Pr. 13, 119991 Moscow, Russia\\
  $^4$also Astronomy Department, Physics Department, and Space
  Sciences Laboratory, University of California, Berkeley 
}

\begin{document}

\date{Received ; in original form }

\pagerange{\pageref{firstpage}--\pageref{lastpage}} \pubyear{2012}

\maketitle

\label{firstpage}

\begin{abstract}
  We report the results of an investigation of particle acceleration
  and electron-positron plasma generation at low altitude in the polar
  magnetic flux tubes of Rotation Powered Pulsars, when the stellar
  surface is free to emit whatever charges and currents are demanded
  by the force-free magnetosphere.  We apply a new 1D hybrid plasma
  simulation code to the dynamical problem, using Particle-in-Cell
  methods for the dynamics of the charged particles, including a
  determination of the collective electrostatic fluctuations in the
  plasma, combined with a Monte-Carlo treatment of the high energy
  gamma rays that mediate the formation of the electron-positron
  pairs. We assume the electric current flowing through the pair
  creation zone is fixed by the much higher inductance magnetosphere,
  and adopt the results of force-free magnetosphere models to provide
  the currents which must be carried by the accelerator. The models
  are spatially 1D, designed to explore the physics, although of
  practical relevance to young, high voltage pulsars.
  
  We observe novel behavior. a) When the current density $j$ is less
  than the Goldreich-Julian value ($0 < j/\GJ{j} < 1$), space charge
  limited acceleration of the current carrying beam is mild, with the
  full Goldreich-Julian charge density being comprised of the charge
  density of the beam, co-existing with a cloud of electrically
  trapped particles with the same sign of charge as the beam.  The
  voltage drops are on the order of $m c^2/e$, and pair creation is
  absent.  b) When the current density exceeds the Goldreich-Julian
  value ($j/\GJ{j} > 1$), the system develops high voltage drops (TV
  or greater), causing emission of curvature gamma rays and intense
  bursts of pair creation. The bursts exhibit limit cycle behavior,
  with characteristic time scales somewhat longer than the
  relativistic fly-by time over distances comparable to the polar cap
  diameter (microseconds). c) In return current regions, where
  $j/\GJ{j} < 0$, the system develops similar bursts of pair creation.
  These discharges are similar to those encountered in previous
  calculations of pair creation when the surface has a high work
  function and cannot freely emit charge, \PapI.  In cases b) and c),
  the intermittently generated pairs allow the system to
  simultaneously carry the magnetospherically prescribed currents and
  adjust the charge density and average electric field to force-free
  conditions.  We also elucidate the conditions for pair creating beam
  flow to be steady (stationary with small fluctuations in the
  rotating frame), finding that such steady flows can occupy only a
  small fraction of the current density parameter space exhibited by
  the force-free magnetospheric model.  The generic polar flow
  dynamics and pair creation is strongly time dependent.  The model
  has an essential difference from almost all previous quantitative
  studies, in that we sought the accelerating voltage (with pair
  creation, when the voltage drops are sufficiently large; without,
  when they are small) as a function of the applied current.
  
  The 1D results described here characterize the dependence of
  acceleration and pair creation on the magnitude and sign of current.
  The dependence on the spatial distribution of the current is a
  multi-D problem, possibly exhibiting more chaotic behavior.  We
  briefly outline possible relations of the electric field
  fluctuations observed in the polar flows (both with and without pair
  creation discharges) to direct emission of radio waves, as well as
  revive the possible relation of the observed limit cycle behavior to
  microstructure in the radio emission. Actually modeling these
  effects requires the multi-D treatment, to be reported in a later
  paper.
\end{abstract}

\begin{keywords}
  acceleration of particles --- 
  plasmas --- 
  pulsars: general --- 
  stars: magnetic field --- 
  stars: neutron
\end{keywords}

\section{Introduction}
\label{sec:introduction}

Young pulsar wind nebulae (PWNe) show that rotation powered pulsars
(RPPs) have dense magnetospheres, at least with regard to those
regions that feed the plasma outflow (e.g.,
\citealt{BucciantiniAronsAmato2011}).  Electron-positron pair creation
in the open field line region that connects to the external world is
the only known candidate for the origin of such outflows, with
acceleration and convertible gamma ray emission occurring either at
low altitude \citep{Sturrock71} or in the outer magnetosphere
\citep{Cheng1986}. High density flows that can feed all the open field
lines can exist only in the low altitude polar cap region (for a
general discussion, see \citealt{Arons2009}).

Theoretical studies of charged particle flow from the magnetic polar
regions of rotation powered pulsars began with the observation by
\citet{GJ}, that an isolated magnetic rotator in vacuum must have a
charged magnetosphere almost corotating with the star. Since RPPs are
strongly gravitationally bound and cool objects (thermal scale height
in any atmosphere orders of magnitude less than the stellar radius)
and have no external source of plasma supply (so far as we know), the
only plasma source is extraction of charged particles from the stellar
surface, leading to a conjectured magnetosphere whose plasma is fully
charge separated, in contrast to all other known astrophysical
systems, whose plasmas are charged but quasi-neutral. \citet{GJ}
speculated that on polar field lines -- those that extend beyond the
light cylinder located at cylindrical radius
$\RLC=cP/2\pi\simeq48,000\,P$ km, $P =$ rotation period in seconds --
a charge separated outflow would form. They argued that the
energy/particle in the outflow would be no more than the gravitational
escape energy $GM_*/R_* \sim 0.3 mc^2 (M_*/1.4 M_{\sun}) (10 \;
\mathrm{km}/R_*)$, $M_*$ and $R_*$ are star's mass and radius
correspondingly -- the particles leave at non-relativistic energies,
in spite of the fact that the electric potential drop \emph{across}
the polar field lines is equal to the full potential of an open
rotating magnetosphere with a dipole magnetic field
$\Vm=\sqrt{W_{\textrm{\tiny{}R}}/c}
\approx10\,(\dot{P}/10^{-15})^{1/2}P^{-3/2}$~TV, $W_{\textrm{\tiny{}R}}
= $ rotational energy loss rate, with $\dot{P} = dP/dt$ .  $\Vm$
vastly exceeds the rest energy and gravitational energy of the
particles, either electrons or protons (or He or other ions populating
the star's crust and atmosphere). The super strong magnetic field
suppresses free acceleration of the particles in the transverse
electric field, whose primary (``zeroth order'') consequence is
corotation of the field lines with the magnetic field embedded in the
neutron star (NS), with field line motion measured by the
$\vec{E}\times\vec{B}$ drift of charged particles across the magnetic
field (which occurs even when the particles have zero Larmor
gyration). The particle loss rate in the conjectured charge separated
scenario is $\dot{N}_{\textrm{\tiny{}R}} = c\Vm/e \approx
2\times10^{30}\,(\dot{P}/10^{-15})^{1/2} P^{-3/2} \; {\rm s}^{-1}$,
orders of magnitude less than that inferred from the injection of
plasma into the young PWNe. The electrodynamics of the magnetosphere
differs drastically, depending on whether the particle loss rate falls
short of or exceeds $\dot{N}_{\textrm{\tiny{}R}}$.  For the young
PWNe, the particle injection rate exceeds
$\dot{N}_{\textrm{\tiny{}R}}$ (by a lot).  In that case, the
magnetosphere's basic state should be one in which
$\vec{E}\cdot\vec{B}$ = 0, with no parallel acceleration sufficient to
generate convertible gamma rays occurring under the pulsar's
rotational control.

The discovery of gamma ray pulsars in the 1970s, and their
proliferation into a population with more than 100 such stars in the
most recent published Fermi pulsar catalog
\citep{FermiPSRCatalogI::2010}, has shown that parallel acceleration
to GeV gamma ray emitting energies (indeed, multi-hundred GeV, in the
Crab pulsar, \citealt{veritas_crab2011}) must occur somewhere, with
energy efficiency exceeding a few tenths of a percent, as measured by
$L_\gamma/W_{\textrm{\tiny{}R}}$, $L_\gamma=$ gamma-ray
  luminosity. If the acceleration is limited by radiation reaction,
as is true in many models, $L_\gamma$ is a good proxy for the energy
put into parallel acceleration. $L_\gamma/W_{\textrm{\tiny{}R}}$ can
approach as much as 50\% at smaller spin down luminosities. Just how
some fraction of the total potential drop gets released in
acceleration along $B$, gamma ray emission and pair creation has been
mysterious since the beginning of pulsar research, made relevant to
the real world by the gamma ray discoveries. Since the polar cap
source is the only one capable of feeding the whole (open)
magnetosphere, its understanding remains of central interest to
modeling pulsar magnetospheres, even though the spectral and beaming
characteristics of the pulsed gamma rays are better modeled by
accelerators in the outer magnetosphere.

Free particle outflow from the NS surface is a common assumption in
most of the current pulsar models (see \S\ref{sec:current_density}).
The polar cap accelerator problem has been studied under that
assumption before
\citep[\emph{e.g.}][]{Michel1974,FawleyAronsScharlemann1977,Mestel1985,shibata97,Beloborodov2008}.
\citet{Michel1974} and \citet{FawleyAronsScharlemann1977} obtained
solutions for the non-neutral space charge limited charged particle
flow for the current density almost equal to the GJ current density.
All these models assume strictly steady flow in the co-rotating frame,
on \emph{all} time scales. In these models, the charge density of the
current carrying beam supplies almost all of the charge density needed
to short out the parallel component of the electric field, while
leaving a residuum $E_\parallel$ sufficient to accelerate the beam --
relativistic energies in a temporally steady flow are found if the
current density $j_\parallel=-(B/P)\cos\chi +
\mathrm{small~corrections} \cong \GJ{j}$;
$\chi=\angle(\vec{\mu},\vec{\Omega})$, the pulsar inclination
angle. \citet{Mestel1985} showed that the velocity of the beam is
monotonically increasing with altitude to relativistic speeds only if
the current density is larger than $\GJ{j}$.  If the current density
is smaller that $\GJ{j}$ the temporally steady velocity of the beam
(assumed to have no momentum dispersion) oscillates spatially,
i.e. particles accelerate and decelerate to a complete halt as they
move outwards into the magnetosphere.  \citet{Beloborodov2008}
rediscovered Mestel \textit{et al.}'s solution and suggested that in
the region of the polar cap where $j_\parallel<\GJ{j}$ particles will
be not accelerated up to high energies as the beam velocity
oscillates, but along the magnetic field lines with
$j_\parallel/\GJ{j}>0$ or $j_\parallel/\GJ{j}<0$ particle acceleration
will be efficient and will lead to pair formation. The quantitative
model which we describe in this paper lends some support to
Beloborodov's speculations, although it does not agree with them in
detail.

In this paper we describe our study of the physics of the polar cap
accelerator in the space charge limited flow regime starting from 
first principles -- assuming free particle outflow from the surface of
a NS we compute the electric field, particle acceleration, gamma-ray
emission, propagation and pair creation simultaneously.  It extends
the study of current flow and pair cascades in neutron star
magnetospheres using the theoretical formulation and self-consistent
numerical techniques introduced in \citet{Timokhin2010::TDC_MNRAS_I}.

The plan of the paper is as follows.  In
\S\ref{sec:current_density}, we review the properties of the
current flow imposed by the magnetosphere, in the force-free model,
pointing out that the current density is the main parameter which
regulates the efficiency of particle acceleration.  In
\S\ref{sec:stationary_sclf} we review the properties of
stationary solutions for the charge separated space charge limited
flow problem.  In \S\ref{sec:num_setup} we briefly describe our
numerical model.  In \S\ref{sec:cold_flow} we describe the
results of numerical modeling for the case of sub-GJ current density,
the regime when particle acceleration is inefficient and no pair
creation is possible.  In \S\ref{sec:j_pairs} we consider flow
regimes with efficient pair creation. In
\S\ref{sec:stationary-cascades} we pay special attention to the
stationary flow regime, which up to now was assumed in (most) works on
pulsar polar cap accelerators, and describe why it has limited
relevance to the force-free model of the pulsar
magnetosphere.  We discuss the implications of our results for the
physics of rotation powered pulsars in \S\ref{sec:discussion}
and summarize our conclusions in \S\ref{sec:conclusions}.

\section{Current density in the polar cap}
\label{sec:current_density}

\begin{figure*}
  \begin{center}
    \includegraphics[width=\textwidth]{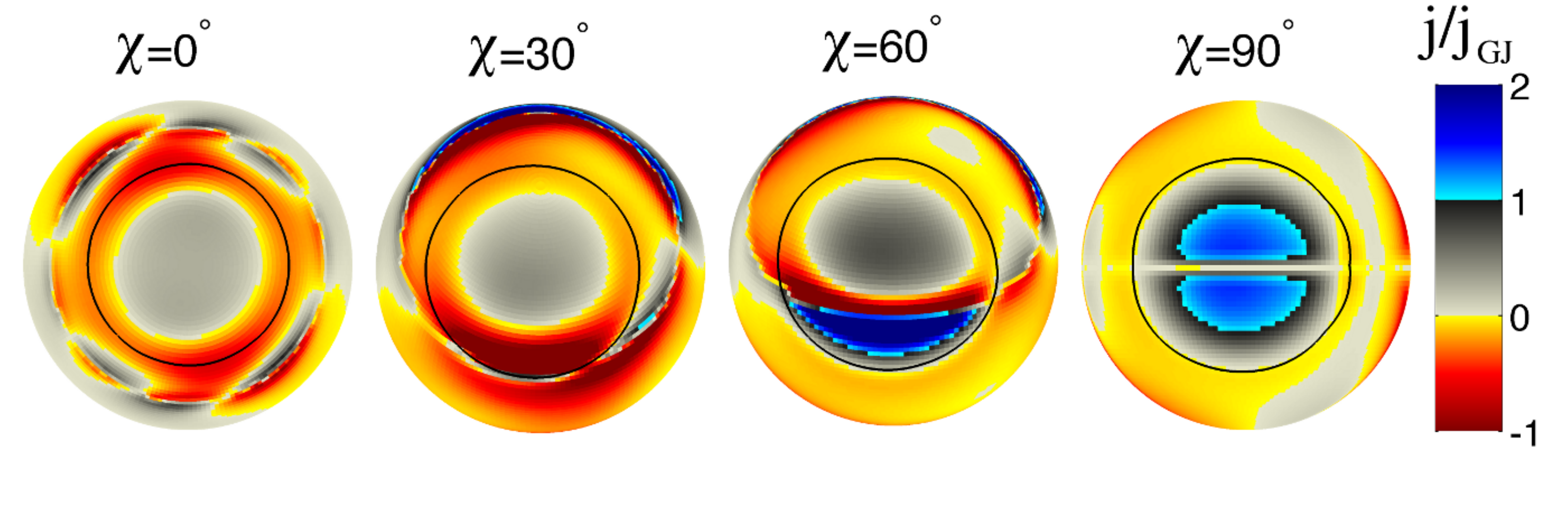}
  \end{center}
  \caption{Field-aligned current density at the polar cap of the
    force-free rotator, with $j_\parallel$ measured in units of the
    Goldreich-Julian current density $\GJ{j}\equiv - \vec{\Omega}
    \cdot \vec{B}/2\pi c$. The black circle is the rim of the polar
    cap - the footprints of the field lines that pass outside the
    light cylinder fall with that circle.  The distributed current is
    shown. The current sheet component coincides with the polar cap
    boundary.  This plot was made by Xuening Bai using results of
    force-free magnetosphere simulations presented in
    \citet{BaiSpitkovsky2010a}.}
  \label{fig:bai}
\end{figure*}

Nebular observations of plasma supply by RPPs suggest the open field
line regions are ``MHD-like'', \textit{i.e.}, having $E_\parallel = 0$
except in special zones (such as the polar cap), which are, in effect,
boundary layers.  There is essentially no observational information on
the properties of the closed magnetosphere, thus the simplest
hypothesis is to follow \citet{GJ} and assume the magnetosphere is
(almost everywhere) filled with plasma that shorts out
$E_\parallel$. The magnetosphere has open and closed magnetic field
line zones.  In the open zone plasma flows away into the pulsar wind;
currents and their associated electromagnetic inertia keep the
magnetic field open, so sustaining the flow. The plasma flows all the
way from the base of the open field line zone in the polar cap of the
pulsar where it is either extracted from the NS surface or generated
by electron-positron cascades (or both).  The distribution of the
current density across the open field line zone, and therefore, across
the polar cap, is determined by the global magnetospheric structure.
Stability of the pulsar mean profiles and sharpness of the peaks in
the spectra of gamma-ray pulsars strongly suggest that on scales
comparable to the light cylinder, the magnetosphere is stationary in
the frame co-rotating with the neutron star with smooth and continuous
plasma outflow. However, the stationary co-rotating magnetosphere
hypothesis demands stationarity only in a statistical sense, fast
local fluctuations which average to a stationary state (and even more
broadly, global variations) can be included within this picture, so
long as they do not smear the beaming profiles.

The polar cap acceleration and possible pair cascade zone -- the main
place where electron - positron plasma that feeds the wind can be
produced -- is much smaller than the characteristic scale $\RLC$
of the magnetosphere. Therefore its inductance is negligible compared
with that of the magnetosphere, and the polar current flow within the
polar cap region must have an average equal to that set by the
magnetosphere's global structure.

In this paper we solve a \emph{local} problem of how the polar cap
cascade zone adjusts itself to the current density required by the
magnetosphere.  On the dynamical timescales typical for the cascade
zone (microseconds) the magnetospherically required current density is
stationary because it could change only on magnetospheric time scales
(tens of milliseconds up to several seconds).  The idea that the
acceleration zone has a magnetospherically determined current appears
in the electrodynamics through the magnetic induction equation
\begin{equation}
  \frac{\partial E_\parallel}{\partial t} = 
  -4\pi j_\parallel + c (\nabla \times \vec{B} )_\parallel
  \approx 
  -4\pi (j_\parallel - \jm),
  \label{eq:dE_par__dt}
\end{equation}
with
\begin{equation} 
   \jm = \frac{c}{4\pi}(\nabla \times \vec{B}_{\mathrm{magnetosphere}})_\parallel 
   \label{eq:jm}
\end{equation}
being the current that sustains the twist to the field lines.  In this
paper we neglect the induced variations in the magnetic field that
accompany variable $E_\parallel$ -- because of the very strong
background magnetic field (which in the region of interest has
$\nabla\times\vec{B}=0$), these have little dynamical effect on the
acceleration and cascade dynamics, see
Appendix~\ref{sec:1D-Electrodynamics} for derivation of
eq.~(\ref{eq:dE_par__dt}).

We study the behavior of the cascade zone under different current
loads, sampling the range of possibilities illustrated in
Fig.~\ref{fig:bai}.  The model has an essential difference from
previous studies, in that we seek the accelerating voltage (with pair
creation when the voltage drops are sufficiently large, without when
they are small) as a function of the applied current.  Previous work
has almost entirely focused on the opposite direction, seeking the
current that emerges from the accelerator when the voltage is fixed,
either by the geometry of by the poisoning of the accelerator by pair
creation. In addition, we allow the system to be fully time dependent.
These generalizations lead to qualitatively different results from
what has appeared before. Our model is one-dimensional, with
  spatial axis along magnetic field lines; from here on we drop
  subscript~$\parallel$ from all quantities.

The characteristic charge density, the Goldreich-Julian charge density,
\begin{equation}
  \label{eq:eta_GJ}
  \GJ{\eta} = -\frac{\vec{\Omega}\cdot\vec{B}}{2\pi{}c}
\end{equation}
sets the characteristic current density
\begin{equation}
  \label{eq:j_GJ}
  \GJ{j} = \GJ{\eta} c \,.
\end{equation}
Following \citet{FawleyAronsScharlemann1977} and
\citet{AronsScharlemann1979}, we assume also that if the star lacks an
atmosphere, the work function for charged particles to leave the NS
surface is small enough that any number of charged particles can be
extracted from the NS surface until the extracting electric field is
screened. In the classical pulsar regime, the theory of charged
particle binding to the crust suggests such free emission to be
likely\ \citep{Medin2010}. More relevantly, X-ray observations of
heated polar caps \citep[][and references therein]{Bogdanov2007}
suggest these stars have atmospheres overlying the solid and ocean
components of the crust, which guarantees free emission of
charges. The opposite case, with complete suppression of particle
emission from the surface, was studied in
\cite{Timokhin2010::TDC_MNRAS_I}, a realization of the scenario
conjectured by \cite{RudermanSutherland1975}

As we will show in the next sections there are three qualitatively
different plasma flow types in the polar cap of pulsar depending on
the ratio of the current density imposed by the magnetosphere to the
GJ current density: (i) $\jm$ has the same sign and its absolute value
is \emph{smaller} that the GJ current density, $0 \le \jm/\GJ{j}<1$,
hereafter sub-GJ current density; (ii) $\jm$ has the same sign and its
absolute value is \emph{larger} that the GJ current density,
$\jm/\GJ{j}\ge{}1$, hereafter super-GJ current density; (iii) $\jm$
has the \emph{opposite} sign to the GJ current density,
$\jm/\GJ{j}<0$, hereafter anti-GJ current density.

The advent of quantitative solutions for the structure of the
force-free model
(\citealp[e.g.][]{CKF,Timokhin2006:MNRAS1,Spitkovsky:incl:06,Kalapotharakos2009,BaiSpitkovsky2010a})
has provided, for the first time, a theory of the current flow
expected as a function of pulsar parameters ($B$, $P$, $\chi$, when
the magnetic field is a star-centered dipole).  Earlier modeling of
polar cap, slot gap and outer gap accelerators adopted the expectation
that the current density is on the order of $\GJ{j}$, and expressed
the hope that the accelerator and pair creation physics do not
sensitively depend on the precise value and spatial distribution of
$j$.  The results reported here show that the magnitude and sign of
the current flow do lead to drastic differences in the open field line
accelerator's behavior in the three regimes (i) - (iii), even though
the order of magnitude of the current is as
expected. Fig.~\ref{fig:bai}, showing $\tilde\jmath\equiv{}j/\GJ{j}$,
reveals that all three flow regimes occur in the force-free
magnetosphere model.  While $|\tilde\jmath\,|$ always has numerical
values on the order of unity, it can be negative (return current) as
well as lying in the separate regimes $0<\tilde\jmath< 1$ and
$\tilde\jmath>1$.  We show these separate regimes have different
dynamical behavior and different implications for pair creation.

We also show that once the constraints of the steady flow models for
space charge limited flow are relaxed, the small departures of the GJ
charge density from the simple estimate $-\Omega B\cos\chi /2\pi c$
created by geometric and general relativistic considerations
\citep{AronsScharlemann1979,Muslimov/Tsygan92} that play an essential
role in the steady flow models%
\footnote{If the GJ charge density were uniform and the beam is
  everywhere relativistic and time stationary, the unique model is no
  acceleration at all \citep{Tademaru1973,FawleyAronsScharlemann1977},
  a severe contradiction.}
have little significance when the flow is
fully time dependent.  Since we consider only the polar cap region,
with altitudes not exceeding the width of the polar flux tube
$\PC{r}= R_* \sqrt{R_*/\RLC} = 0.145 P^{-1/2} \; {\rm km} \ll R_* = 10 \;
{\rm km}$, spatial variation of $B$ is mostly unimportant.  If we do
not say so explicitly otherwise, throughout the paper we assume that
the GJ charge density is constant, independent of the distance along
$B$.

\section{Time-Stationary Space Charge Limited Flow}
\label{sec:stationary_sclf}

\begin{figure}
  \begin{center}
   \includegraphics[width=\columnwidth]{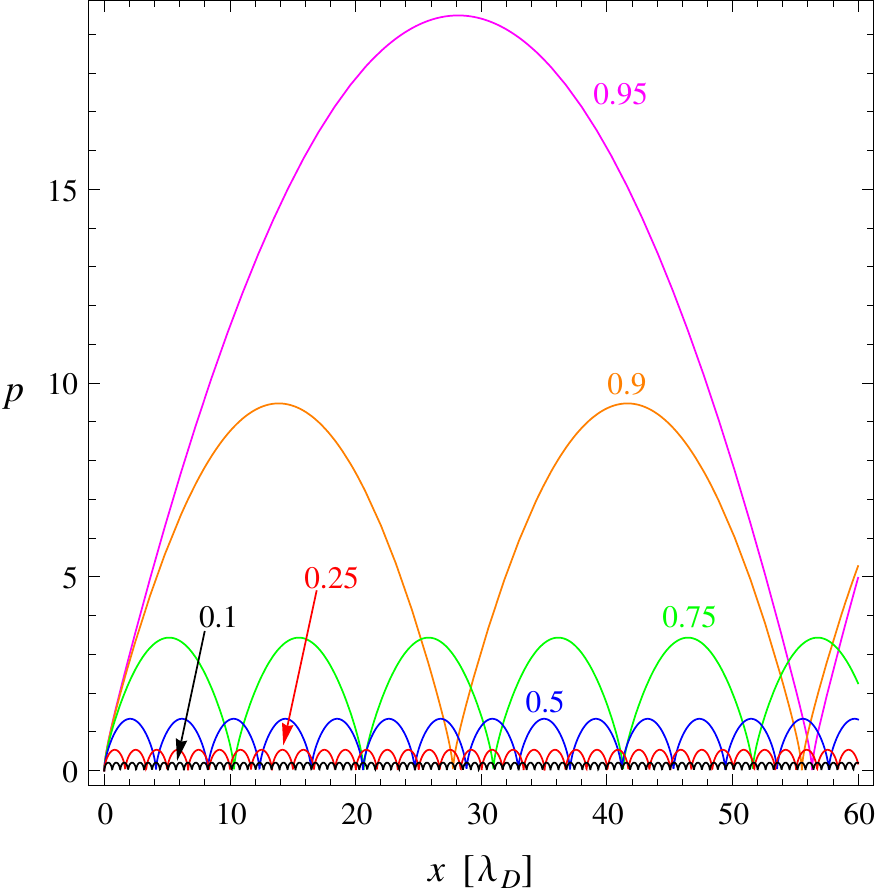}
  \end{center}
  \caption{Phase space trajectories (4-velocity $p$
    vs. distance normalized to $\lambdaDGJ$) for stationary space charge
    limited flow for current densities
    $\jm/\GJ{j} = 0.1,0.25,0.5,0.75,0.9,0.95$}
  \label{fig:posc}
\end{figure}

\begin{figure}
  \begin{center}
   \includegraphics[width=\columnwidth]{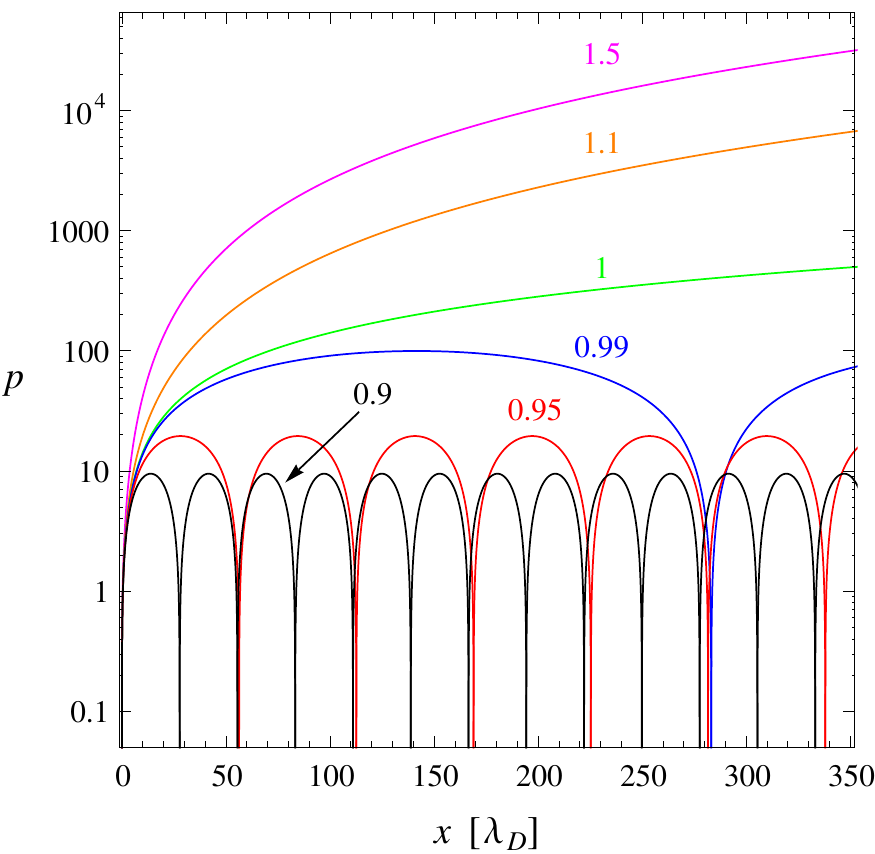}
      \end{center}
  \caption{Phase space trajectories for stationary space charge
    limited flow for current densities
    $\jm/\GJ{j}=0.9,0.95,0.99,1,1.1,1.5$. Note that in contrast to
    Fig.~\ref{fig:posc} the vertical axis on this plot is
    logarithmic.}
  \label{fig:posc_pacc}
\end{figure}

\begin{figure}
  \begin{center}
    \includegraphics[width=\columnwidth]{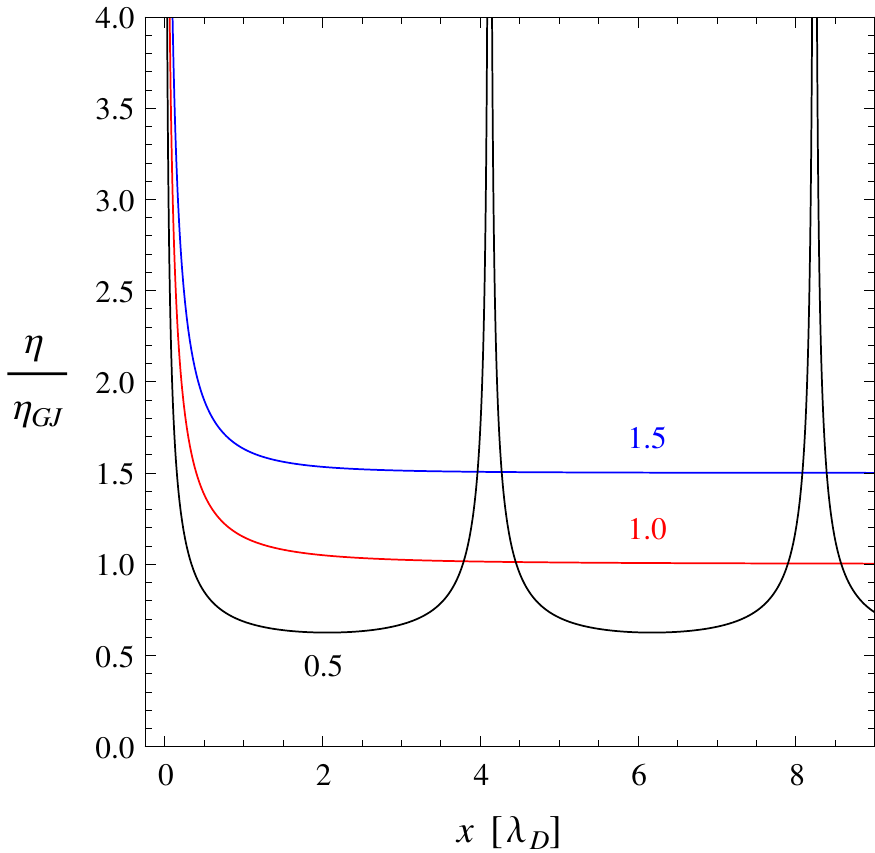}
  \end{center}
  \caption{Charge density of stationary space charge limited flow
    normalized to the GJ charge density $\GJ{\eta}$ as a function of
    distance $x$ normalized to $\lambdaD$ for current densities
    $\jm/\GJ{j}=0.5,1.0,1.5$.}
  \label{fig:eta_osc_acc}
\end{figure}

Copious pair creation occurs when there is a sufficiently large
accelerating potential difference along $B$ when pairs are
absent. Such potential drops, typically $\gtrsim 10^{12}$ Volts
\citep{Sturrock71}, readily exist in the absence of current flow, that
is, in a vacuum. Under pulsar conditions, if current does flow, pairs
appear when the current flow co-exists with TV potential
drops\footnote{TV potential drops are required if curvature radiation
  from charges accelerated along a locally dipole magnetic field is
  the emission mechanism for the gamma rays that convert to pairs. The
  required potential drops are smaller if the emission mechanism is
  inverse Compton scattering (either resonant or non-resonant) of
  softer photons from the surface \citep[e.g.][and references
  therein]{Hibschman/Arons:pair_multipl::2001}.}, that is the current
is a relativistic beam. Therefore, the starting place is the
properties of time stationary, space charge limited, charge separated
flow.  In this section we give an overview of these properties while
detailed derivation of the equations used in this section is given in
Appendix~\ref{sec:stat-sclf} together with some useful
asymptotics. These review and extend a variety of results already in
the literature.

For definiteness, we consider pulsars with an acute angle $\chi$
between $\vec{\Omega}$ and $\vec{B}$ (``acute''
pulsars). These objects have $\GJ{\eta} < 0$ at the polar cap, and
require electron emission to supply $\GJ{\eta}$.

Let us consider an electron beam starting with zero velocity at $x=0$
and let the current density $\jm$ imposed by the magnetosphere
(eq.~(\ref{eq:jm})) be a fraction $\xi$ of the GJ charge density
\begin{equation}
  \label{eq:j_ksi_jGJ}
  \jm\equiv\xi\GJ{j}\,.
\end{equation}
In our case the GJ charge density is negative.  In stationary flow the
current density is constant in both space and time and is equal to the
imposed current density $j\equiv\jm$, thus $dE/dt=0$ in
eq.~(\ref{eq:dE_par__dt}).  The stationary electric field
$E_{\mathrm{s}}$ is then given by Gauss's law which in the frame
corotating with the NS takes the form \citep[e.g.][and references
therein]{FawleyAronsScharlemann1977}
\begin{equation}
  \label{eq:dE_dx}
  \frac{dE_{\mathrm{s}}}{dx} = 4\pi(\eta-\GJ{\eta})\,.
\end{equation}
The magnitude of the electric field increases with distance if the
magnitude of the charge density $\eta$ is larger than the GJ charge
density and decreases otherwise.

The charge density at any given point of the flow is
\begin{equation}
  \label{eq:eta}
  \eta = j/v = \frac{\jm}{c} \frac{\sqrt{p^2+1}}{p} \, ,
\end{equation}
where $v$ is the flow velocity and $p\equiv{}\gamma{}v/c$ is the
4-velocity = momentum in units of $mc$.  $p$ of a charge separated
stationary beam is given by the solution of the equation
(\ref{eq:d_p_ds_Appendix})
\begin{equation}
  \label{eq:d_p_ds}
  \left(\frac{dp}{ds}\right)^2 = 
    2\, \frac{p^2+1}{p^2} 
    \left( 1+ \xi{}p - \sqrt{p^2+1}  \right)\,.
\end{equation}
where the distance $s$ is measured in units of the Debye length
$\lambdaDGJ$ of a cold electron plasma with GJ number density
\begin{equation}
  \label{eq:lambdaDebye}
  \lambdaDGJ = c \left( \frac{4\pi\GJ{\eta}e}{m}\right)^{-1/2} \simeq 
  2\,B^{-1/2}_{12}P^{1/2} \mbox{cm}\,.
\end{equation}
$B_{12}$ is the pulsar magnetic field in $10^{12}$~G, $P$ is the
pulsar period in seconds.

The numerical solutions of (\ref{eq:d_p_ds}) for different imposed
current densities are shown in
Figs.~\ref{fig:posc},~\ref{fig:posc_pacc} for different values of
$\xi$; these results confirm the work of \cite{shibata97}.  For
$0<\jm/\GJ{j}<1$ the steady flow oscillates spatially, with particle
momenta oscillating in the interval $[0,p_{\max}]$, with
\begin{equation}
  \label{eq:p_max}
  p_{\max} =\frac{2\xi}{1-\xi^2}.
\end{equation}
$dp/ds = 0$ at $p=p_{\max}$ and so is the RHS of
eq.~(\ref{eq:d_p_ds}).  The value of $p_{\max}$ and the spatial period
of oscillations $s_0$ increase with increasing $\xi$,
see 
Figs.~\ref{fig:posc},~\ref{fig:posc_pacc}.  For $\xi\ge{}1$
acceleration is monotonic with $p$ increasing to infinity, with
asymptotic behavior
\begin{equation}
  \label{eq:p_large}
  p = \sqrt{2}s +\frac{\xi-1}{2}s^2\,. 
\end{equation}
See Fig.~\ref{fig:posc_pacc} for $\xi=1,1.1,1.5$.

The reason for such behavior is as follows.  The flow starts with zero
initial velocity at $x=0$ where the electric field is zero.  When the
particle velocity is small the imposed current density is produced by
high particle density moving slowly. In such places the absolute value
of the beam charge density (cf. eq.~\ref{eq:eta}) is larger that that
of the GJ charge density, $|\eta|>|\GJ{\eta}|$ -- see
Fig.~\ref{fig:eta_osc_acc} where we plot the ratio $\eta/\GJ{\eta}$
for flows with $\xi=0.5,1,1.5$.  The charge density is negative and
according to eq.~(\ref{eq:dE_dx}) the electric field in this region is
decreasing towards more negative values, thus accelerating electrons.
If the imposed current density exceeds the GJ current density,
$\xi>1$, the absolute value of the beam charge density -- whose
maximum value is $|\jm/c|=\xi|\GJ{\eta}|$ -- never becomes smaller
than $|\GJ{\eta}|$, hence, $dE_{\mathrm{s}}/dx<0$ and acceleration
continues up to infinity -- the electric field and potential are
monotonic.  For $\xi<1$, on the other hand, particles' velocities
increase to values such that the imposed current density can be
sustained by particle number density smaller than the GJ number
density. Then $|\eta|<|\GJ{\eta}|$ (see the line for $\jm=0.5\GJ{j}$
in Fig.~\ref{fig:eta_osc_acc}), $dE_{\mathrm{s}}/dx>0$ and the
accelerating electric field weakens, changes sign and decelerates
particles - in cold flow, the particles decelerate to zero velocity,
and the cycle repeats.

The model of space charge limited flow outlined here provides a
physical framework for the expected particle energetics.  It is based
on the approximation of one cold fluid and an assumption of complete
stationarity of the flow.  It can be extended to a two-fluid model to
account for presence of positrons and pair creation \citep[as
in][]{Arons1983}.  However kinetic effects such as particle trapping
can not be included in a cold fluid approximation, although certain
aspects {\it can} be modeled if momentum dispersion (``pressure'') is
included, with an assumed equation of state.  A kinetic model
incorporates momentum dispersion in the collisionless medium without
having to make arbitrary assumptions about the equation of state. As
we show in the following sections, particle trapping and pair creation
profoundly affect the plasma dynamics, with momentum dispersion being
essential to the dynamics behind the simultaneous adjustment of the
charge density to the condition of low voltage drop along $B$, modeled
as $\vec{E} \cdot \vec{B} = 0$ in the force- free global model, and
the adjustment of the field aligned current $j$ to the
magnetospherically imposed $j_m$.

The study of plasma kinetics in a general regime -- without relying on
stationarity or perturbation theory -- is possible only by means of
numerical simulations.  In following sections we describe our study of
plasma, both fully non-neutral and quasi-neutral when pair cascades
form, with the help of a self-consistent hybrid numerical model
incorporating both charged particles and photons code.

\section{Numerical Setup}
\label{sec:num_setup}

We use the same one-dimensional hybrid Particle-In-Cell/Monte Carlo
hybrid code described in \PapI{} modified for the space charge limited
flow regime.  Below we briefly describe the main equations, notations
and numerical algorithms; a detailed description can be found in
\PapI, \Ss2, 3.

We solve the evolutionary equation for the electric field $E$
\begin{equation}
  \label{eq:dE_dt}
  \frac{\partial E(x,t)}{\partial t} = -4\pi\left( j(x,t)-\jm \right)\,,
\end{equation}
where $j(x,t)$ is the actual current density and $\jm$ is the current
density imposed on the cascade zone by the magnetosphere.  This
equation is the Ampere's law, eq.~(\ref{eq:dE_par__dt}).  We are
solving an initial value problem, thus an initial distribution of the
electric field $E(x,t=0)$ must be supplied.  At the start of the
simulation we construct the initial distribution of the electric field
by solving the Gauss equation for the electric potential $\phi$
assuming some initial charge density distribution
$\eta_{\mathrm{start}}$
\begin{eqnarray}
  \frac{d^2\phi}{dx^2} &=& -4\pi(\eta_{\mathrm{start}}-\GJ{\eta})  \label{eq:d2Phi_dx2}
  \\
  E &=& -\frac{d\phi}{dx} \label{eq:E_dPhi_dx}
\end{eqnarray}
We proceed with the time integration of eq.~(\ref{eq:dE_dt}) using a
charge conserving scheme
\citep[e.g.][]{VillasenorBuneman92,Birdsall1985}, so the Gauss
equation is satisfied at each successive time step up to machine
precision.  The GJ charge density enters in the
equation~(\ref{eq:d2Phi_dx2}) for the initial configuration of the
electric field; this information is then ``carried on'' in time by
eq.~(\ref{eq:dE_dt}).

To model the space charge limited flow at every time step we inject
electrons and protons just outside the numerical domain used for
electric field calculation and let the system pull the necessary
amount of particles into that domain.  We \emph{do not} set $E(x=0,t)$
to zero as a boundary condition but rather allow the plasma in the
system to enforce this condition as part of the simulated physics.  A
detailed description of our algorithm for reproducing the space charge
limited flow condition at the NS surface is given in
Appendix~\ref{sec:bc_sclf}.  When pair creation cascades occur, we
take into account only curvature radiation as the gamma-ray emission
mechanism; pairs are created by single photon absorption in the strong
magnetic field \citep[e.g.][]{Erber1966}.

We performed many numerical experiments starting from different
initial conditions: (i) computational domain filled by plasma with
charge density equal, less, and higher than the GJ charge density as
well as starting with vacuum; (ii) different initial potential drop
over the domain; (iii) different length of the computation domain.  In
all cases without exception after initial relaxation on the time scale
of the order of the flyby time of the domain the system settled down
to a configuration which depends only on the imposed current
density $\jm$.

\section{Low Energy Charge Separated, sub-GJ Flow:
  $0<\jm/\GJ{\MakeLowercase{j}}<1$}
\label{sec:cold_flow}

\begin{figure*}
\begin{center}
  \includegraphics[width=\textwidth]{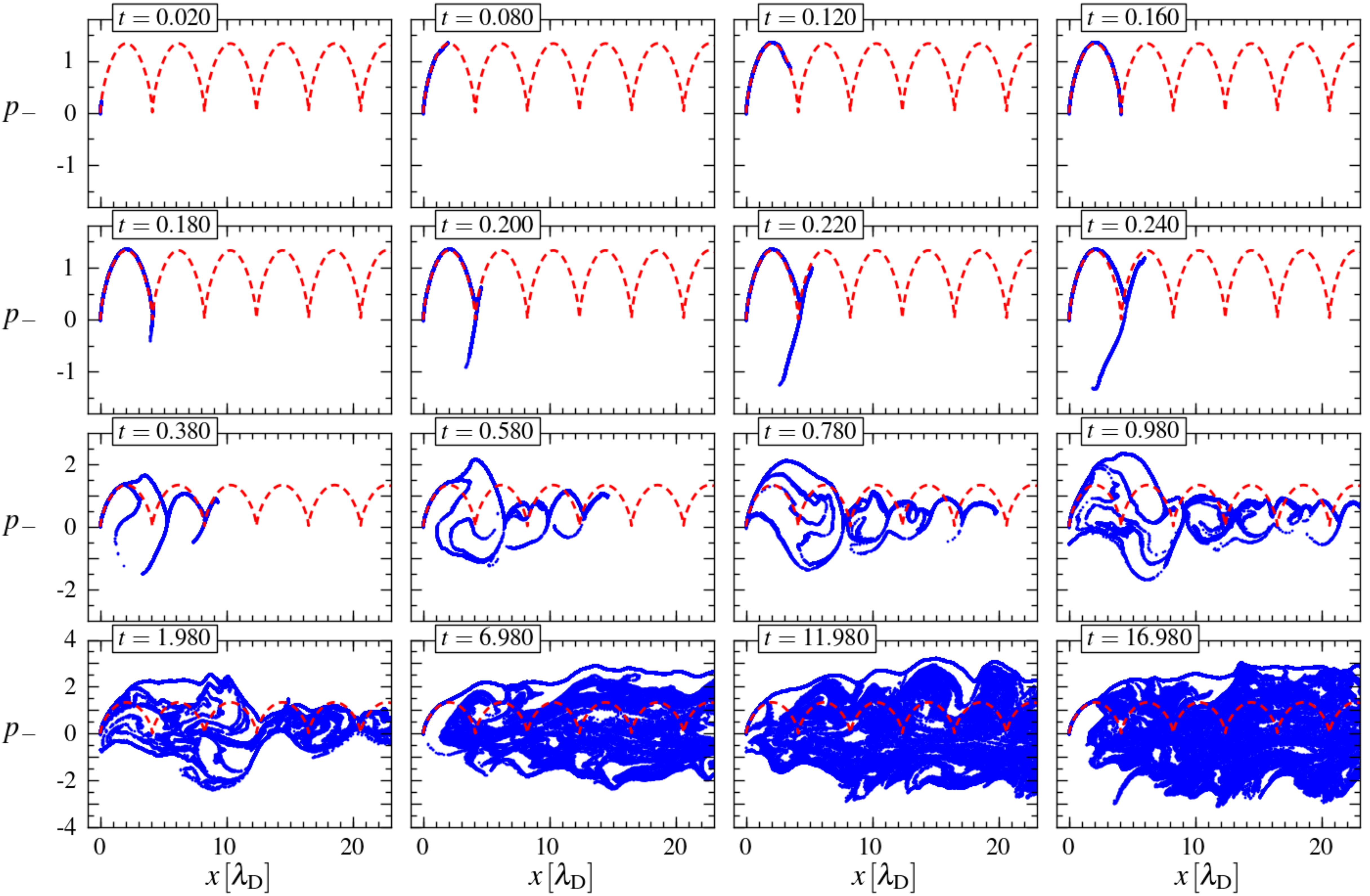}
\end{center}
\caption{Development of ``trapped particle'' flow when the space
  charge limited flow which starts into vacuum.  Phase space portrait
  of particles are shown for 16 moments of time indicated in small
  boxes on the top of each plot.  The current density $\jm=0.5\GJ{j}$.
  The distance $x$ is measured in units of the Debye length.  Red
  dashed lines show the analytical solution for stationary flow.
  Particle momenta $p_{-}$ are normalized to $m_ec$.  The total length
  of the computational domain $L=50\lambdaDGJ$, only part of it
  ($x<50\lambdaDGJ$) is shown here.  Time is measured in flyby time
  ($L/c$) of the whole domain.  Snapshots in the same row have the
  same time interval between them, but these time intervals are
  different for different rows, they increase towards the bottom row.}
  \label{fig:j0.5_timeseries}
\end{figure*}

\begin{figure*}
\begin{center}
  \includegraphics[width=\textwidth]{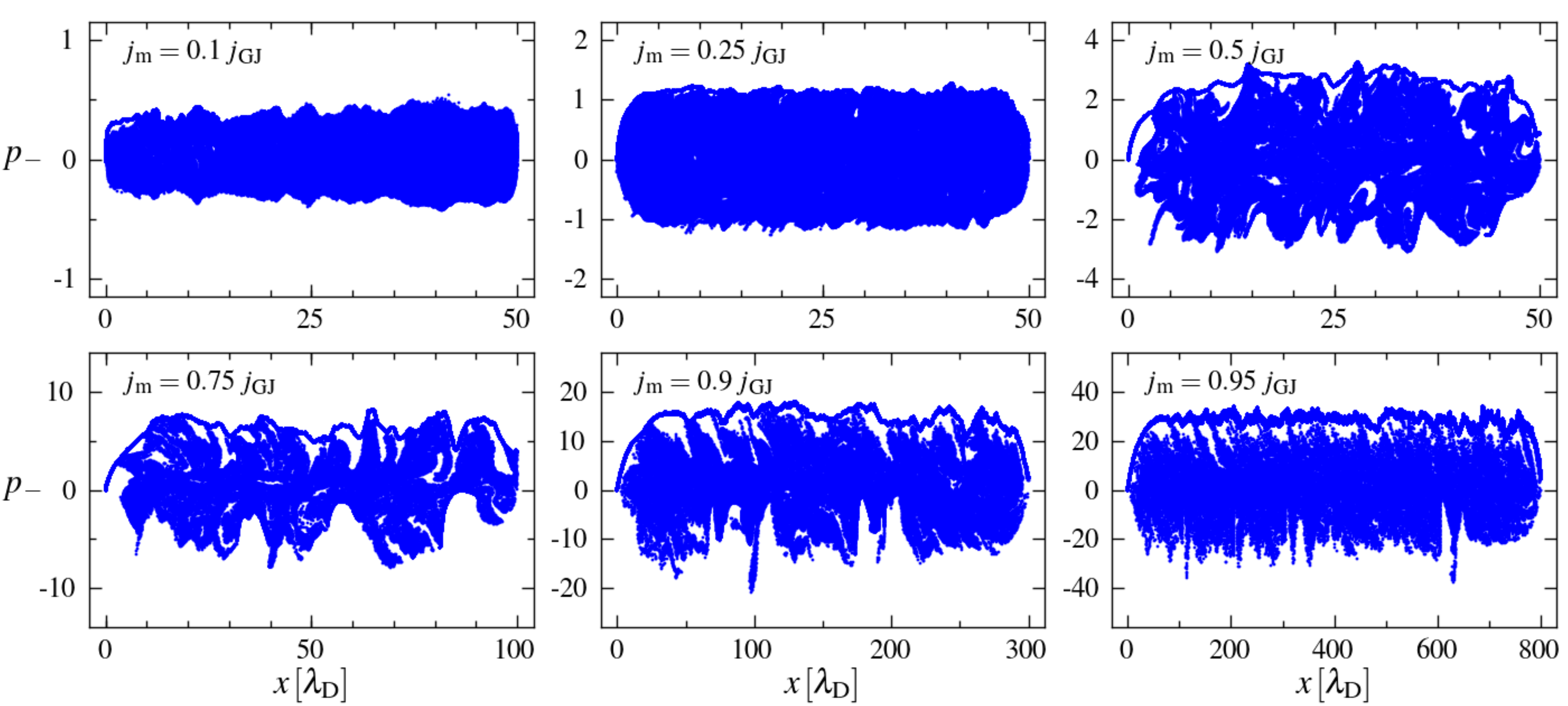}
\end{center}
\caption{Phase space portraits of well developed space charge limited
  flows for 6 different current densities: $\jm/\GJ{j}=0.1, 0.25, 0.5, 0.75,
  0.9, 0.95$.  Current density $\jm$ is indicated in the left upper
  part on each plot.  Distance is measured in units of $\lambdaD$,
  particle momenta $p_{-}$ are normalized to $m_ec$.  Note that the
  lengths of the computation domain differ between these plots.}
  \label{fig:xps_cold}
\end{figure*}

\begin{figure*}
\begin{center}
  \includegraphics[width=\textwidth]{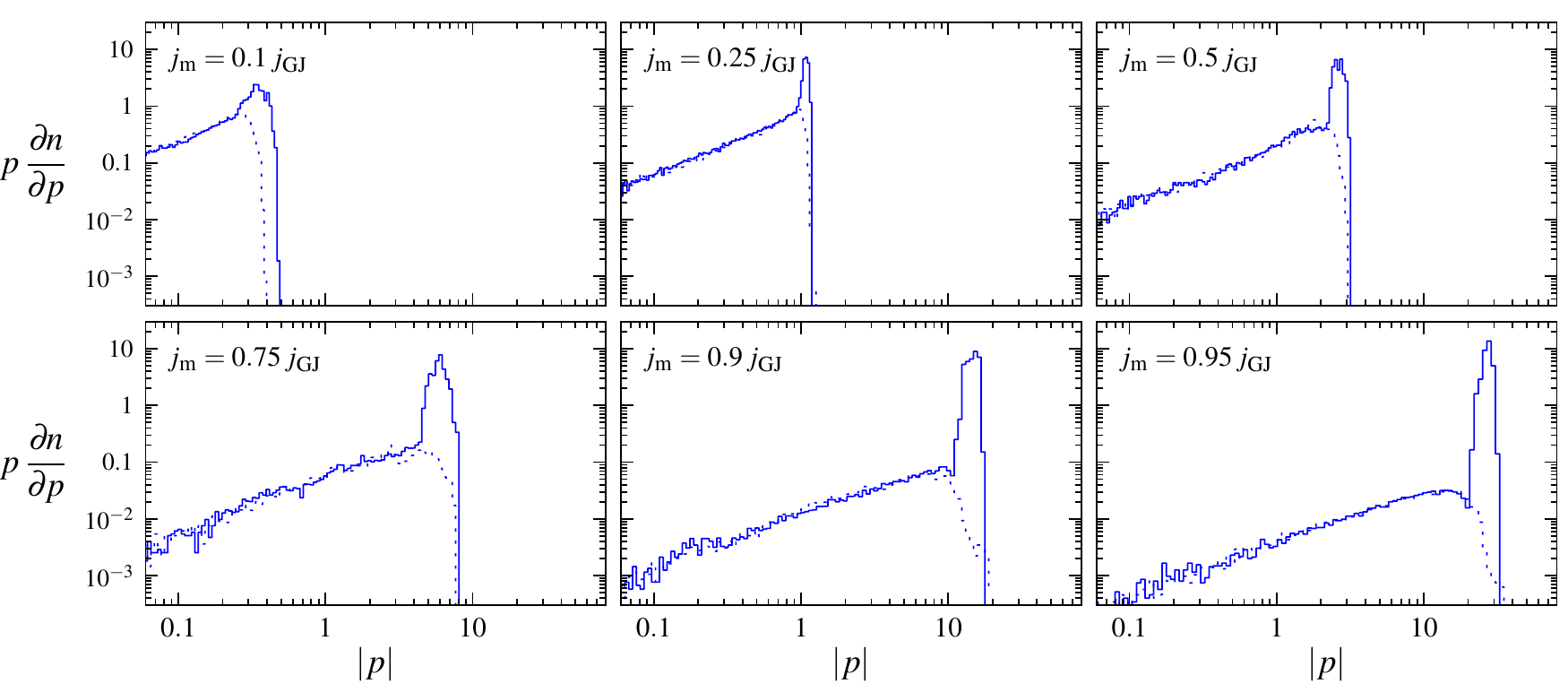}
\end{center}
\caption{Particle distribution functions $p\,(\partial{n}/\partial{p})$
  for well developed space charge limited flows from
  Fig.~\ref{fig:xps_cold}.  Distribution functions of particles with
  positive momenta (moving toward the magnetosphere) are shown by
  solid lines, distribution functions of particles with negative
  momenta (moving toward the NS) by dashed lines.}
  \label{fig:seds_cold}
\end{figure*}

\begin{figure*}
\begin{center}
  \includegraphics[width=\textwidth]{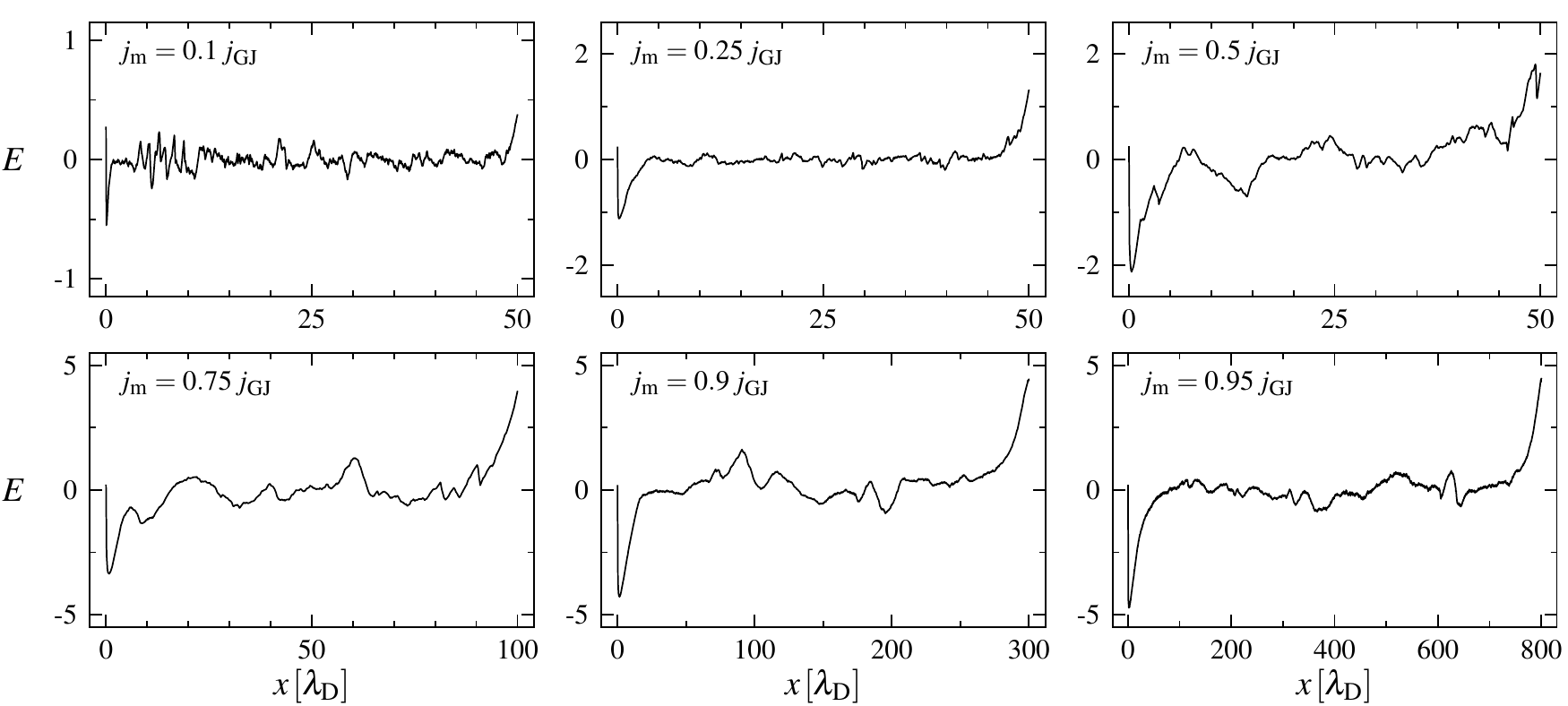}
\end{center}
\caption{Snapshots of electric field $E$ for well developed space
  charge limited flows taken at the same moment as the phase space
  portraits from Fig.~\ref{fig:xps_cold}. $E$ is normalized to
  $\pi\GJ{\eta}\lambdaDGJ$. }
\label{fig:e_acc_cold}
\end{figure*}

\begin{figure*}
\begin{center}
  \includegraphics[width=\textwidth]{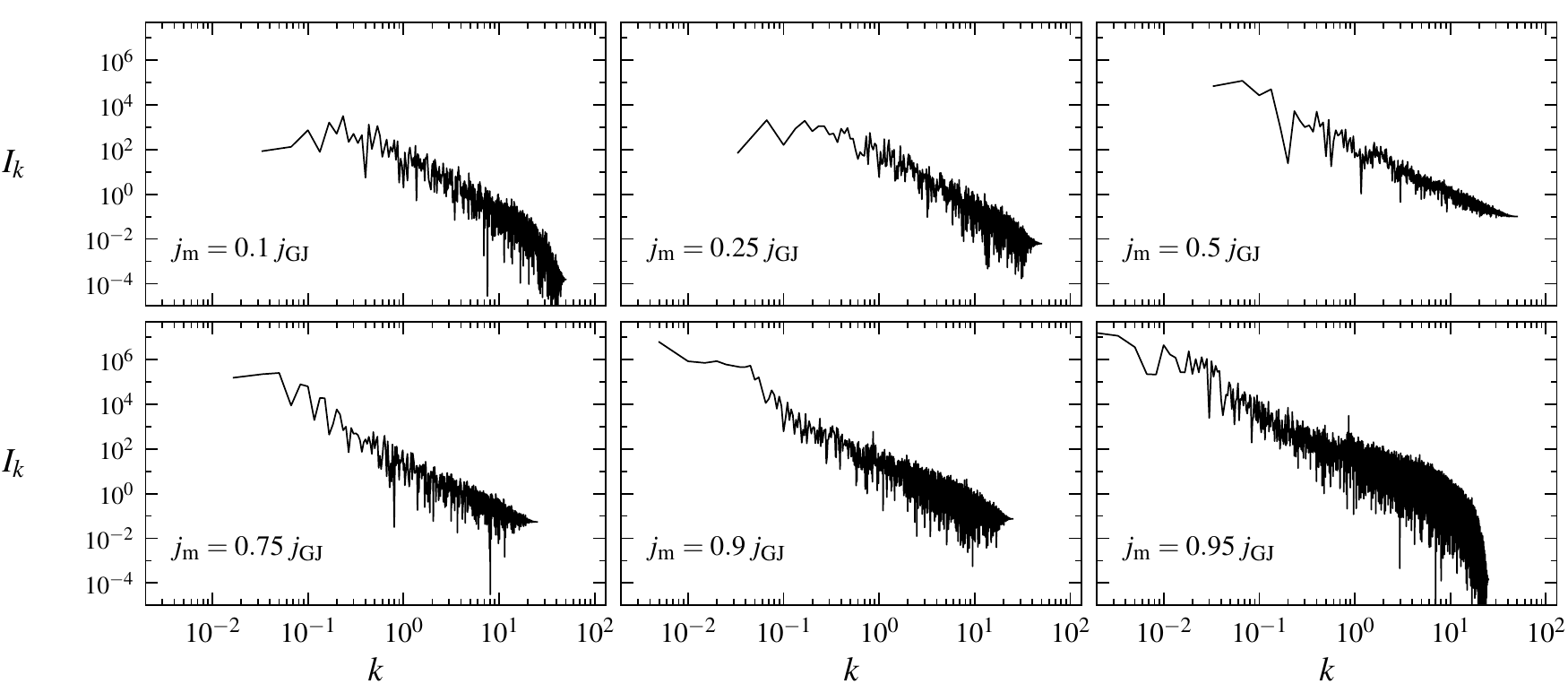}
\end{center}
\caption{Power spectra of the electric field $I_k=|E_k|^2$ for well
  developed space charge limited flows from
  Fig.~\ref{fig:xps_cold}. $k$ is normalized to the corresponding
  $1/\lambdaDGJ$. }
  \label{fig:ffts_cold}
\end{figure*}

As it turns out, there is no pair formation for $0<\jm/\GJ{j}<1$ and 
the only characteristic spatial scale of the flow is 
the Debye length $\lambdaDGJ$.  In this section we will discuss the
properties of such flow using simulations in domains with the length
$L$ up to several hundreds of $\lambdaDGJ$, which is much less than the
width of the polar cap.  Using such a small domain we well resolve the
characteristic spatial scale; increasing the domain length does not
change the results.

According to the stationary solution from \S\ref{sec:stationary_sclf}
there is no relativistic particle acceleration when $0<\jm/\GJ{j}<1$.
The flow is spatially oscillatory and the maximum momentum of
particles $p_{\max}$ is by far too small for emission of pair
producing photons.  It is unlikely that such oscillatory cold flow can
exist -- near the stagnation points it may be a subject to
``instability'' with characteristics of the wave breaking to which
nonlinear waves in cold fluids with velocity stagnation are subject,
which could destroy the spatial oscillation and create an effective
resistor, across which a substantial fraction of the perpendicular (to
$B$) voltage drop might appear. The available voltage drop across $B$
is huge and in principle it might happen that the system ends up in a
state with highly oscillatory electric field and bursts of pair
formation as predicted in the two-fluid plasma model of
\citet{Levinson05}.  On the other hand, finite amplitude electrostatic
waves containing similar stagnation points show wave breaking, with
alteration of cold to hot flow with momentum dispersion no more than
comparable to the flow velocities found in the originally constructed
cold oscillation \citep{AkhiezerPolovin1956, TajimaDawson1979}. That
would create a warm but non-relativistic or mildly relativistic
charge-separated outflow.

We find that in the sub-GJ regime, the non-neutral flow is indeed
low-energy, with particle energies orders of magnitude below the
energy/particle required for pair production.  However, the final
state differs drastically from the oscillatory flow from the
stationary solution shown in
Figs.~\ref{fig:posc},~\ref{fig:posc_pacc}.  Even if at the beginning
of the simulations particles follow trajectories of the stationary
solution, the standing non-linear wave structure breaks quickly -
particles at the velocity zeros of the cold flow go both up and down.

A good example of this inherent instability is shown in
Fig.~\ref{fig:j0.5_timeseries}.  We start from a vacuum configuration
-- when there are no particles in the domain -- and let the system
evolve.  In Fig.~\ref{fig:j0.5_timeseries} we show snapshots of the
phase space portraits of the flow with $\jm=0.5\GJ{j}$, the distance
from the NS, normalized to $\lambdaDGJ$, is along $x$-axis, particle
momenta, normalized to $m_ec$, are along $y$-axis.  The whole domain
has the length $L=50\lambdaDGJ$ and we show only a part of it here.
Time $t$ is measured in flyby time of the whole domain $L/c$.  With
the dashed red line we show phase space trajectories of particles from
stationary solution.  Particles coming from the surface at first
follow the trajectories of the stationary solution.  However, after
coming to the first stagnation points some of the particles are turned
back and the flow start to randomize.  After several tens of plasma
periods $\lambdaDGJ/c$ the flow reaches its final configuration.

Examples of final configurations for space charge limited flow with
different current densities are shown in Fig.~\ref{fig:xps_cold},
where we plot phase space portrait of particles in the whole
computation domain.  The flow has two components: a warm beam of
particles with highest momentum which produce the required current
density and a cloud of charged particles circulating in the domain --
these compose an electrically trapped, ``thermal'' component.  In the
cloud component there is roughly equal number of particles moving in
opposite directions, these particles do not contribute to the current
but contribute $\GJ{\eta}-\jm/c$ to the charge density keeping
$\eta_{\mathrm{total}}$ equal to the GJ charge density.  The
distinction between these components is not absolute as some particles
from the beam go into the cloud and vise versa, although the fraction
of mixing particles is small.  Some of the particles in the cloud have
very low momenta and so they can adjust to any given charge density.
Hence, in sub-GJ space charge limited flow, $0\le\jm/\GJ{j}<1$, the
electric field is not sensitive to variation of the GJ charge density%
\footnote{We performed simulations for $\jm/\GJ{j}<1$ with variable GJ
  charge density, and, as expected saw the electric field to be just
  as screened as in the uniform GJ density case.} -- in contrast to
the large importance of variation of the GJ charge density for
relativistic acceleration of space charge limited cold beams in the
polar cap cascade models of \citet{AronsScharlemann1979} and
\citet{Muslimov/Tsygan92}.

Plasma flow in the sub-GJ regime can be described as a beam of mildly
relativistic particles propagating through a cloud of trapped
particles with near-thermal distribution.  In Fig.~\ref{fig:seds_cold}
we plot particle distribution functions.  The beam component is
visible as a bump on the distribution function at the high momentum
side.  The cloud component has a near-thermal (Maxwell-Juttner)
momentum distribution (at least in its low-energy part)
$\partial{n}/\partial{p}\propto{p}$ -- such quasi-thermalization is a
common consequence of the phase mixing between the particles and
fields built into the wave breaking process.

\begin{figure}
  \begin{center}
    \includegraphics[width=\columnwidth]{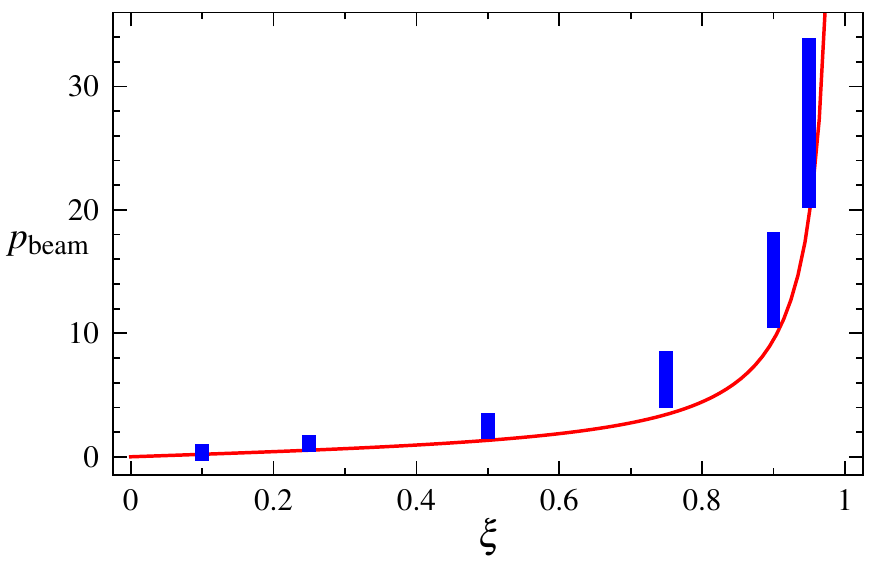}
  \end{center}
  \caption{Momenta of the current currying particle beam.  The widths
    of the peaks at Fig.~\ref{fig:seds_cold} as the function of
    $\xi=\jm/\GJ{j}$ are shown by the blue bars.  The solid red line
    is the maximum achieved momentum for the stationary SCLF solution.}
  \label{fig:beam}
\end{figure}

When particles leave the NS they are non-relativistic and their charge
density $\eta=\jm/v=\xi\GJ{j}/v$ is larger than the GJ charge density
and so they form a charge sheet near the surface generating
accelerating electric field, just as in the idealized stationary case
(see Fig.~\ref{fig:eta_osc_acc}).  When particles reach the velocity
such that $|\jm|/v<|\GJ{\eta}|$ the electric field derivative $dE/ds$
changes sign and after some distance the electric field can start
decelerating particles.  At the point where $E=0$ particles reach
their maximum momentum.  Above the gap particles from the cloud
component add additional charge and so the change density there is
equal to $\GJ{\eta}$ and the electric field is screened.  The length
of this gap is the order of $s_{\max}=s_0/2$ (half the spatial
  period of cold flow oscillations) and so the maximum momentum
particles gain in this gap is comparable to $p_{\max}$ (both
$s_{\max}$ and $p_{\max}$ depend on $\jm$).  In Fig.~\ref{fig:beam} we
plot momenta of particles in the beam component as functions of
$\xi=\jm/\GJ{j}$ superimposed of the theoretical dependence of
$p_{\max}$ given by eq.~(\ref{eq:p_max}); the agreement is pretty
good.  The current density in the final configuration is close to
$\jm$ throughout the domain, with small fluctuations around this value
$\delta{j}\ll\jm$.

In Fig.~\ref{fig:e_acc_cold} we plot electric field in the calculation
domain at the same moments of time as the phase space portraits shown
in Fig.~\ref{fig:xps_cold}. The electric field at any given point
fluctuates, but the relative fluctuation at the beginning and at the
end of the domain (accelerating and decelerating regions, see below)
are much smaller than those in the center of the domain (the region of
the charge cloud).  The electric field in the gap near the NS launches
the beam component.  The electric field at the other end of the
computation domain supports the cloud components by reversing the
momenta of most of the particles in the cloud moving away from the NS,
sending them back.  This electric field is not strong enough to
reflect most of the beam particles.  The mixing between the beam and
cloud components is the strongest here.  This electric field appears
self-consistently when the flow reaches its steady configuration%
\footnote{The \emph{formation} of the cloud component is not linked to
  the appearance of the electric field at the end of the domain, see
  e.g. Fig.~\ref{fig:j0.5_timeseries}}, because it is needed to
sustain the cloud component necessary to match both the charge and the
current density in the domain. In our small scale 1D simulations we
impose the current density on the domain, which finally gives rise to
the electric field at the right end of the domain.  In reality it is
the magnetosphere which sets the current density by twisting the
magnetic field lines and generating electric field reversing some of
the particles.  This second region with unscreened electric field at
the magnetosphere end of the domain in our model may be a
``compressed'' version of some parts of the outer magnetosphere.  For
example, on field lines that pass through the null surface where $
\vec{\Omega} \cdot \vec{B} = 0$, the plasma cloud and beam, composed
of only one sign of charge, cannot freely enter the outer
magnetosphere \citep{Scharlemann/Arons/Fawley/78} in the absence of
other sources of electric field in other parts of the magnetosphere
(Goldreich \& Julian's ``hanging charge clouds''), offering the
possibility of opening a vacuum gap within the otherwise force-free
structure. On there other hand, on polar field lines that never cross
the null surface -- most of them, in the aligned rotator -- the charge
separated beam and cloud can extend outwards ``forever'', in
principle.  Multi-dimensional particle simulations, analogous to those
of \citet{SpitkovskyArons2002}, are required to see if indeed this
speculation is true (as well as determine how this essentially 1D
model might fit together with the other, more ``lively'' aspects of
the polar flow outlined in \S\ref{sec:j_pairs}).

In Fig.~\ref{fig:ffts_cold} we show power spectra of the fluctuating
electric field in the central parts of the calculation domain, outside
of the acceleration zones.  $I_k=|E_k|^2$, where $E_k$ is the spatial
Fourier amplitude of the electric field and the wave vector $k$ is
normalized to the $\lambdaDGJ^{-1}$.  These spectra can be fit with
the power law $I_k\propto{}k^{-\alpha}$ with $\alpha$ between 2 and 3
for all current densities.

Thus, along magnetic field lines where the current density is sub-GJ
there is no pair formation, so long as strong electric fields in
neighboring, more active flow zones do not leak into the cloud.  In an
aligned rotator, for example, no pairs are forming above most of the
polar cap area.  Plasma flowing along such magnetic field lines is
mildly relativistic, consists of particles of only one sign
(electrons) and its density is low, equal to the GJ number density.
However, observational evidence for ongoing pair formation in pulsars
is very strong and, hence, there must be regions in the pulsar
magnetosphere where electron-positron plasma is created.

\section{Plasma flow with pair formation - super-GJ and return current
  regions}
\label{sec:j_pairs}

Let us now consider what happens along magnetic field lines where the
current density has either (i) the opposite sign to the GJ current
density, $j/\GJ{j}<0$, anti-GJ flow, or (ii) the same sign and
absolute value larger than the GJ current density, $j/\GJ{j}>1$,
super-GJ flow.  Case (i) includes the regions of return current,
including the current sheet, while case (ii) is relevant for most of
the magnetic field lines in a nearly-orthogonal rotator.  As we show
in this section, in both of these cases a strong accelerating electric
field is generated.  When the resulting potential drop is sufficient
to accelerate particles up to the energies such that they can emit
photons that convert to pairs within the accelerator, the plasma flow
will be highly-nonstationary with intermittent pair creation.

In a real pulsar the magnetic field strength falls with the distance (
$\propto r^{-3}$ for a dipole field) and pair creation is possible
only at sufficiently low altitude (if gamma ray interaction with
thermal low energy photons is neglected).  In order to imitate the
effect of pair creation attenuation with the distance we set the
magnetic field strength to zero starting at distance $x_B$ from the NS
-- the magnetic field is given by
\begin{equation}
  \label{eq:B_model}
  B(x) = \left\{
  \begin{array}{rl}
    B_0,& \mathrm{if} \, x\le{} x_B \\
    0, & \mathrm{if} \,  x> x_B.
  \end{array}
  \right.
\end{equation}

If the charge density were formed from steadily flowing relativistic
beams, it would vary with the height as $\eta(x)\propto{}B(x)$. The GJ
charge density changes in a slightly different way,
$\eta(x)\propto{}B(x)f(x)$ due to inertial frame dragging
\citep{Muslimov/Tsygan92,Beskin1990} or/and field line curvature
\citep{Scharlemann/Arons/Fawley/78}.  If one neglects the latter
effects, scaling $\propto{B(x)}$ can be incorporated into the spatial
(for eq.~(\ref{eq:d2Phi_dx2})) or temporal (for eq.~(\ref{eq:dE_dt}))
coordinates.  The the electrodynamics of the cascade zone can be
modeled in a 1D problem with constant GJ charge density; the only
effect of the GJ charge density variation will be in changing the
spatial (and temporal) scales.  The same scaled 1D model can also be
used to study the effects on the electrodynamics of the cascade zone
of deviation of the GJ charge density scaling from being
$\propto{B(x)}$ by considering a problem where \GJ{\eta} depends on
the distance; the variation of the GJ charge density will be given by
$f(x)$ as the dependence on $B(x)$ is already incorporated in the
model.  First, in \Ss\ref{sec:j_negative},~\ref{sec:j_greater}, we
consider the case when the GJ charge density is constant, i.e. this
case corresponds to a model where variation of $\GJ{\eta}/B$ is
neglected.  Then in \S\ref{sec:j_negative_rho_variable} we address the
influence of the GJ charge density variation on the physics of the
cascade zone.

While simulations of space charge limited flow with sub-GJ current
densities described in \S \ref{sec:cold_flow} can be considered as
directly related to more complete pulsar models -- the distance over
which wave breaking and trapping control the flow is much smaller then
the width of the polar cap, and the 1D approximation is well motivated
-- the pair creation models presented in this section can be
considered only as illustrative of the physics but not fully
applicable to pulsars.  The spatial scales over which pair creating
photons are absorbed, even when they are emitted at very low altitude
({\emph e.g.} the attenuation length of the magnetic field) are much
larger that the width of the polar cap, making transverse structure
essential for modeling the pulsar environment -- such effects are
necessary for a full evaluation of the pair yield, since much of the
pair creation occurs in regions beyond the acceleration zone.
Nevertheless, from our 1D models we obtain insight into the basic
cascades physics in a regime never previously modeled.
Multi-dimensional models will be considered elsewhere.

Within the context of the 1D model, the domain length $L$, and the
value of $\xb$ (the height above which the magnetic field is too weak
to support pair creation) are the parameters having the largest
departure from what would appear in multi-D context.  These lengths in
our simulations are much smaller that in real pulsars, but that choice
allowed us to construct models utilizing reasonable amounts of CPU
time and to explore a broad parameter space.  $L$ and $\xb$ were
chosen in such a way that the minimum size of the accelerating region
necessary to start pair creation is at least 2-3 times less than
$\xb$.

We performed numerous simulations for different initial conditions and
physical parameters in order to study qualitative behavior of
cascades.  We performed simulations with different values of the
numerical parameters (spatial resolution, time steps, particle
injection rate, number of particles per cell) in order to check the
numerical convergence.  In all physical cases presented in this
section plasma flow is quasi-periodic, and this behavior does not
depend on initial conditions -- after a short relaxation time,
comparable to the flyby time of a relativistic particle through the
computational domain, the system settles down to a limit cycle sort of
behavior.  We describe here a particular set of simulations which is
representative for all other models.  In
\Ss\ref{sec:j_negative},~\ref{sec:j_greater} we use simulations with
the following physical parameters: the length of the domain
$L=2.4\times10^4$~cm, the potential drop in vacuum across the domain
$\Delta{V}=10^{14}$~Volts\footnote{Such vacuum potential drops over
  the domains of this size are realistic in young, high magnetospheric
  voltage pulsars, $\Vm>10^{15}$~V.  This choice allows studying pair
  formation over distances small compared to the polar flux time
  diameter. In more common pulsars with $\Vm<10^{13}$~V, pair
  formation happens over (much) larger distances. However, as we
  mentioned before, our simulations represent a toy model addressing
  the general behavior of the polar cap acceleration zone and our
  choice of parameters is motivated by convenience of simulations,
  rather than by attempt to model a real
  pulsar.\label{fn:large_voltage}},
the radius of curvature of the magnetic field lines with respect to a
photon orbit $\rho_c=10^6$~cm $\sim $ the stellar radius (small
compared to the pure star centered dipole value $\sqrt{R_*
  \RLC}\sim10^{7.8}P^{1/2}$~cm but easily attained in offset dipole
geometry (\citealt{Arons1998, Harding2011}; Arons, in preparation);
magnetic field strength $B_0=10^{12}$~G.  The distance $x_B$ marking
the transition to the outer magnetosphere with small magnetic field is
set to $x_B=0.7L$.  The particular simulations described in these two
sections differ only in the imposed current density $\jm$.

We illustrate the flow dynamics with series of snapshots for different
physical quantities shown in
Figs.~\ref{fig:ctss_jp05__1}-\ref{fig:ctss_jp15__3},
\ref{fig:ctss_jm15__1}-\ref{fig:ctss_jm15__3}.  In those figures each
column shows detailed information about physical conditions in the
computation domain at a given moment of time: the number densities of
electrons and positrons $n_\pm$ (shown as charge densities of
electrons and positrons $\eta_\pm$, $n_\pm=|\eta_\pm|$), total charge
density $\eta$, current density $j$, the accelerating electric field
$E$, phase portraits of electrons, positrons, and pair producing
photons, and (in Figs.~\ref{fig:ctss_jp15__1}-\ref{fig:ctss_jp15__3})
-- protons.  Particles with positive values of the 4-velocity $p$ are
those which move away from the NS (toward higher altitude), particles
with negative $p$ move toward the NS.  These plots are similar to ones
in \PapI, with the only difference that now we use semi-logarithmic
scale for particle momenta (linear for $-5<p<5$ and logarithmic
everywhere else) on phase space portraits that show dynamics of high
and low energy particles on the same plot.

The number density, charge density, and the current density are
normalized to the corresponding GJ values: $\eta_\pm$ and $\eta$ are
normalized to $|\GJ{\eta}^0|$,  $j$ to $|\GJ{\eta}^0|c$, where
$\GJ{\eta}^0$ is the GJ charge density at the NS surface (the
distinction between $\GJ{\eta}$ and $\GJ{\eta}^0$ will be important in
\Ss\ref{sec:j_negative_rho_variable},~\ref{sec:stationary-cascades}).
The electric field is normalized to $E_0\equiv|\GJ{\eta}^0|\pi{}L$.  The
distance $x$ on those plots is normalized to $L$, much larger than the
Debye length $\lambdaDGJ$.  The time $t$ is normalized to the
relativistic flyby time of the computational domain $L/c=0.8\,\mu$sec
for the parameters chosen.  The time is counted from the start of a
particular simulation, so only time intervals between the snapshots
have physical meaning.

In all cases plasma flow and pair formation has limit cycle behavior.
In each case we illustrate this behavior by 3 series of snapshots
taken within one typical cycle.  These 3 series show 3 phases of
plasma flow: cascade ignition (Figs.~\ref{fig:ctss_jp05__1},
\ref{fig:ctss_jp15__1}, \ref{fig:ctss_jm15__1}), development of the
cascade (Figs.~\ref{fig:ctss_jp05__2}, \ref{fig:ctss_jp15__2},
\ref{fig:ctss_jm15__2}), and filling the domain with dense pair plasma
(Figs.~\ref{fig:ctss_jp05__3}, \ref{fig:ctss_jp15__3},
\ref{fig:ctss_jm15__3}).  On each figure time intervals between
snapshots are equal, but these intervals are different on different
figures.

\subsection{Flow with $\jm/\GJ{j}<0$}
\label{sec:j_negative}

\begin{figure*}
  \begin{center}
    \includegraphics[width=\textwidth]{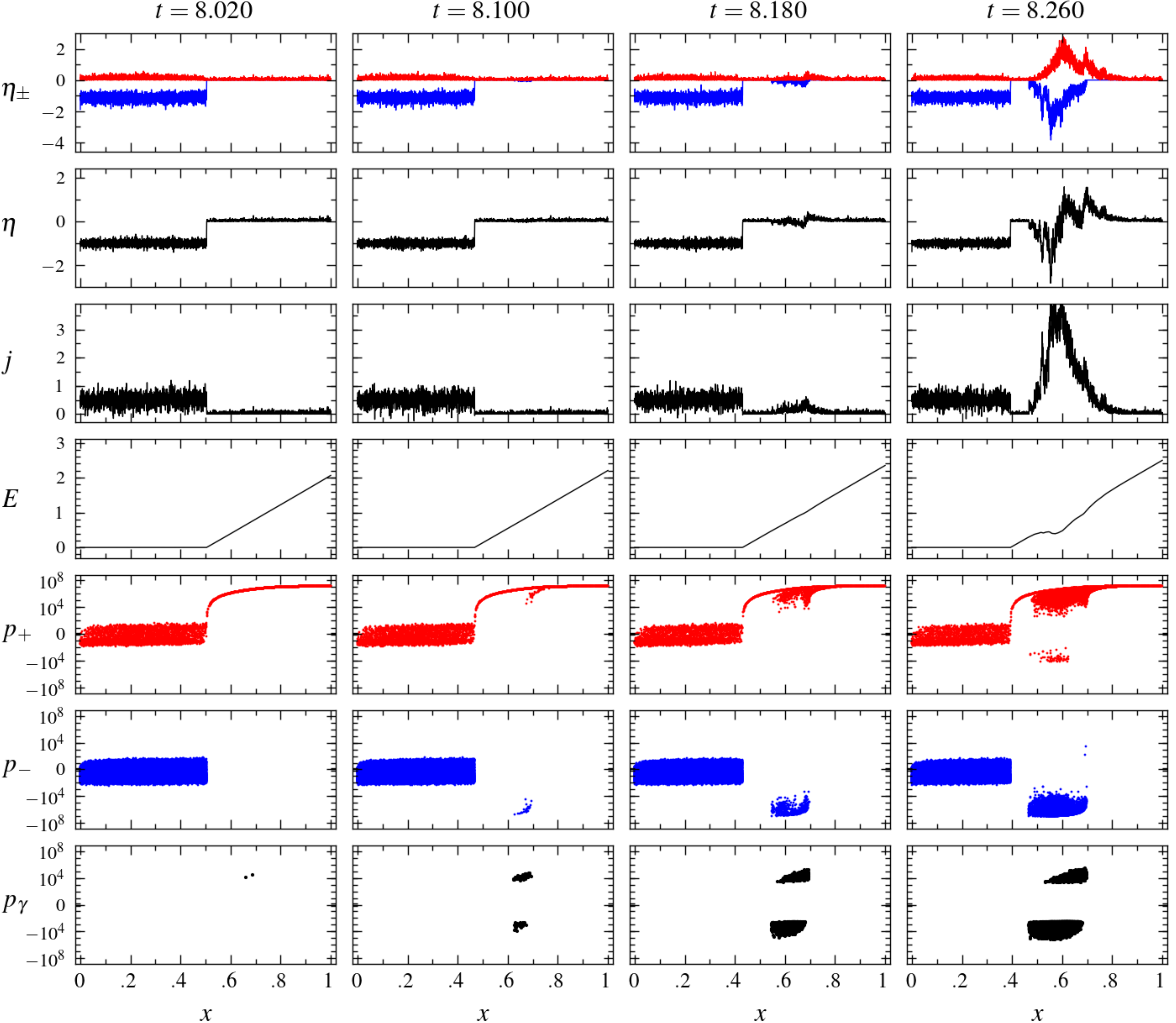}
  \end{center}
  \caption{Ignition of pair formation in anti-GJ flow with
    $\jm=-0.5\GJ{j}$.  Several physical quantities are shown as
    functions of the distance $x$ from the NS; $x$ is normalized to
    the domain size $L$.  Plots in each column
    (for the same time $t$) are aligned -- they share the same values
    of $x$.  The following quantities are plotted: %
    \textbf{$\mathbf 1^{st}$ row:} $\eta_{\pm}$ -- charge density of
    electrons (negative values, blue line) and positrons (positive
    values, red line); $\eta_{\pm}$ is normalized to the absolute
    value of the Goldreich-Julian charge density $|\GJ{\eta}|$.  %
    \textbf{$\mathbf 2^{nd}$ row:} the total charge density $\eta$
    normalized to the absolute value of the Goldreich-Julian charge
    density $|\GJ{\eta}|$.  %
    \textbf{$\mathbf 3^{nd}$ row:} current density $j$ normalized to
    the absolute value of the Goldreich-Julian current density
    $|\GJ{j}|\equiv|\GJ{\eta}c|$.  %
    \textbf{$\mathbf 4^{rd}$ row:} accelerating electric field $E$
    normalized to the ``vacuum'' electric field $E_0\equiv|\GJ{\eta}|\pi{}L$.  %
    \textbf{$\mathbf 5^{th}$ row:} phase space portrait of positrons
    (horizontal axis -- positron position $x$, vertical axis --
    positron momentum $p_{+}$ normalized to $m_ec$). The vertical axis
    is logarithmic except for the region around zero momentum
    ($-5<p_{+}<5$), where the scale is linear.  %
    \textbf{$\mathbf 6^{th}$ row:} phase space portrait of electrons
    (horizontal axis -- electron position $x$, vertical axis --
    electron momentum $p_{-}$ normalized to $m_ec$).
    \textbf{$\mathbf 7^{th}$ row:} phase space portrait of
    pair-producing photons (horizontal axis -- photon position $x$,
    vertical axis -- photon momentum $p_{\gamma}$ normalized to
    $m_ec$). }
  \label{fig:ctss_jp05__1}
\end{figure*}

\begin{figure*}
  \begin{center}
    \includegraphics[width=\textwidth]{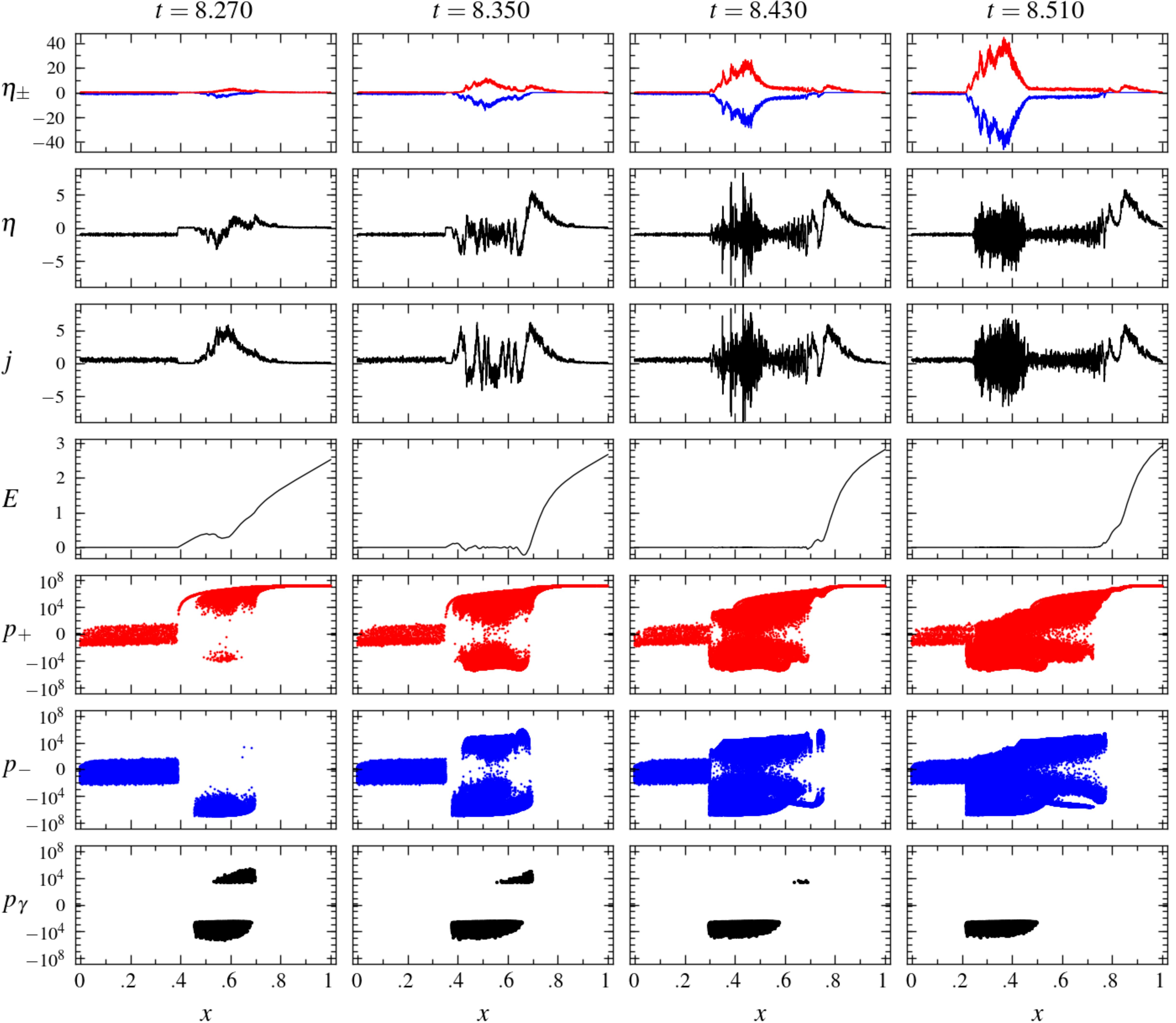}
  \end{center}
  \caption{Screening of the electric field in anti-GJ flow with
    $\jm=-0.5\GJ{j}$.  The same quantities are plotted as in
    Fig.~\ref{fig:ctss_jp05__1}.}
  \label{fig:ctss_jp05__2}
\end{figure*}

\begin{figure*}
  \begin{center}
    \includegraphics[width=\textwidth]{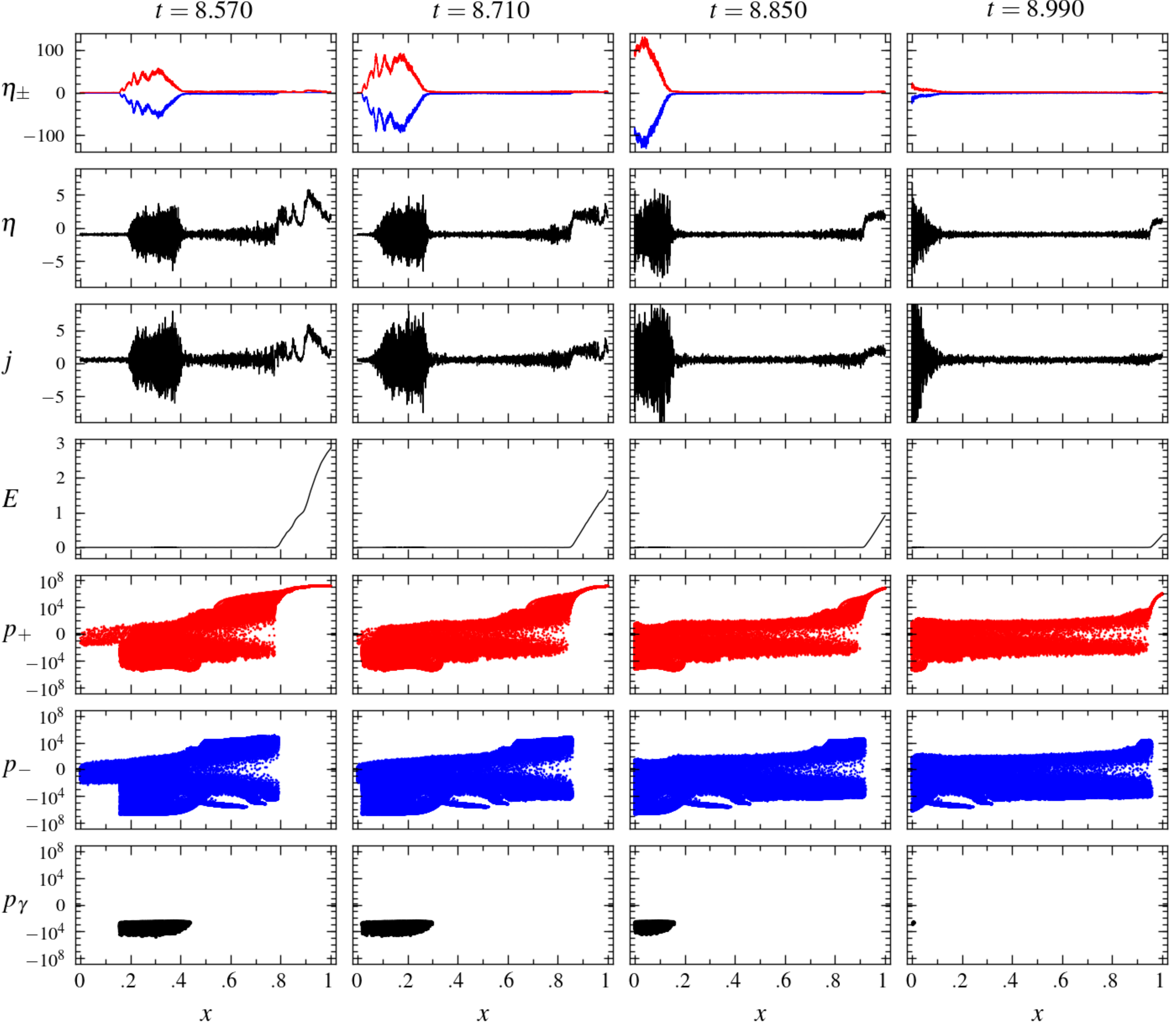}
  \end{center}
  \caption{Filling of computation domain with dense pair plasma in
    anti-GJ flow with $\jm=-0.5\GJ{j}$.  The same quantities are
    plotted as in Fig.~\ref{fig:ctss_jp05__1}.}
  \label{fig:ctss_jp05__3}
\end{figure*}

\begin{figure*}
  \begin{center}
    \includegraphics[width=\textwidth]{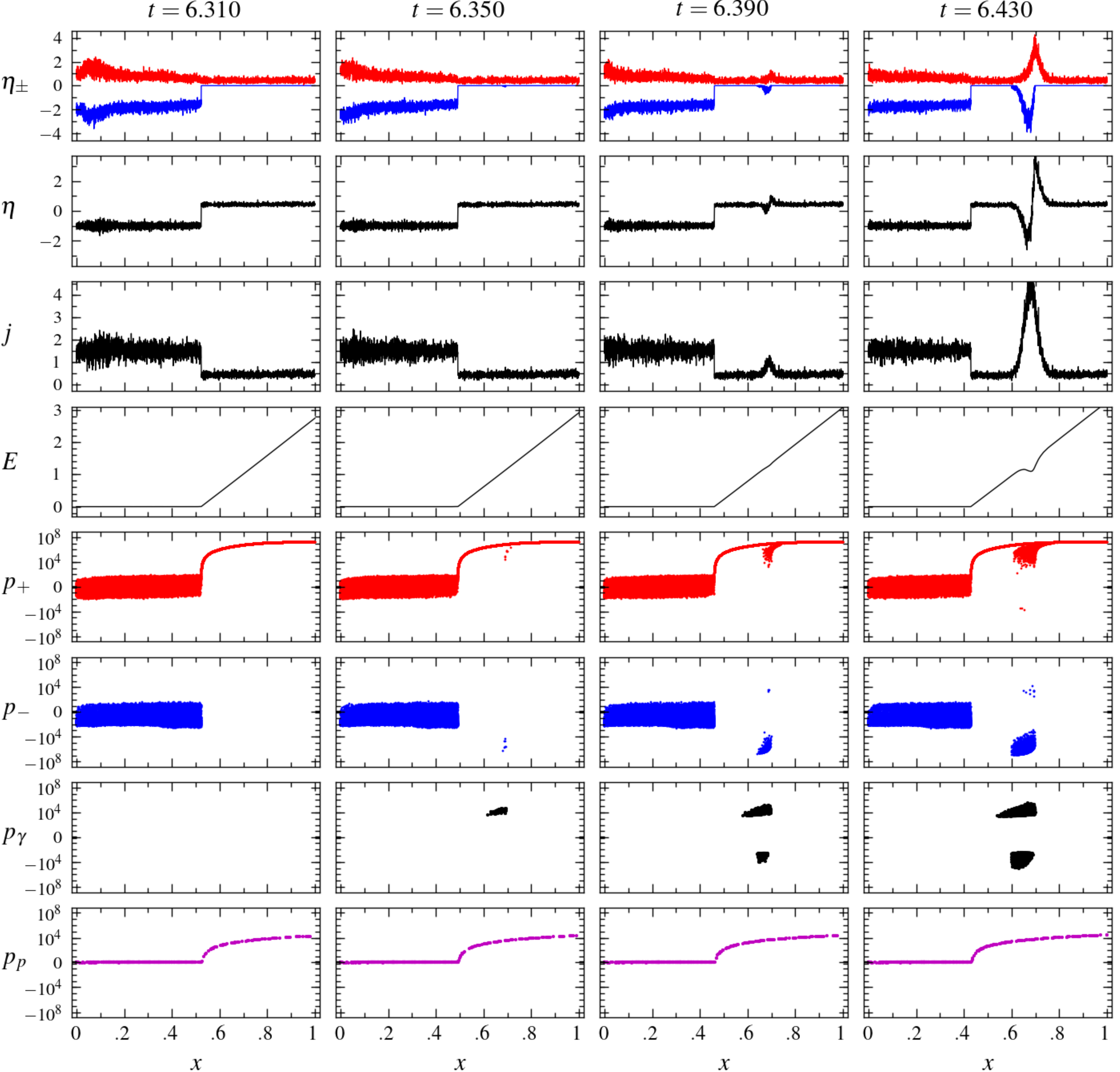}
  \end{center}
  \caption{Ignition of pair formation in anti-GJ flow with
    $\jm=-1.5\GJ{j}$.  The same quantities are plotted as in
    Fig.~\ref{fig:ctss_jp05__1} with addition of the phase space
    portraits for protons in \textbf{$\mathbf 8^{th}$ row:}
    (horizontal axis -- proton position $x$, vertical axis -- proton
    momentum $p_p$ normalized to $m_pc$).  }
  \label{fig:ctss_jp15__1}
\end{figure*}

\begin{figure*}
  \begin{center}
    \includegraphics[width=\textwidth]{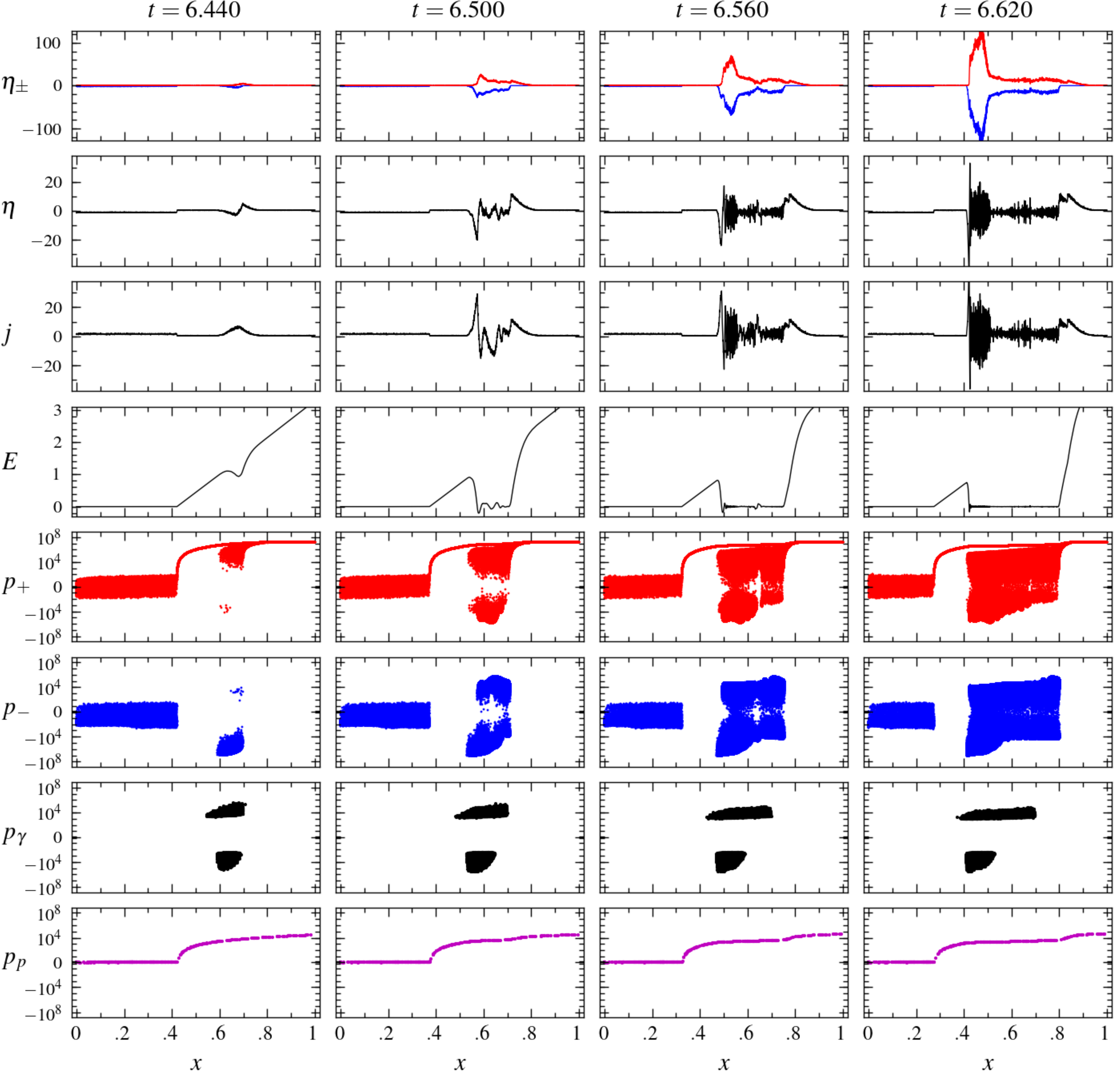}
  \end{center}
  \caption{Screening of the electric field in anti-GJ flow with
    $\jm=-1.5\GJ{j}$.  The same quantities are plotted as in
    Fig.~\ref{fig:ctss_jp15__1}.}
  \label{fig:ctss_jp15__2}
\end{figure*}

\begin{figure*}
  \begin{center}
    \includegraphics[width=\textwidth]{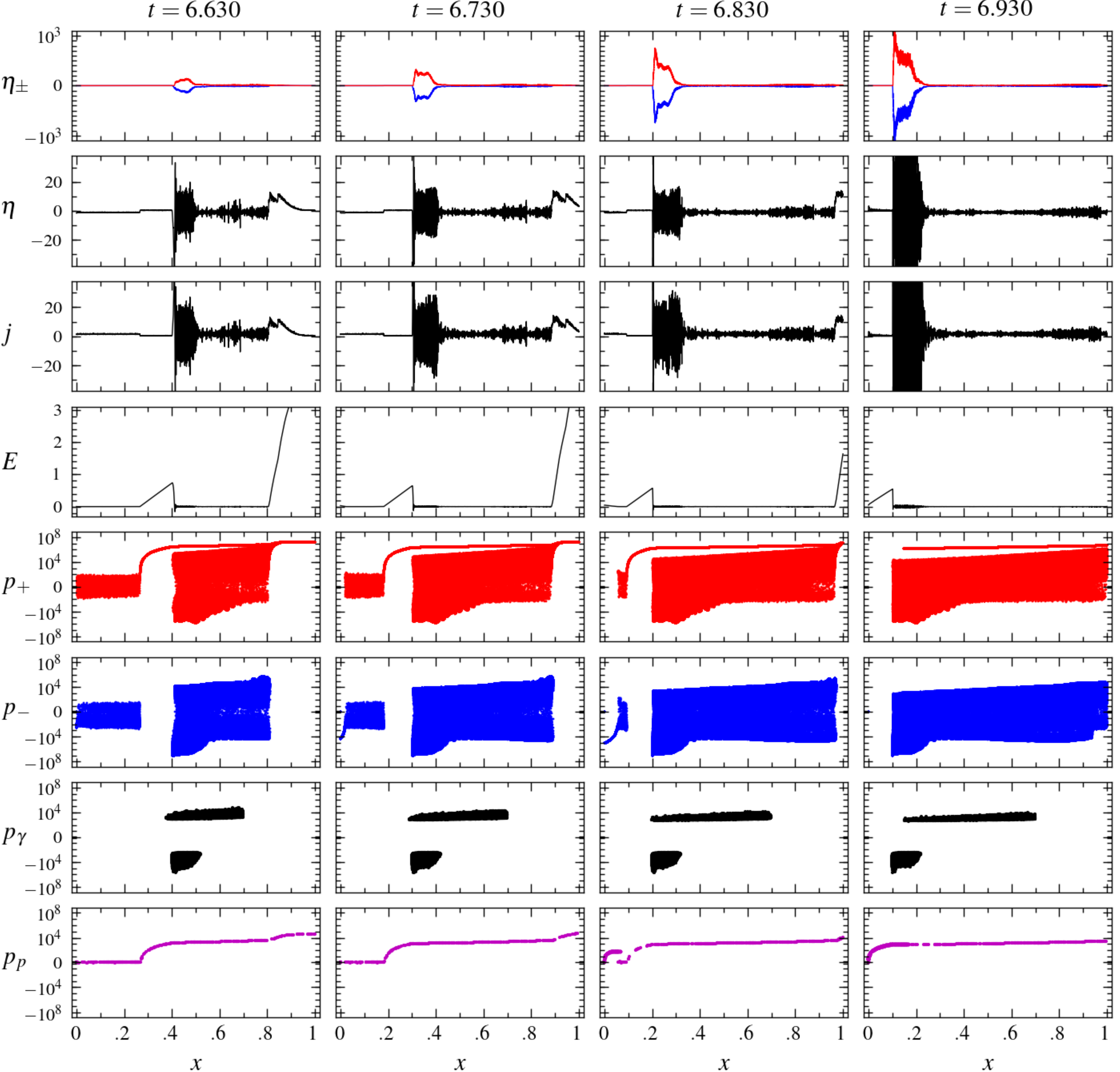}
  \end{center}
  \caption{Filling of computation domain with dense pair plasma in
    anti-GJ flow with $\jm=-1.5\GJ{j}$.  The same quantities are
    plotted as in Fig.~\ref{fig:ctss_jp15__1}.  }
  \label{fig:ctss_jp15__3}
\end{figure*}

\begin{figure*}
\begin{center}
  \includegraphics[width=\textwidth]{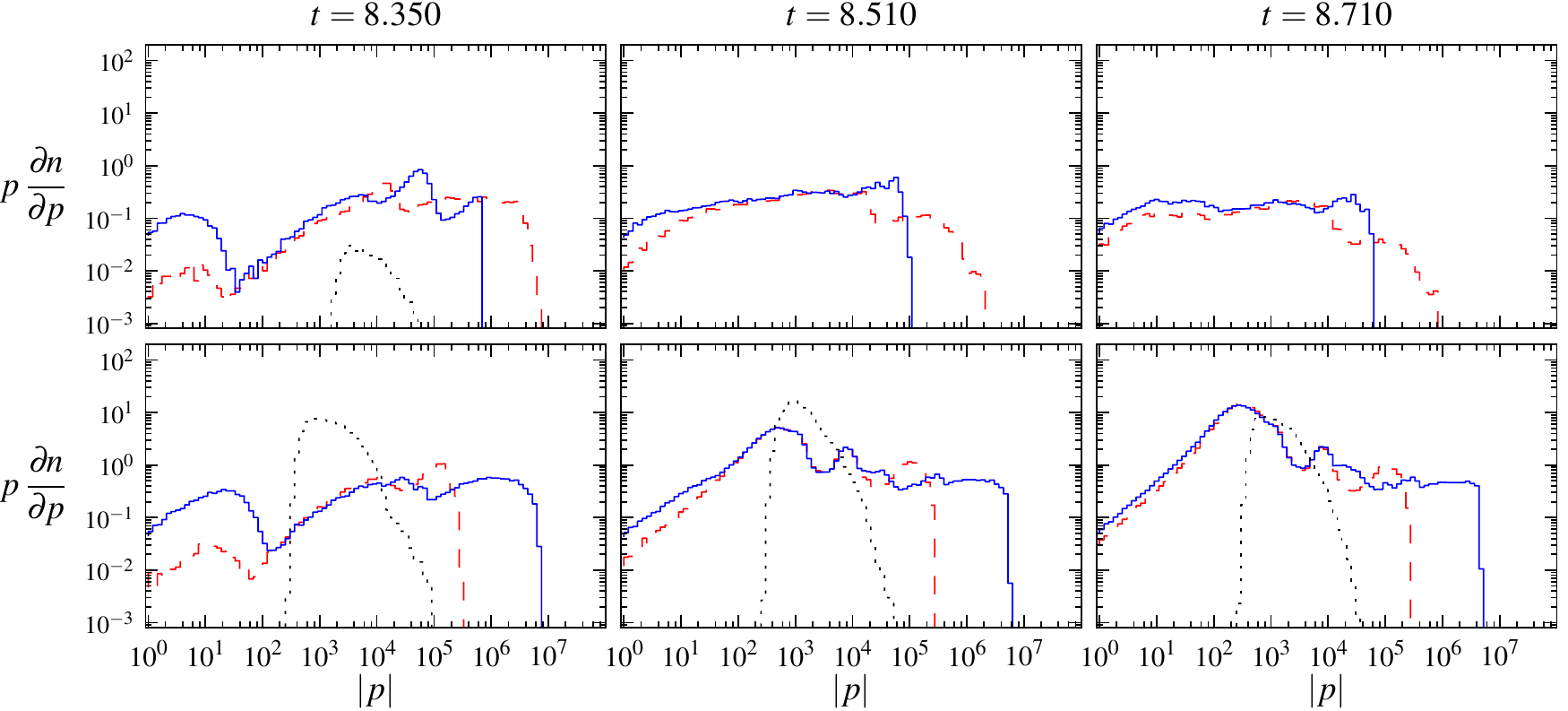}
\end{center}
\caption{Momentum distributions of particles at three moments of time
  for cascade with anti-GJ current density $\jm=-0.5\GJ{j}$.  Positron
  distributions are shown by solid blue lines, electron distributions
  -- by dashed red lines, distribution of pair producing photons -- by
  dotted black lines.  Plots in the top row show distributions for
  particles moving toward the magnetosphere ($p>0$), plots in the
  bottom row -- distributions for particles moving toward the NS
  ($p<0$).  Each column corresponds to the same moment of time shown
  above the plots.  Plots in each columns are aligned and share the
  same values of $|p|$.  The following spacial regions were used for
  plotting the average distribution functions: $x\in[0,0.7]$ for
  $t=8.350$, $x\in[0,0.75]$ for $t=8.510$ and $x\in[0,0.85]$ for
  $t=8.710$. }
  \label{fig:seds_jp05}
\end{figure*}

\begin{figure*}
\begin{center}
  \includegraphics[width=\textwidth]{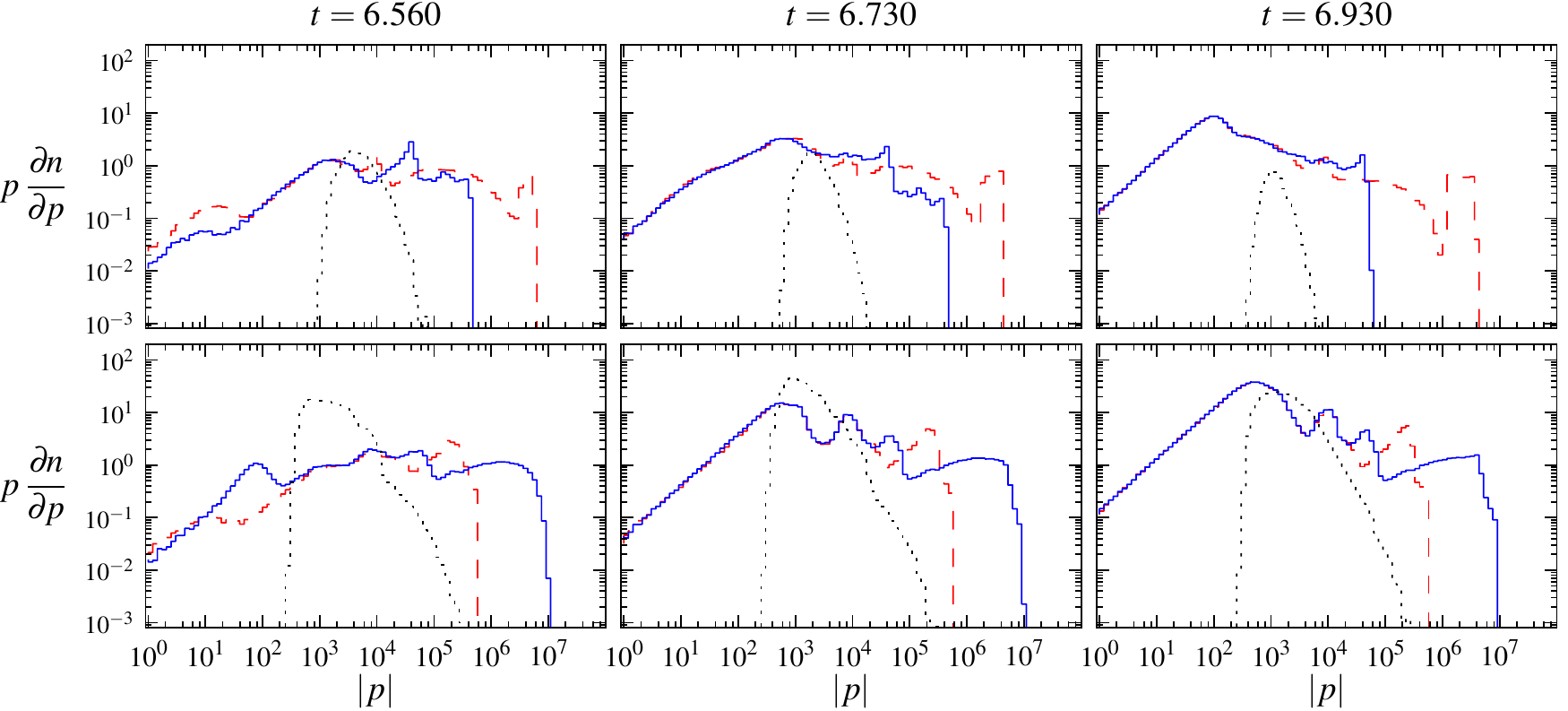}
\end{center}
\caption{Momentum distributions of particles at three moments of time
  for cascade with anti-GJ current density $\jm=-1.5\GJ{j}$.
  Notations are the same as in Fig.~\ref{fig:seds_jp05}.  The
  following spacial regions were used for plotting the average
  distribution functions: $x\in[0,0.75]$ for $t=6.560$, $x\in[0,0.88]$
  for $t=6.730$ and $x\in[0,1]$ for $t=6.930$.}
  \label{fig:seds_jp15}
\end{figure*}

\begin{figure*}
  \begin{center}
    \includegraphics[width=\textwidth]{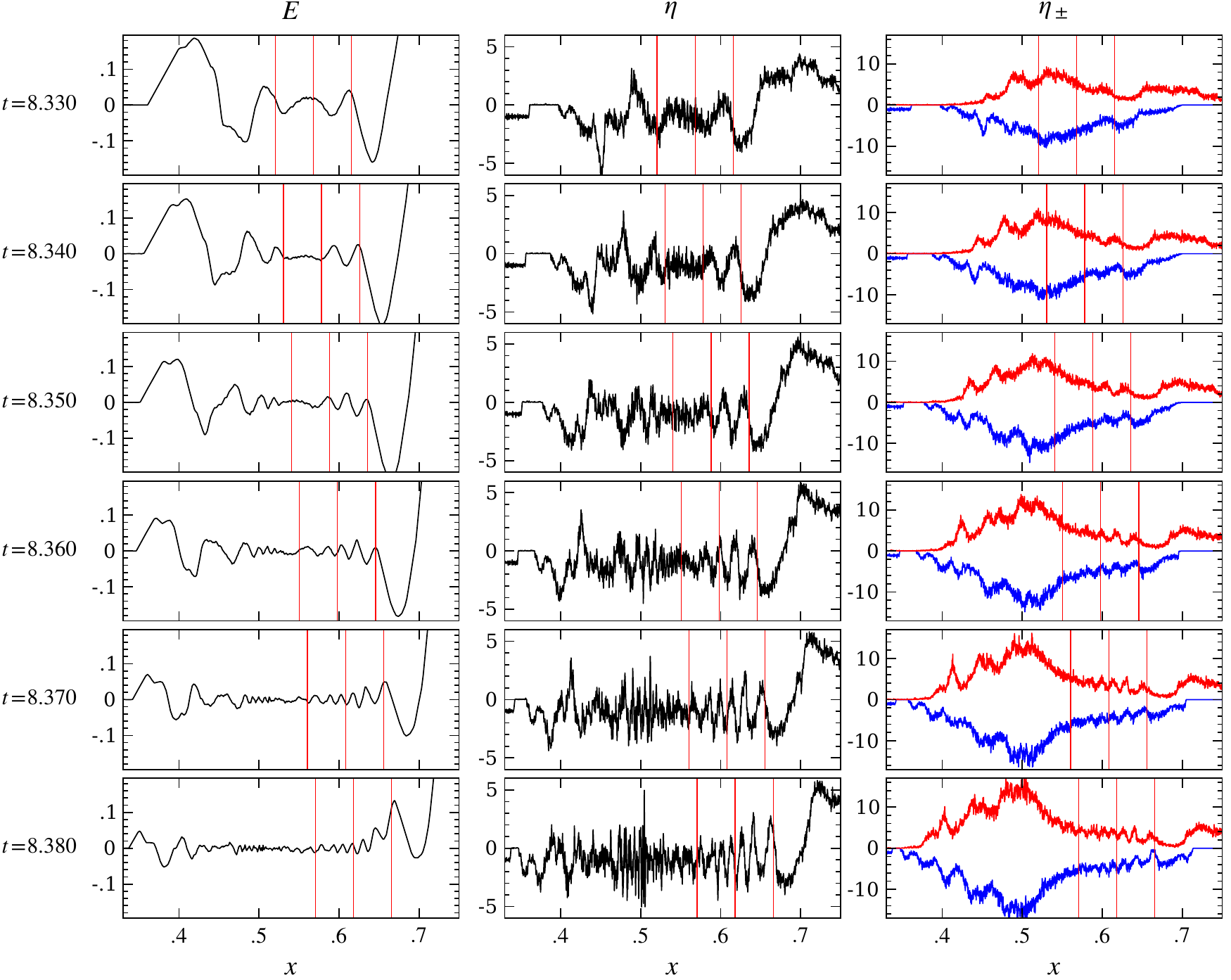}
  \end{center}
  \caption{Screening of the electric field and formation of
    superluminal electrostatic wave for cascade with $\jm=-0.5\GJ{j}$.
    There are six snapshots for the electric field $E$, the total
    change density $\eta$ and the charge density of electrons
    (negative values, blue line) and positrons (positive values, red
    line) $\eta_{\pm}$. All quantities are plotted as functions of
    distance $x$ for the part of the calculation domain with intense
    pair formation.  Snapshots are taken at equally separated time
    intervals.  Plots in each column are aligned and share the same
    values of $x$.  The same normalizations for physical quantities
    are used as in Fig.~\ref{fig:ctss_jp05__1}.  The three thin red
    vertical lines in each plot mark fiducial points moving with the
    speed of light toward the magnetosphere. }
    \label{fig:wave_jp05}
\end{figure*}

In this section we describe cascade development for two cases when the
imposed current density is anti-GJ (i.e. has the opposite sign to the
GJ current density), for $\jm=-0.5\GJ{j}$ and $\jm=-1.5\GJ{j}$.  To
support such current electrons on average must move toward smaller
$x$, down toward the NS, positrons toward larger $x$, up into the
magnetosphere.  Electrons could be freely emitted from the NS surface,
but because the imposed current causes the growing electric to point
away from the surface, electrons at the surface are accelerated
downwards and none are extracted from the star%
\footnote{Due to numerical noise the electric field fluctuates and
  sometimes electrons ``from the NS surface'' enter the domain;
  however, the number of such electrons is well below the fluctuation
  of electron number density due to numerical noise.}.
This situation resembles the case of the
\citet{RudermanSutherland1975} cascades studied in \PapI, in the sense
that all the particles are produced during the burst of pair
formation.  Hence, the physics of pair cascades with anti-GJ imposed
current density discussed in this section is also applicable to
Ruderman \& Sutherland cascades as well.

Particles leave the domain and some time after the burst of pair
formation the domain becomes depleted of electrons.  This process
starts at the right end of the domain -- at the ``magnetosphere'' end
-- as electrons are moving down toward the NS.  In the region depleted
of electrons, the positrons support the current density $j<\jm$ with
charge density $\eta>\GJ{\eta}$.  This gives rise to the electric
field which accelerates positrons toward the magnetosphere - see the
phase portraits of positrons in
Figs.~\ref{fig:ctss_jp05__1},\ref{fig:ctss_jp15__1}.  The electric
field in this region at any given point is growing with time and the
size of the region with unscreened electric field is getting bigger as
the remaining electrons are moving toward the NS.  The electric field
grows linearly with the distance because positrons are relativistic
and so their charge density remains constant (see plots for $E$ in
Figs.~\ref{fig:ctss_jp05__1},\ref{fig:ctss_jp15__1}).

Positrons in the region with the unscreened electric field are
accelerated up to higher and higher energies and begin emitting pair
producing gamma-rays.  Positrons remaining from the previous burst of
pair formation are the particles which ignite the next burst.  As
positrons are moving up (and so do the first pair producing photons),
the first pairs are produced at the largest distance from the NS where
pair formation is still possible, in our case near $x=\xb=0.7L$.  The
injected pairs are picked up by a very strong electric field and are
accelerated to high energies in less time that the first
generation positrons.  They emit pair producing capable gamma-rays,
but now both secondary electrons and positrons are emitting pairs.  As
electrons and positrons move in opposite directions so do the
gamma-rays -- see snapshots for gamma-rays phase space at $t>8.100$ in
Fig.~\ref{fig:ctss_jp05__1} and $t>6.390$ in
Fig.~\ref{fig:ctss_jp15__1}.

Secondary electrons and positrons are moving in opposite directions
and the plasma gets polarized -- see plots for $\eta_\pm$ and $\eta$
at $t=8.260$ in Fig.~\ref{fig:ctss_jp05__1} and at $t=6.430$ in
Fig.~\ref{fig:ctss_jp15__1}.  When their number density become
comparable to the GJ number density these particles start screening the
electric field (see plots for $E$ at $t=8.260,8.270$ in
Figs.~\ref{fig:ctss_jp05__1},~\ref{fig:ctss_jp05__2} and at
$t=6.430,6.440$ in
Figs.~\ref{fig:ctss_jp15__1},~\ref{fig:ctss_jp15__2}).  The screening
starts in the region where the first pairs were injected because pairs
have been injected here for the longest time and their number density
is larger.

The rate at which particles left from the previous burst of pair
formation are leaving the domain depends on the imposed current
density.  The average bulk motion of electrons for $\jm=-0.5\GJ{j}$ is
less than $0.5c$, while for $\jm=-1.5\GJ{j}$ it is close to $c$%
\footnote{Note that the time intervals between the snapshots in
  Fig.~\ref{fig:ctss_jp05__1} is two times larger than that in
  Fig.~\ref{fig:ctss_jp15__1}},
and the size of the electron-depleted region grows slower in the
former case.  Second generation electrons, which mark the upper
boundary of the gap, move with ultrarelativistic velocity toward the
NS.  Because of this in the case of $\jm=-0.5\GJ{j}$ the gap with the
accelerating electric field quickly disappears -- at $t=8.430$ in
Fig.~\ref{fig:ctss_jp05__2} secondary particles have already caught up
with the particles from the previous burst of pair formation and the
electric field is screened everywhere where pair formation is
possible.  In the case of $\jm=-1.5\GJ{j}$ the gap moves toward the NS
approximately retaining its size -- see
Figs.~\ref{fig:ctss_jp15__2},~\ref{fig:ctss_jp15__3}.  This behavior
is generic.  Namely, for imposed current densities with less-than-GJ
absolute values, $|\jm|<|\GJ{j}|$, the average bulk motion of
particles from the previous burst of pair formation is
non-relativistic; that leads to quick closure of the accelerating gap.
If the absolute value of the imposed current density is
larger-than-GJ, $|\jm|>|\GJ{j}|$, the average bulk motion of particles
from the previous burst of pair formation is relativistic and the gap
propagates in the direction of this bulk motion.  The same behavior
was observed also in the case of Ruderman-Sutherland cascades studied
in \PapI.

In the case of $\jm=-0.5\GJ{j}$ the gap with the accelerating electric
field closes and particles in the region where pair formation is
possible are not accelerated anymore.  The pair formation continues
only because of particles accelerated before the gap disappears.
There are comparable numbers of particles accelerated in both
directions, but electrons moving toward the NS propagate in the region
where pair formation is possible and so after screening of the
electric field most of the pairs are created with the initial momenta
directed toward the NS (see phase portraits for pair producing
gamma-rays for $t>8.180$ in
Figs.~\ref{fig:ctss_jp05__1},~\ref{fig:ctss_jp05__2},~\ref{fig:ctss_jp05__3}
and particle distribution functions in Fig.~\ref{fig:seds_jp05}).  In
the case of $\jm=-1.5\GJ{j}$ the gap propagates down to the NS
surface.  While the gap is moving, positrons left from the previous
bursts of pair formation, which are still present at the NS side of the
domain, enter the gap, are picked up by the electric field, and emit
pair producing gamma-rays.  In this case the pair production is
sustained not only by electrons accelerated in the initial screening
event but also by positrons continuously accelerated in the moving gap
toward the magnetosphere.  Phase portraits of gamma-rays in
Figs.~\ref{fig:ctss_jp15__1},\ref{fig:ctss_jp15__2},\ref{fig:ctss_jp15__3}
clearly show the presence of gamma-rays moving towards the
magnetosphere during all stages of cascade development (see also the
plot for particle distribution functions in Fig.~\ref{fig:seds_jp15}).
Gamma-rays with positive momenta occupy a larger and large fraction of
the space as the gap propagates toward the NS.

Regions far from the NS surface experience charge starvation first.
The pair formation is possible only in the regions with strong
magnetic field and until the starved region extends into the strong
field domain the electric field remain unscreened.  After the start of
pair formation, the dense plasma propagates also in the direction of
the magnetosphere screening the accelerating electric field there.  In
our simulations the region with $x>0.7L$ represents the rest of the
magnetosphere, and in snapshots with $t=8.430-8.990$, in
Figs.~\ref{fig:ctss_jp05__2},~\ref{fig:ctss_jp05__3} and
$t=6.560-6.930$ in
Figs.~\ref{fig:ctss_jp15__2},~\ref{fig:ctss_jp15__3}, one can clearly
see how the dense pair plasma fills the domain with $x>0.7L$.  Filling
of that domain with plasma is forced by a small electric field induced
in the dense pair plasma by the imposed current.  The rate at which
these regions are filled with the dense plasma depends on the imposed
current -- in the case of $\jm=-0.5\GJ{j}$ the average bulk motion of
the pair plasma is about $0.5c$, while for $\jm=-1.5\GJ{j}$ it is
relativistic.

The accelerating electric field is very strong in the domain where
pairs can not be produced until electrons generated in the discharge
arrive, and positrons (either primary or secondary ones) entering this
domain before electron arrival are accelerated up to high energies.
In our model pairs are created only by single photon absorption in
strong magnetic field, what automatically restricts pair formation to
the polar cap region.  However electron-positron creation via
$\gamma$-$\gamma$ absorption can be possible at much higher distances
from the polar cap, similar to the outer gap scenario
\citep{Cheng1986}.  In our model between the bursts of polar cap pair
formation in the magnetospheric region above the low altitude pair
producing zone, at $x>\xb$, particles can be accelerated up to
radiation-reaction limited energies and emit very high energy
gamma-rays.  In a more realistic model, when $\gamma$-$\gamma$
absorption is taken into account, this should give rise to cascades in
the outer magnetosphere resembling to some extend the outer gap
scenario, except that the accelerating zone would not be limited to
the region around the null surface, where the GJ charge density
changes sign. Due to intermittency of plasma flow in the return
current region \emph{both} polar cap and the outer gaps like cascades
might exist along the same magnetic field lines.  However, without
higher dimensional simulations it is not clear how the interaction
between two such cascade zones would play out.

Screening of the accelerating electric field during each burst of
pair formation gives rise to a superluminal electrostatic wave.  In
Fig.~\ref{fig:wave_jp05} we show screening of the electric field in
the gap for the flow with $\jm=-0.5\GJ{j}$, for flow with
$\jm=-1.5\GJ{j}$ the situation is very similar.  Plots in
Fig.~\ref{fig:wave_jp05} show in more detail than in
Fig.~\ref{fig:ctss_jp05__2} how the accelerating electric field is
being screened by newly created pair plasma about $t=8.350$.
Screening of the electric field starts in the middle of the blob of
newly created plasma and spreads to its edges.  This spreading
occurs in the form of an electrostatic wave which phase velocity is
larger that $c$.  This process is almost identical to what was
observed for cascades described in \PapI{} where more detailed
discussion of the screening process can be found in \S4.2 (cf. our
Fig.~\ref{fig:wave_jp05} with Fig.~5 in \PapI).

As in the simulations described in \PapI, second generation particle
momentum distributions are very broad, extending towards
non-relativistic energies.  At least one of the reasons for momentum
broadening of the particle spectra is trapping of particles in the
strong electrostatic wave excited at the beginning of each burst of
pair formation.  The dynamics of these discharges is very similar to
what is seen in simulations of the Ruderman \& Sutherland cascade;
more details on momentum spreading can be found in \S4.3 of \PapI.
The appearance of the low energy component in the pair plasma is
visible in the electron and positron phase portraits -- initially at
distances where pairs are injected, there are no particles in the
phase space with low momenta, but later particles fill in the low
momentum regions (see plots at $t\ge8.350$ in
Figs.~\ref{fig:ctss_jp05__2},~\ref{fig:ctss_jp05__3} and for
$t\ge6.500$ in Figs.~\ref{fig:ctss_jp15__2},~\ref{fig:ctss_jp15__3}).
In Figs.~\ref{fig:seds_jp05},~\ref{fig:seds_jp15} we plot the particle
momentum distribution $p\:(\partial{}n/\partial{}p)$ for three
different moments of time.  In the upper panel we plot the momentum
distribution of particles moving toward the magnetosphere, $p$ is
positive; in the lower panel -- the momentum distribution of particles
moving toward the NS, $p$ is negative.  These distributions are
averages for regions where the electric field at the magnetosphere end
is screened.  On these plots one can see that the low energy plasma
component is present at all stages of cascade development.

There is another important effect in cascades along field lines
carrying return current -- positive ions (in our model protons)
can be pulled from the NS.  Both electrons and protons can be
extracted from the NS surface, and in our simulations we allow
injection of both species.  The imposed current density requires
positively charged particles to move toward the magnetosphere,
i.e. protons could be pulled from the NS.  However, for less-than-GJ
current densities ($|\jm|<|\GJ{j}|$) there are always electrons and
positrons near the NS surface left from the previous burst of pair
formation.  The gap's accelerating electric field disappears before it
reaches the NS, while pair plasma appears after every burst of pair
formation keeping the electric field near NS screened.  In our
simulations of cascades with less-than-GJ current densities we did not
see injection of protons.  In the case of larger-than-GJ current
density ($|\jm|>|\GJ{j}|$) protons are indeed pulled from the NS, at
least at some stage of cascade development, and this cannot be
attributed only to numerical effects.  In this case we see some
protons pulled from the NS at any time, even if pair plasma is present
near the NS.  The number density of these protons never exceeds
$\sim0.2\GJ{n}$, at least $10$ times less than the number density of
electrons and positrons; protons do not play a substantial role in the
electrodynamics of discharges.  The number density of protons pulled
from the NS while electrons and positrons are still present near the
NS surface depends on the numerical resolution and, hence, our
simulations are inconclusive about whether this effect is real.
However, there is a stage in cascade development when the region near
the NS gets depleted of pair plasma; at those times proton extraction
from the NS is certainly a real effect.  The number of particles left
from the previous burst of pair formation near the NS is not enough to
support the imposed current density and screen the electric field
until the new portion of the pair plasma arrives.  A gap with the
electric field appears at the NS surface -- see snapshots at
$t=6.730,6.830$ in Fig.~\ref{fig:ctss_jp15__3}.  This gap is best
visible on phase portraits of positrons as a second region (the one
close to the NS) depleted of those particles.  The electric field in
the gap starts accelerating protons from the NS surface, as is clearly
visible in phase space portrait for protons in
Fig.~\ref{fig:ctss_jp15__3}.  In our simulations this second gap does
not become large enough to accelerate electrons to energies sufficient
to start another burst of pair formation near the NS surface, but we
cannot exclude this possibility in all cases.  Finally, the main gap
reaches the star and pulls out protons as well -- see snapshot at
$t=6.930$ in Fig.~\ref{fig:ctss_jp15__3}.

\subsection{Flow with $\jm/\GJ{j}>1$}
\label{sec:j_greater}

\begin{figure*}
\begin{center}
  \includegraphics[width=\textwidth]{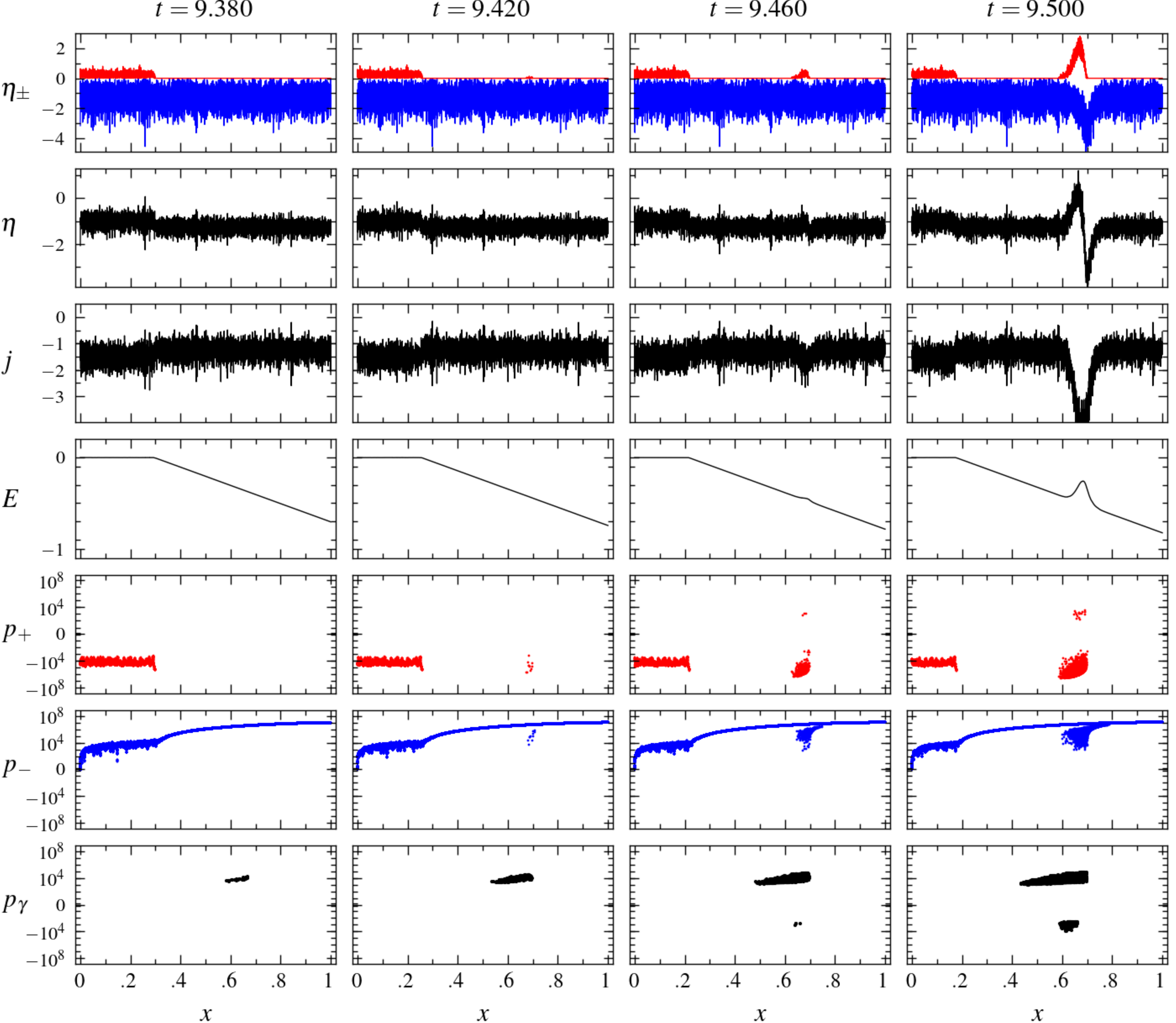}
\end{center}
\caption{Ignition of pair formation in super-GJ flow with
  $\jm=1.5\GJ{j}$. The same quantities are plotted as in
  Fig.~\ref{fig:ctss_jp05__1}.}
  \label{fig:ctss_jm15__1}
\end{figure*}

\begin{figure*}
  \begin{center}
    \includegraphics[width=\textwidth]{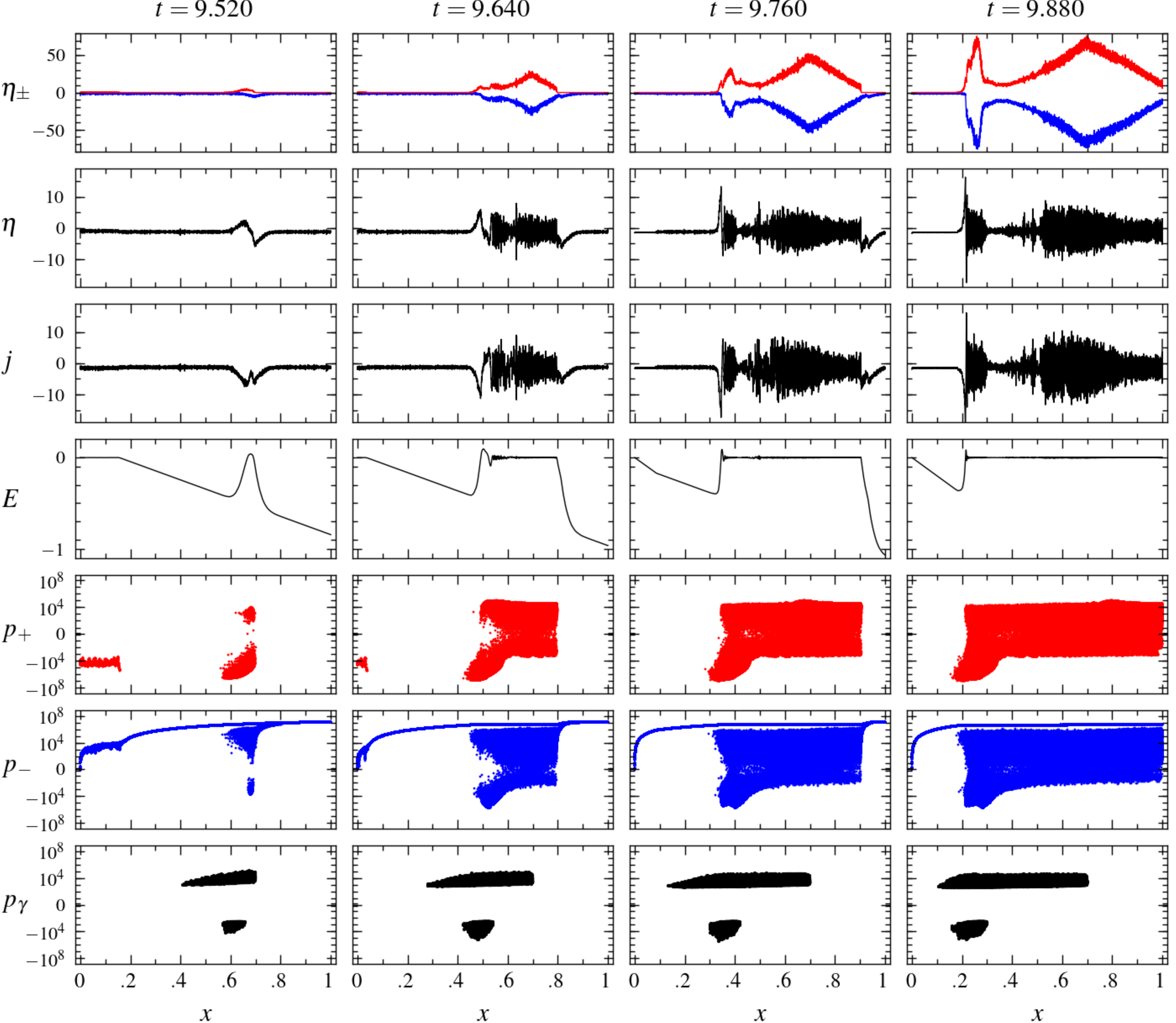}
  \end{center}
  \caption{Screening of the electric field in super-GJ flow with
    $\jm=1.5\GJ{j}$.  The same quantities are plotted as in
    Fig.~\ref{fig:ctss_jp05__1}.}
  \label{fig:ctss_jm15__2}
\end{figure*}

\begin{figure*}
  \begin{center}
    \includegraphics[width=\textwidth]{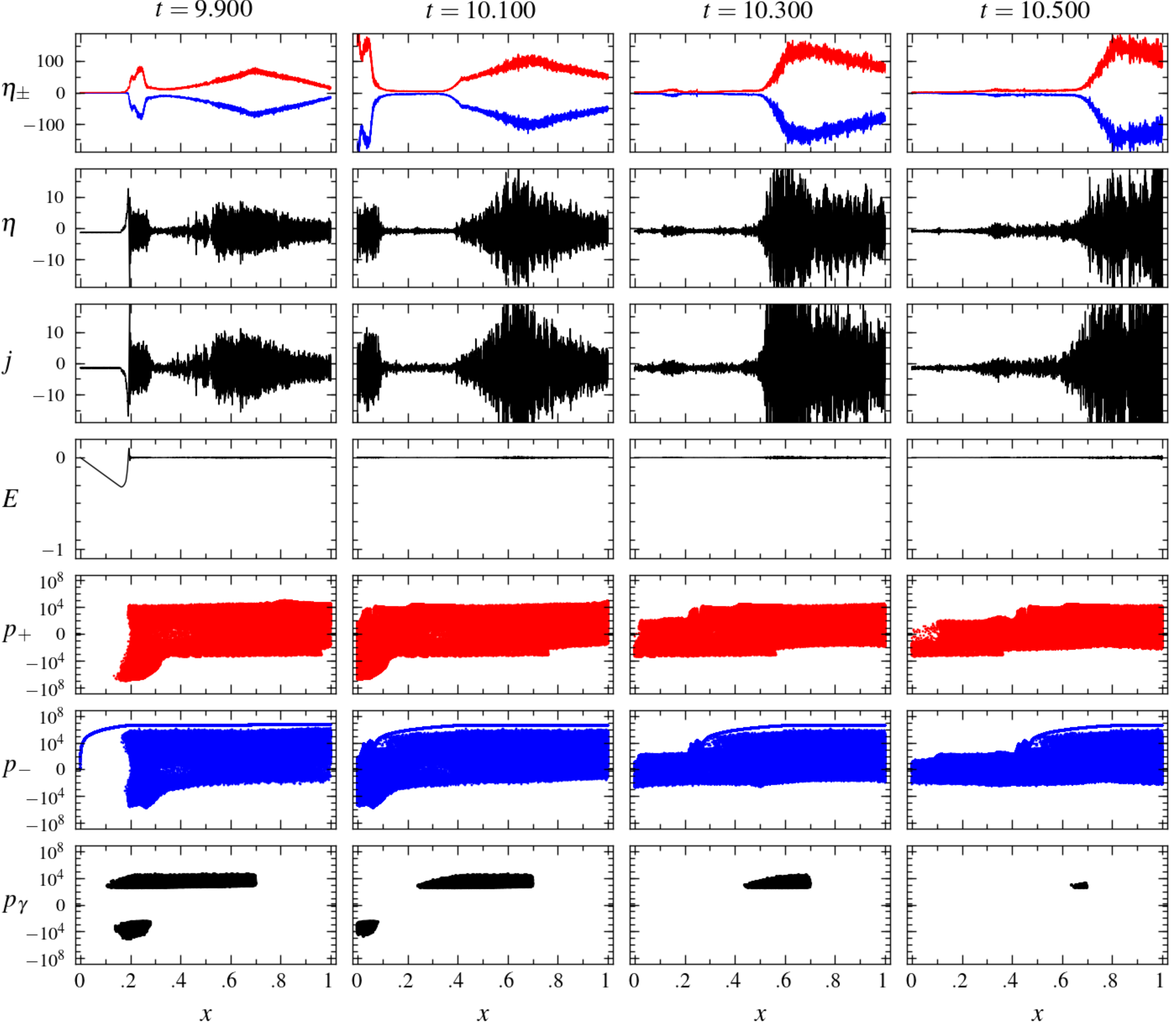}
  \end{center}
  \caption{Filling of computation domain with dense pair plasma in
    super-GJ flow with $\jm=1.5\GJ{j}$.  The same quantities are
    plotted as in Fig.~\ref{fig:ctss_jp05__1}.}
  \label{fig:ctss_jm15__3}
\end{figure*}

\begin{figure*}
  \begin{center}
    \includegraphics[width=\textwidth]{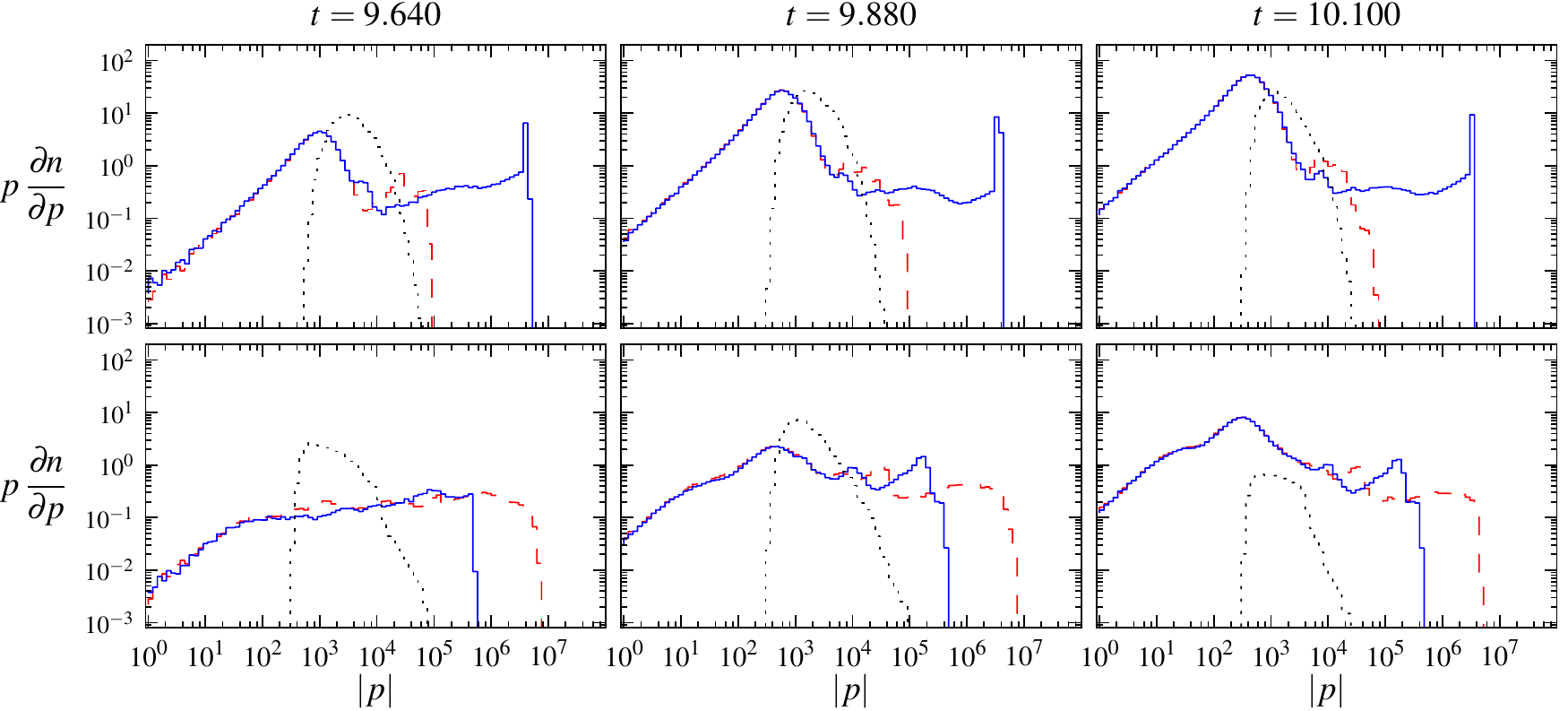}
  \end{center}
  \caption{Momentum distributions of particles at three moments of
    time for cascade with super-GJ current density $\jm=1.5\GJ{j}$.
    Notations are the same as in Fig.~\ref{fig:seds_jp05}.  The
    following spatial regions were used for plotting the average
    distribution functions: $x\in[0,0.795]$ for $t=9.640$, $x\in[0,1]$
    for $t=9.880$ and $x\in[0,1]$ for $t=10.100$.}
  \label{fig:seds_jm15}
\end{figure*}

In this section we describe cascade development for the case when the
imposed current density is super-GJ.  As an example of such flow we
consider the case with $\jm=1.5\GJ{j}$.  To support this current
density electrons must move up, towards the magnetosphere, and
positrons down, toward the NS.  There is a continuous source of
electrons at the surface of the NS, $x=0$.  Positrons appear during the
bursts of pair formation and there is no source of positrons during
the quiet phase, between successive bursts of pair formation.  To keep
the electric field screened there \emph{must} be both electrons and
positrons present at each point -- electrons with larger-than-GJ
number density moving up and positrons moving down.  Positrons
compensate for the larger-than-GJ number density of electrons keeping
the total charge density equal to the GJ charge density.

Positrons are leaving the domain and as there is no source of
positrons, some time after the burst of pair formation the domain
becomes depleted of positrons.  This depletion starts at the upper
end of the domain -- at the ``magnetosphere'' -- as positrons are
moving toward the NS.  In the region depleted of positrons electrons
produce current density $|j|<|\jm|$ and the charge density
$|\eta|>|\GJ{\eta}|$, because their number density is set at regions
closer to the NS where positrons are still present.  This gives rise
to the accelerating electric field which grows as the remaining
positrons are moving toward the NS, see Fig.~\ref{fig:ctss_jm15__1}.
For time shots at $t=9.380-9.500$ on the plots for the charge density
$\eta$ and the current density $j$ there are small jumps in both
$\eta$ and $j$ at the point where the number density of positrons
drops to zero.  The electron number density remains the same, but
positrons are leaving, enlarging the region depleted of positive
charges.

Electrons are accelerated up to higher and higher energies in this gap
and start emitting pair producing gamma-rays.  In this case, electrons
extracted from the NS surface, with the number density of the order of
$|\jm|/e$, are the particles which ignite the next burst of pair
creation.  As electrons are moving up (as do the first pair producing
photons), the first pairs are produced at the farthest distance where
pairs formation starts to be possible, near $x=\xb=0.7L$ in our
example.  The injected pairs are picked up by the very strong electric
field and are accelerated to high energies in less time than the
primary electrons.  Soon they start emitting pair producing capable
gamma-rays which decay into pairs, see snapshots for $t\ge9.420$ in
Fig.~\ref{fig:ctss_jm15__1}.  Electrons and positrons are moving in
opposite directions, the pair plasma becomes polarized and starts
screening the electric field, see plots for $\eta_\pm$, $\eta$, and
$E$ at $t=9.500,9.520$ in
Figs.~\ref{fig:ctss_jm15__1},\ref{fig:ctss_jm15__2}.

The screening of the electric field proceeds similarly to that in the
case of anti-GJ flow with $\jm=-1.5\GJ{j}$ described in
\S\ref{sec:j_negative}.  The electric field is being screened
first near $x=\xb$, this creates a finite size gap with accelerating
electric field which moves toward the NS.  The bulk motion of
positrons left from the previous burst of pair formation is
relativistic and the newly created pairs are relativistic too, so the
boundaries of the regions where electric field is screened are moving
in the same direction (toward the NS) with the same speed.  The region
in between with non-screened field -- the gap -- is therefore also
moving, retaining its size.  Electrons are continuously extracted from
the NS to sustain the imposed current density.  They enter the gap,
are accelerated and emit pair producing gamma-rays.  Pairs are
produced by (i) primary electrons extracted from the NS surface
accelerated by the moving gap and (ii) secondary positrons moving
toward the NS accelerated in the initial screening event.  The beam of
primary electrons accelerated in the gap is clearly seen on plots for
particle momentum distribution, Fig.~\ref{fig:seds_jm15}, as a spike
in the electron distribution function for positive values of $p$ (the
upper plots in Fig.~\ref{fig:seds_jm15}).  In contrast to the case
with $\jm=-1.5\GJ{j}$, where the number of particles which were
continuously accelerated in the gap was small and most of the pair
were produced with momenta directed toward the NS, here more pairs are
injected with momenta directed toward the magnetosphere -- see
Fig.~\ref{fig:seds_jm15} and plots for particle number density
$\eta_\pm$ at $t>9.640$ in
Figs.~\ref{fig:ctss_jm15__2},~\ref{fig:ctss_jm15__3}.  As in the case
of anti-GJ flow secondary particles have broad momentum spectra -- the
low energy plasma components is present at at all stages of cascade
development, Fig.~\ref{fig:seds_jm15}.

The pair creation ends when the down flow component of the pairs
reaches the stellar surface - the gap closes and the accelerating
electric field is poisoned throughout the strong magnetic field
region, while the beam component that emits the primary pair creating
photons and its daughter pairs moves up at the speed of light, into
the magnetosphere, as can be seen in the $e^+$ and $e^-$ phase
portraits with advancing time in
Figs.~\ref{fig:ctss_jm15__2},~\ref{fig:ctss_jm15__3}.  Positrons' bulk
motion is toward the NS and some time after the plasma burst leaves --
the wake of the departing burst being the source of positrons -- the
upper regions become deplete of positrons and the cycle starts again.

Both in the case of super-GJ flow with constant GJ charge density
described here and in anti-GJ flow described in
\S\ref{sec:j_negative}, the cascade starts at the upper boundary of
the strong field region when the upper regions become depleted of
positrons or electrons correspondingly, as the latter move down.  The
rate at which these regions become charge depleted depends on the
imposed current density, the higher the absolute value of $\jm$ the
faster is the bulk motion of the charge component moving toward the
NS. For $|\jm|$ comparable or larger than $|\GJ{j}|$ this bulk
velocity is close to $c$.  Thus the limit cycle period is larger than
the relativistic fly-by time over the length of the strong field
region.  In our simulations, that region has size $0.7L = 168$~meters,
with fly by time $0.56\,\mu$sec.  However, in a real pulsar, the optical
depth for pair creation by curvature photons emitted by the beam
exceeds unity all the way out to heights comparable to the stellar
radius.  Then the fly-by and recurrence times might be as long $\sim
30\,\mu$sec, with the intermittent plasma starved regions taking the
form of long filaments.  Assessing these structures and accounting for
competition between neighboring filamentary gaps is an intrinsically
2D issue, and will be treated elsewhere.

Because the cascade starts far from the NS, at the largest distance
where pair formation becomes possible, in our simulations the
accelerating electric field is unscreened between bursts of pair
formation above the pair formation zone.  In this regard in the case
of constant GJ charge density super-GJ flow is similar to the anti-GJ
flow.  Namely another cascade zone in the outer magnetosphere might
coexist with the polar cap accelerating zone.  However, as we will
show in the next section variation of the GJ charge density with the
distance can significantly change the behavior of cascades in the
super-GJ flow.

\subsection{Effects of spatially varying Goldreich-Julian charge density}
\label{sec:j_negative_rho_variable}

\begin{figure*}
\begin{center}
  \includegraphics[width=\textwidth]{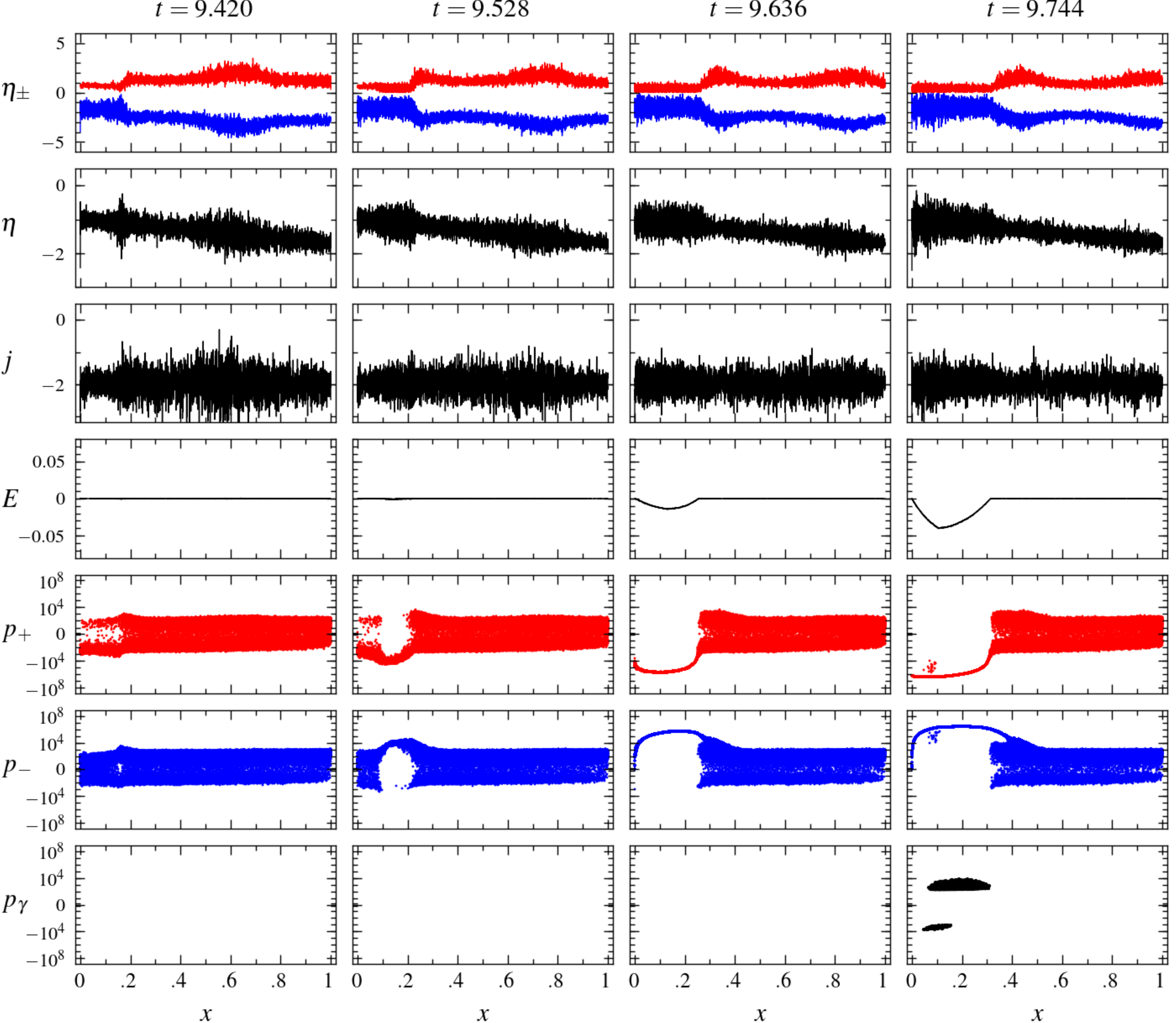}
\end{center}
\caption{Ignition of pair formation in the flow with linearly varying
  GJ charge density and the imposed current density $\jm=2\GJ{j}$. The
  same quantities are plotted as in Fig.~\ref{fig:ctss_jm20__1}}
  \label{fig:ctss_jm20__1}
\end{figure*}

\begin{figure*}
\begin{center}
  \includegraphics[width=\textwidth]{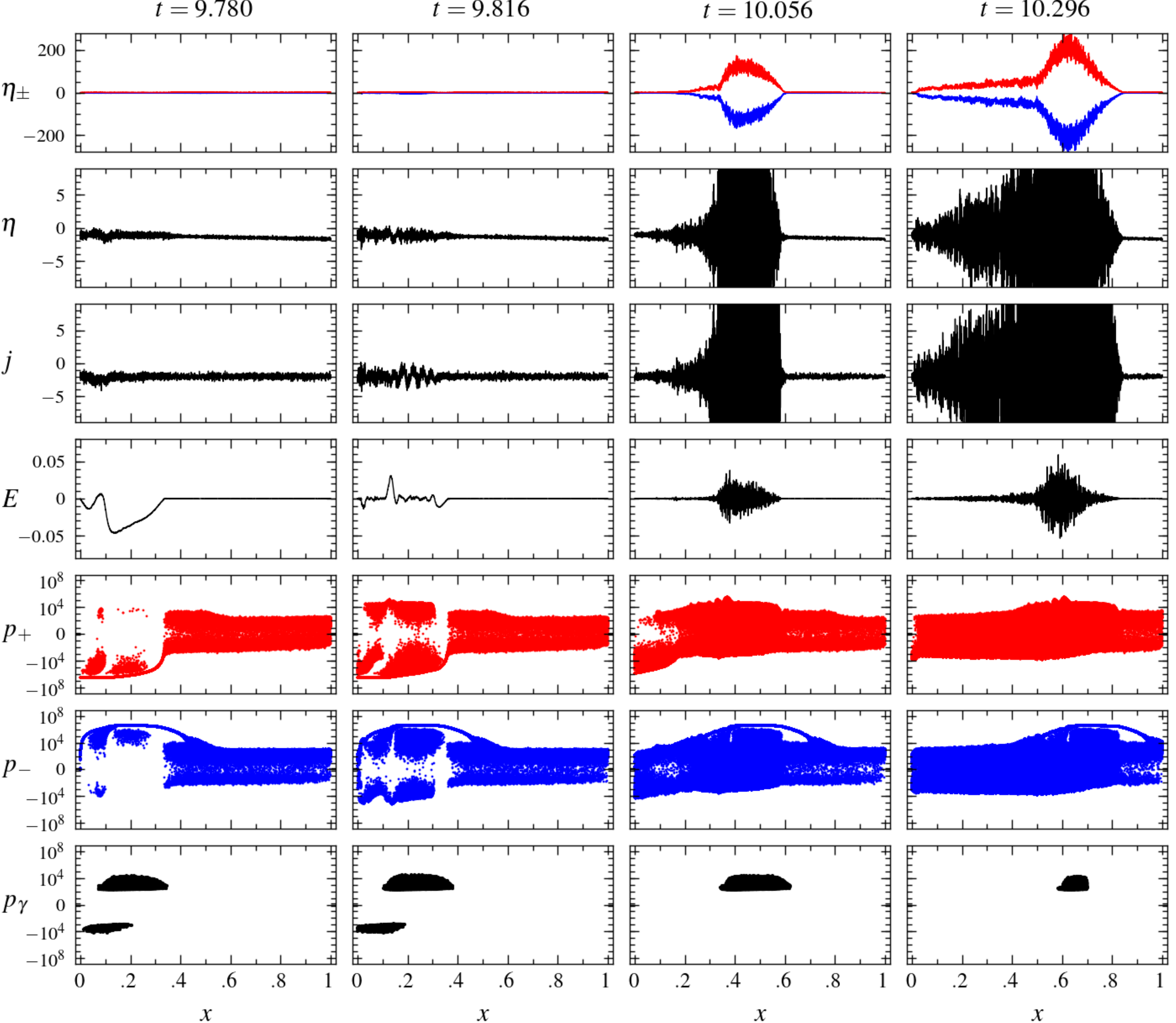}
\end{center}
\caption{Screening of the electric field and final stages of pair
  formation in the flow with linearly varying GJ charge density and
  the imposed current density $\jm=2\GJ{j}$. The same quantities are
  plotted as in Fig.~\ref{fig:ctss_jp05__1}. Note that the time
  intervals between the first two snapshots are two times smaller that
  those between the rest of the plots.}
  \label{fig:ctss_jm20__2}
\end{figure*}

\begin{figure*}
  \begin{center}
    \includegraphics[width=\textwidth]{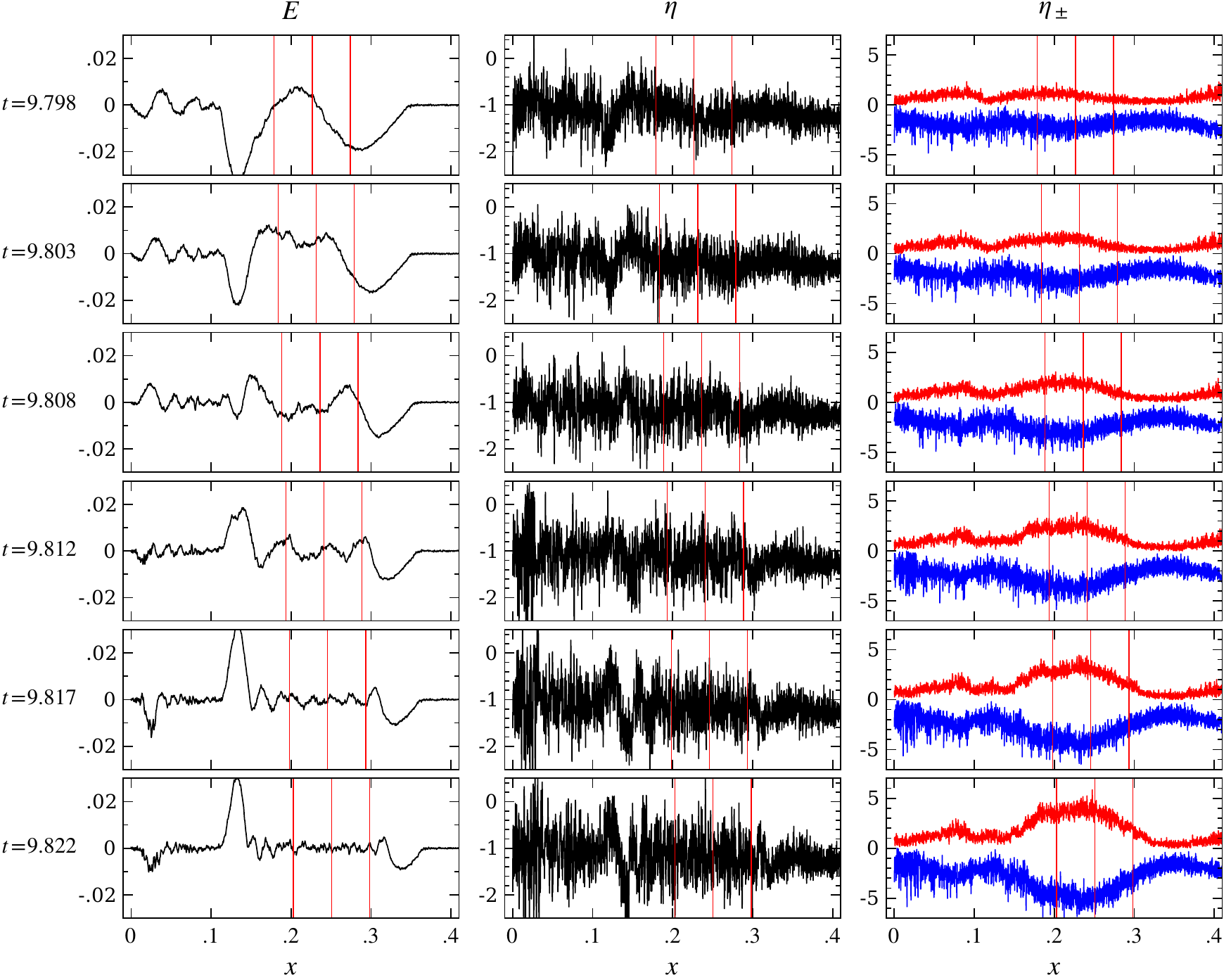}
  \end{center}
  \caption{Screening of the electric field and formation of
    superluminal electrostatic wave in the flow with linearly varying
    GJ charge density and the imposed current density $\jm=2\GJ{j}$c.
    The same quantities are plotted as in Fig.~\ref{fig:wave_jp05}.}
    \label{fig:wave_jm20}
\end{figure*}

In this section we address the effect of the mismatch between the
charge density of the space charge limited beam component and the GJ
charge density on the cascade development.  In the widely adopted
steady flow model of polar cap cascades
\citep{AronsScharlemann1979,Muslimov/Tsygan92} this mismatch is the
reason for the appearance of the accelerating electric field.

We performed simulations with different scaling of the GJ charge
density, both in shape (linear, power-law, exponential,
decreasing/increasing with the distance) as well as in amplitude for
all the cases considered in the previous sections.  For the space
charge limited flow with $0<\jm/\GJ{j}<1$ and $\jm/\GJ{j}<0$ we found no
qualitative difference in cascade behavior due to variation of the GJ
charge density.  Namely, for anti-GJ flows, $\jm/\GJ{j}<0$, the
regions far from the neutron star became charge-depleted first and the
cascade started at the largest distance where pair formation is
possible, with the acceleration zone propagating toward the NS.
Sub-GJ flows, with $0<\jm/\GJ{j}<1$, -- if the imposed current density
was less that the \emph{local} value of the GJ current density
$\GJ{\eta}(x)\,c$ everywhere -- remained low energetic and had
two-component: a moderately relativistic beam propagating through a
cloud of trapped particles.  The main properties of space charge
limited flow for these cases described in
\Ss\ref{sec:cold_flow},~\ref{sec:j_negative} are not affected by
variations of the GJ charge density with the distance.

The flow with super-GJ current density $\jm/\GJ{j}>1$, however, can be
strongly affected by variations of the GJ charge density -- we
observed different flow behavior when the absolute value of the
$\GJ{\eta}/B$ increased with the distance from the NS.  In that case
the gap can appear close to the NS surface, in contrast to all other
cases when it appeared at large distances.  If such gaps can generate
pairs, the regions above such gaps, far from the NS, remain filled
with dense pair plasma at all times.

Below we describe in detail an example of such flow using simulations
with exaggerated charge density contrast in order to clearly
demonstrate the above mentioned effects.  We assume that the GJ charge
density changes with the distance $x$ as
\begin{equation}
  \label{eq:eta_x}
  \eta(x) = \GJ{\eta}^0\left[ 1 + a \left(\frac{x}{L}\right)  \right]\,,
\end{equation}
where $\GJ{\eta}^0$ is the GJ charge density at the surface of the NS
and $a$ is a positive number.

Very close to the surface ($x \ll R_*$) all the sources of
inhomogeneity of $\GJ{\eta}/B$ can be approximated in this manner.
For example, if variations of the GJ charge density are because of the
field line curvature \citep{AronsScharlemann1979} the parameter $a$ in
eq.~(\ref{eq:eta_x}) for dipolar magnetic field would be given by the
following expression
\begin{equation}
  \label{eq:a_AS}
  a_{\textrm{\tiny{}AS}} \approx 
  -\frac{3}{4}\,
  \frac{\theta_* \cos\phi_*  \sin\chi}
  {\cos{\chi} - (3/2)\,\theta_* \cos\phi_* \sin\chi}\,
  \frac{L}{R_*}\,,
\end{equation}
where $\theta_*$ and $\phi_*$ are colatitude and azimuth of the
field line at the NS surface, cf. eq.~(12) in
\citet{AronsScharlemann1979}.  Note that for favorably curved
magnetic field lines $\cos\phi_*<0$ and so
$a_{\textrm{\tiny{}AS}}>0$. If GJ charge density variations are
due to inertial frame dragging \citep{Muslimov/Tsygan92} then
\begin{equation}
  \label{eq:a_MT}
  a_{\textrm{\tiny{}MT}} \approx 3\kappa_g\cos\chi\frac{L}{R_*}\,,
\end{equation}
$\kappa_g \equiv 2GI/R_*^3 c^2 \sim 0.15$, $I=$ NS's moment of
inertia, cf. eq.~(32) in \citet{Muslimov/Tsygan92}.

Physical parameters of our simulations are chosen in such a way that
the length of the accelerating gap is smaller that $\xb$.  This setup
differs from those described in
\Ss\ref{sec:j_negative},~\ref{sec:j_greater} in that the vacuum
potential drop over the the domain is larger --
$\Delta{V}=9\times10^{14}$~Volts%
\footnote{see footnote on p.~\ref{fn:large_voltage}}.
The parameter in the expression~(\ref{eq:eta_x}) for the GJ charge
density is $a=0.7$, so that the GJ charge density varies linearly from
$-|\GJ{\eta}^0|$ to $-1.7|\GJ{\eta}^0|$ throughout the domain%
\footnote{This amount of variation is huge compared to what actually
  occurs -- for example, if the variation is due to inertial frame
  dragging, for realistic parameters $a \sim 0.01$, if
  $L\sim\PC{r}$.}.
The imposed current density is
$\jm=2\GJ{j}^{\,0}\equiv{}2\GJ{\eta}^0c$, so that everywhere in the domain
the imposed current density is larger than the local value of the GJ
current density, $\jm>\GJ{\eta}(x)\,c$.  As the variation of the GJ
charge density with the distance (after accounting for decrease in
$B$, see the second paragraph of the \S\ref{sec:j_pairs}) are
not larger than $15\%$ \citep{Hibschman/Arons:pair_multipl::2001}, our
setup can be considered as a toy model for cascades in young pulsars
with $\jm>1.15\GJ{j}^{\,0}$.  All other parameters of the model are the
same as those of the models in
\Ss\ref{sec:j_negative},~\ref{sec:j_greater}.

In contrast to all other cases considered in previous sections the
dense pair plasma is always present at large distances from the NS --
the intermittent temporal gaps separating periods when the domain is
full of plasma disappear.  So, we illustrate our example with two
series of snapshots in
Figs.~\ref{fig:ctss_jm20__1},~\ref{fig:ctss_jm20__2}.

When the whole domain is filled with pair plasma the current density
is constant, $j=\jm$.  The number density of electrons is larger than
$|\GJ{\eta}|/e$ and positrons (which are left from the previous burst
of pair formation) are necessary to keep the charge density equal to
the local value of the GJ charge density.  The absolute value of the
GJ charge density is smaller closer to the NS surface, $\GJ{\eta}$ is
negative, and more positrons are necessary to screen the electric
field there than at larger altitudes.  Positrons, on average, move
toward the NS, and, as they slam on the NS surface and leave the
domain, positrons from higher altitudes should come closer to the NS
in order to keep the electric field screened.  At some altitude the
required flux of positrons towards the NS becomes larger than the
positron flux that can be extracted from an adjacent higher altitude
region without making that region charge starved -- the resulting
charge starvation causes a starvation electric field to appear close
to the NS (see snapshot at $t=9.528$ in Fig.~\ref{fig:ctss_jm20__1},
the gap is best visible in the phase portraits on electrons and
positrons).  In the region above the gap there are still enough
positrons to screen the electric field -- less positrons are needed
here as the absolute value of the GJ charge density is higher there
and so the mismatch in charge density $|\jm/c-\eta(x)|$ is smaller.
At lower altitudes, after all positrons which had been there before
the gap appeared leave the domain, the influx of positrons from above
is unable to keep the electric field screened and so the gap extends
up to the NS surface (snapshot at $t=9.636$ in
Fig.~\ref{fig:ctss_jm20__1}).

The upper end of the gap extends toward higher altitudes as positrons
move toward the NS, depleting higher altitude regions of positive
charge.  While the gap develops, until the ignition of the new
cascade, the current density remains close to the imposed current
density -- electrons are being extracted from the NS surface and
provide the required current density (see plots for $j$ in
Fig.~\ref{fig:ctss_jm20__1}).  The current density is super-GJ,
$\jm>\GJ{j}$, there are too few positrons in the gap and
$|\eta|>|\GJ{\eta}|$ what gives rise to a strong accelerating electric
field in the gap (see plot for $E$ at $t=9.636$ in
Fig.~\ref{fig:ctss_jm20__1}).  Both electrons extracted from the NS
surface and positrons flowing from higher altitudes are being
accelerated in the gap and as the gap widens their maximum energies
get higher and they become the primary particles igniting the new
discharge cycle (snapshot at $t=9.744$ in
Fig.~\ref{fig:ctss_jm20__1}).

Filling of the gap with pair plasma proceeds quickly, as the pairs are
injected throughout the whole length of the gap (time shots at
$t=9.780,9.816$ in Fig.~\ref{fig:ctss_jm20__2}).  This happens because
both electrons and positrons, moving in opposite directions, are
emitting pair creation capable photons, and their number densities are
comparable.  The gap is filled with plasma before the particles
created in the previous burst of pair formation leave the higher
altitude region, and so, in contrast to the cases described in
\Ss\ref{sec:j_negative},~\ref{sec:j_greater} the higher altitude
regions are always filled with dense pair plasma.

Screening of the accelerating electric field during each burst of pair
formation proceeds slightly different than in cases described in
previous sections.  Particles are injected throughout the whole length
of the gap and not at the gap's one end.  Simulations shown in
Fig.~\ref{fig:ctss_jm20__2} have two ``centers'' within the gap where
screening starts.  In Fig.~\ref{fig:wave_jm20} we show in more detail
than in Fig.~\ref{fig:ctss_jm20__2} how the accelerating electric
field is being screened by newly created pair plasma about the time
$t=9.816$.  Fluctuations of charge and particle number densities are
significantly less pronounced than those in Fig.~\ref{fig:wave_jp05}
because the accelerating field was due to mismatch of the charge
density $|\jm/c-\eta|<|\GJ{j}|$.  The region of screened electric
field spreads from two centers towards the edges of the gap.  In
Fig.~\ref{fig:ctss_jm20__2} we show spreading of the second zone with
the center around $x\simeq0.2$; the first ``center'' was at
$x\simeq0.1$, the field screening started there earlier and electric
field is already poisoned here.  This spreading occurs again in the
form of an electrostatic wave which phase velocity is larger that $c$
and amplitude decreases with time.

The position where the gap appears depends on the imposed current
density and on the variation of the GJ charge density -- the higher
$\jm$ and the larger the  variation of $\GJ{\eta}$, the closer to the NS the
gap starts developing.  The accelerating electric field in this gap is
smaller than that in the case with the constant GJ charge density from
\S\ref{sec:j_greater} because the number density of positrons is
high and so not the whole charge density $\jm/c$ is contributing to
the electric field.  If this electric field is not enough to produce
pairs, then the gap grows up to high altitude until most positrons
leave the domain;  then a vacuum-like gap, with the charge density
of the order of $\jm/c$, starts developing at the outer end of the
domain as in the case of constant GJ charge density from
\S\ref{sec:j_greater}.

The reason why variation of GJ charge density does not affect space
charge limited flow with sub-GJ current density is the presence of the
trapped particle ``cloud'' component which can adjust to any charge
density variation (assuming the current density remains sub-GJ), as
was mentioned in \S\ref{sec:cold_flow}.  For flow with anti-GJ current
density the accelerating electric field appears in the region depleted
of electrons, which on average move toward the NS.  In this case
regions closer to the NS always have the larger number density of
particles of the same sign as $\GJ{\eta}$ and will become charge
starved last, and so the gap develops at higher altitudes.  For flows
with the super-GJ current density when the absolute value of the GJ
charge density \emph{decreases} with altitude, $a<0$, fewer positrons
are necessary at lower altitudes to poison the electric field than at
higher altitudes.  Positrons are moving toward the NS and the low
altitude region becomes charge starved after all the others and the
gap develops at higher altitudes as in the case of the flow with a
constant GJ charge density.

Therefore, the regions with super-GJ flow can have two qualitatively
different behaviors depending on the value of the imposed current
density and the variation of the GJ charge density.  The cascade
either starts close to the NS and high altitude regions are always
filled with dense pair plasma, or the cascade starts at high altitude
and between the bursts of pair formation, the high altitude regions
are charge starved with vacuum-like accelerating electric field which
could give rise to an outer magnetosphere cascade zone.  In the first
case the period between successive bursts of pair formation should be
of the order the the gap's fly-by time and can be as short as
$\sim\PC{r}/c\sim$ fractions of microseconds; in the second case the
repetition rate will be larger than the fly-by time of the strong
field zone, $\sim{}R_*/c\sim$ tens of microseconds. From our toy model
we can not make quantitative prediction about the values of the
imposed current density and magnitudes of the GJ charge density
variations which separate those behaviors.

\subsection{On stationary cascades}
\label{sec:stationary-cascades}

\begin{figure*}
\begin{center}
  \includegraphics[width=\textwidth]{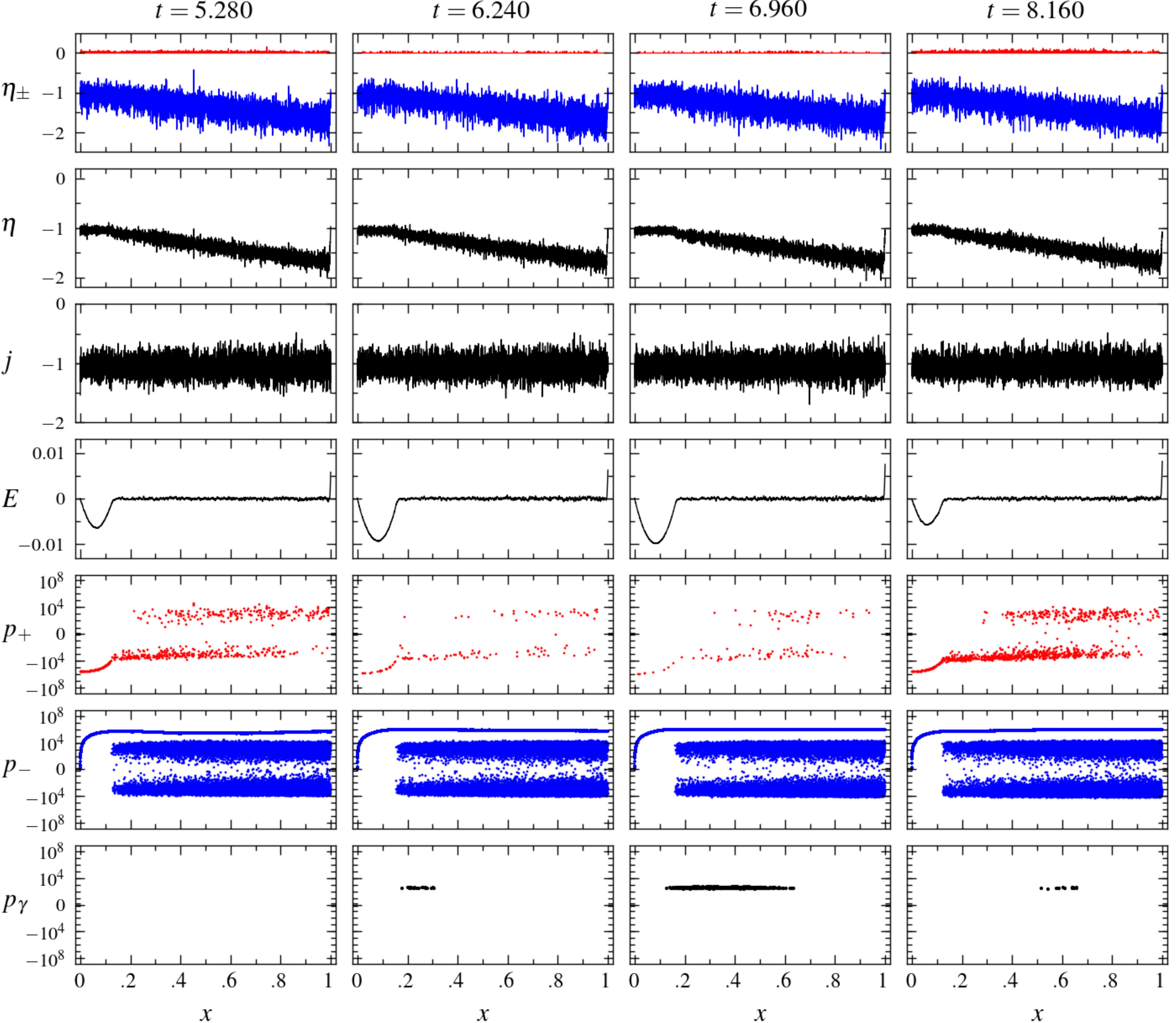}
\end{center}
\caption{Plasma flow with quasi-stationary cascade.  The imposed
  current density is $\jm=1.059\GJ{j}$, it is ``fine-tuned'' in such
  way that the accelerating electric gap near the NS surface does not
  disappear. All other physical parameters are the same as in the case
  shown in Figs.~\ref{fig:ctss_jm15__1},\ref{fig:ctss_jm15__2}. The
  same quantities are plotted as in Fig.~\ref{fig:ctss_jp05__1}. Note
  that the time intervals between snapshots are not equal.}
  \label{fig:ctss_jm1059}
\end{figure*}

In all models previously considered in the literature, the size of the
accelerating gap was limited by the opacity of the magnetosphere to
the gamma-rays: pairs were injected when the optical depth to pair
creation reached unity and those pairs screened the electric field.
In our simulations with the constant GJ charge density, the electric
field in the gap monotonically increases with distance, and the gap
grows with time.  That electric field could be screened only by
injection of particles with the charge sign opposite to the charge
sign of primary particles, and those particles, when injected, were
accelerated in the direction opposite to the direction of motion of
the primary particles by a strong electric field.  They start emitting
pair producing gamma rays after traveling a short distance, injecting
pairs and so screening the electric field in their way.  That led to
the motion of the gap as a whole or to the gap closure, if the other
end of the gap moved with subrelativistic velocity.

In the case of a flow with varying GJ charge density, described in
\S\ref{sec:j_negative_rho_variable}, the initial field was not a
monotonic function of the distance, but the gap was growing and,
again, the only way to limit the gap and screen the electric field was
by injection of particles of the opposite sign.  Once injected those
particles ultimately destroyed the gap, since their flux exceeded that
of the primary beam -- stationary equilibria with counter streaming
particles of the opposite sign exist only when the counter streaming
flux is less than the primary flux, as in the
\citet{AronsScharlemann1979} model.  Stationary particle acceleration
and pair creation would be possible if pairs are injected (mostly)
outside the gap, so that there will be not too many particles with
charge sign opposite to that of primary particles within the gap.

For example, the stationary flow model of \citet{AronsScharlemann1979}
constructs a finite length acceleration zone bounded by the stellar
surface below and the abrupt onset of pair creation above, which
screens the electric field in a thin layer (the pair formation front,
PFF). The space charge limited (electron) beam extracted from the
surface provides the primary particles that radiate pair creating
gamma rays. Almost all of the pair creation occurs above the PFF. A
small amount of trapping within the transitional PFF layer sends
charges of sign opposite to the primary beam (positrons) back down
toward the surface with flux small compared to that of the primary
beam, while extra electrons are added to the beam, restoring the
charge density to equality with the Goldreich-Julian density. This
restoration causes the screening (``poisoning'') of the accelerating
field. In that model, $E \rightarrow 0$ is imposed as a boundary
condition, using the conclusion that dense pairs ($n_\pm \gg
|\GJ{\eta}/e|$) flash into existence at a rather well defined height
-- a physically correct conclusion when curvature radiation dominates
the gamma ray emission and one photon absorption in the magnetic field
is the major opacity source, since each primary beam particle emits
many pair creating gamma rays and opacity is a rapidly varying
exponential function of the photon energy.  The application of that
boundary condition, along with the space charge limited flow
condition, has the consequence that the charge density of the beam has
a unique value $\eta_{\mathrm{beam}}$.  Since $j =
c\eta_{\mathrm{beam}}$, this model becomes discrepant with the
magnetospheric current density $\jm$, in general.

We tried to find an Arons \& Scharlemann like solution with the
acceleration gap limited by the pair formation front.  For the
linearly varying GJ charge density given by eq.~(\ref{eq:eta_x}) and
model parameters from \S\ref{sec:j_negative_rho_variable} we explored
the parameter space by running different simulations for different
values of the imposed current density starting from
$\jm=2\GJ{j}^{\,0}$ and decreasing it up to $\jm\simeq\GJ{j}^{\,0}$
when pair production ceased.  In each subsequent simulation with
smaller $\jm$ the position of the first pair injection moved further
from the NS because the accelerating electric field was also smaller.
In all cases when pair injection occurred inside the gap the gap
boundary starts moving -- we never saw a stationary PFF (nor did we
see a stationary PFF in any other simulations).  So, the conclusion
implied by the simulations is that the time asymptotic state is not a
steady flow similar to Arons \& Scharlemann's. Even though
acceleration in the charge starved gaps is limited by pair creation
with thin boundaries between pairs and quasi-vacuum, reminiscent of
Arons \& Scharlemann's PFFs, the gaps move, either up into the
magnetosphere or down toward the star -- there is no truly stationary
state, and fully developed limit cycles appear to be the answer, at
least in 1D.

We now exhibit a novel steady flow model, which takes advantage of
the properties of non-neutral beam flow when $\GJ{\eta}/B$ is
nonuniform, that does not require poisoning by the pairs to produce $E
= 0$ at the gap's top.  This model exploits the variation of the
charge density with distance, in the case when the current density is
larger than the local GJ current density only up to some height $h_s$,
then at distances larger than $h_s$ -- where the non-neutral flow
become sub-GJ -- the electric field will be screened on the scale of
the local Debye length, as discussed in \S\ref{sec:cold_flow}.  In
such a situation the accelerating electric field exists only up to the
distance $h_s$, above which it is shorted out by a mixture of trapped
electrons - those with two turning points in their orbits - and free
pairs.

We assume the GJ charge density varies linearly with the distance,
eq.~(\ref{eq:eta_x}), as in \S \ref{sec:j_negative_rho_variable}.  By
hypothesis, the flow is steady and the current density is equal to the
imposed current density $\jm = \xi \GJ{j}^{\,0}$, $\xi>1$. (see
eq.~(\ref{eq:dE_dt})).  If $\xi-1<a$, at some point the non-neutral
beam flow becomes sub-GJ.  From the Poisson equation for the electric
field~(\ref{eq:dE_dx}), in the absence of pairs we have
\begin{equation}
  \label{eq:E_stationary_gap}
  E_s = 
  4\pi\GJ{\eta}^0 L
  \left[ 
    \left( \xi-1 \right) \frac{x}{L} - 
    \frac{a}{2} \left(\frac{x}{L}\right)^2 
  \right]\,.
\end{equation}
At the distance
\begin{equation}
  \label{eq:h_s}
  h_s=2\frac{\xi-1}{a} L
\end{equation}
$E_s$ changes sign as the flow become sub-GJ.  Identifying $h_s$ with
the gap height yields the gap's potential drop to be
\begin{equation}
  \label{eq:phi_stationary_gap}
  \Delta{}V_{s,\,\mathrm{gap}}=\frac{8}{3}\pi\GJ{\eta}^0L^2\frac{(\xi-1)^3}{a^2}\,
\end{equation}
increasing in proportion to the imposed current density.  If the
resulting potential drop in the gap is large enough for particles to
inject pairs within the gap with number large enough to form a
positron back flux larger than the primary electron flux, the flow
will be non-stationary, like that described in
\S\ref{sec:j_negative_rho_variable}.  But if the potential drop is
such that copious pairs will be injected at distances larger than
$h_s$, positrons will be subjected only to a fluctuating, relatively
small electric field, which sustain the cloud component at altitudes
higher than $h_s$, where the flow is sub-GJ.  Only a fraction of the
positrons will be advected into the gap, and, if this fraction will be
much smaller than the GJ charge density the flow in the gap will be
not disturbed enough to make it non-stationary.

As an example of a flow with $E(h_s) = 0$ in the non-neutral current
flow, pair creation \emph{and} non-neutral trapping, we show snapshots
in Fig.~\ref{fig:ctss_jm1059} of the flow with the same parameters as
in \S\ref{sec:j_negative_rho_variable}, except with a carefully chosen
current density which was set large enough to allow pair formation
within the domain (at $x<x_B$), but small enough so that the particle
back flux does not destroy the gap.  The imposed current density in
this model is $\jm=1.059\GJ{j}^{\,0}$.  Above the altitude
$x=h_s\approx0.17L$ the flow is sub-GJ and clearly has beam-cloud
structure qualitatively similar to that in sub-GJ flows from
\S\ref{sec:cold_flow} (see phase portraits for electrons).  The
charged cloud of trapped electrons screens the electric field at
$x>h_s$.  Primary electrons, when accelerated up to high energies,
emit gamma-rays which decay into electron positron pairs \emph{above}
the gap.

Both secondary electrons and positrons mix with the cloud component
and some positrons are advected toward the gap.  The gap shrinks a bit
and the resulting potential becomes smaller than a critical value
necessary to sustain pair production.  When the number density of
positrons drops, the gap gets bigger and the pair formation starts
again.  The gap, however, never disappears and the cascade is close to
a stationary configuration.  This configuration is quite sensitive to
the imposed current density; large gap fluctuations appear at imposed
current densities only a few per cent larger than
$\jm=1.059\GJ{j}^{\,0}$.  Our model, however, has an exaggerated
charge density gradient and a very high voltage and we expect that the
mixing of positrons and their advection for real pulsar parameters
would be less efficient and, hence, there might be a larger parameter
range (i.e. an interval of the imposed current densities) where this
gap plasma flow can sustain quasi-stationary cascades.

Such stationary configurations require the imposed current density
being larger that the local GJ current density at the NS surface but
then becoming less than the local GJ current density at some distance
from the NS.  Variations of the GJ charge density (after accounting
for the decreasing magnetic field) would be not larger that
$\simeq15\%$
\citep{Muslimov/Tsygan92,Hibschman/Arons:pair_multipl::2001} and,
hence, the imposed current densities for stationary cascades are
within the range $|\jm-\GJ{j}^{\,0}|\lesssim{}0.15, \jm/\GJ{j}>1$.  Taking into
account that the resulting gap should not be very large to prevent
pair injection within the gap, that range is even smaller.

Fig.~\ref{fig:bai} makes clear that in the force-free theory, the
current density mostly does not fall in this narrow range, lending
support to the conclusion that limit cycle and trapping behavior are
the generic physics for polar flow, with and without pairs.

\section{Discussion}
\label{sec:discussion}

These calculations, and those reported in
\cite{Timokhin2010::TDC_MNRAS_I}, were designed to investigate the
theoretical issue of how the pulsar magnetosphere with current flow
determined by the magnetosphere's global dynamics couples to the
neutron star through the polar cap beam acceleration and pair creation
zone.  This problem was investigated 30-40 years ago, primarily using
order of magnitude estimates and analytical, steady flow models of
charge separated beams,  the
  accelerator was modeled as having
voltage fixed by pair creation and geometry. That led to a
determination of the charge density in the accelerator with the
current then simply being charge density times the speed of
light\footnote{In differing degrees, the charge density was made up of
  counter streaming relativistic beams, leading to current density
  lying between $\GJ{\eta}c$ and $2\GJ{\eta}c$.}.
No effort was made to show that the current density estimated actually
matched that required by the global structure -- the models yielded
currents with the correct order of magnitude, and in the absence of
actual solutions for the global structure of the magnetosphere, that
answer was regarded as good enough, although skepticism was expressed
that operating the accelerator model as having fixed voltage actually
could produce the correct answer (\citet{Mestel1999} and references
therein).  The calculations reported here show that indeed, when the
accelerator is operated with current fixed rather than voltage, the
polar cap accelerator does work, and works in a fully time-dependent
manner, as first conjectured by \citet{GJ} and \citet{Sturrock71}, but
with behavior different from previously published models%
\footnote{Following \citep{Shibata91}, when the pulsar is not near
  ``death valley'' in $P-\dot{P}$ space, the load inserted into the
  magnetospheric circuit by the pair creation discharges creates a
  small perturbation of the magnetospheric current $j_m$, thus
  justifying our treatment of $j_m$ as an external parameter in the
  description of the discharges.}

Our results show that when the acceleration and cascades are one
dimensional, acceleration and pair creation exhibits limit cycle
behavior, with cycle time somewhat larger
  than the relativistic flyby time over the length of the
accelerator.  When the 1D approximation is realistic (very young
pulsars), the length of the accelerator is less than the polar cap
diameter $\PC{r}\approx144 P^{-1/2}$ meters, suggesting
quasi-periodicity in the acceleration and pair creation on time scales
$\gtrsim 10^{-6} P^{-1/2}$ seconds.  As was suggested long ago, such
variations, if reflected in the radio emission, might be the origin of
the microstructure in the radio emission
\citep{RudermanSutherland1975}.  If limit cycle behavior is robust
with respect to multi-dimensional considerations, this dynamical
behavior might turn out to be a useful model for microstructure (in
contrast to non-linear optical phenomena in the transfer of radio
waves through the pair plasma, for example).

When charges are fully bound to the surface (solid surface with high
binding energy and no atmosphere, \citealt{Medin2010}), current flow
always adjusts to the magnetospheric load through pair creation
discharges, as was shown in \PapI.  The lowest energy particles in the
pair discharge exhibit transient trapping in the fluctuating electric
field.  Those reversals of particle momenta allow both current and
charge densities to adjust to the presumed force-free conditions, for
any value of the magnetospherically imposed current density.

At long periods/weak $B$, pair creation becomes impossible, the
magnetosphere is a vacuum with no conduction current flow and spin
down occurs through generation of strong vacuum waves
\citep{Pacini1967}.  Pulsars enter this regime by crossing the pair
creation ``death valley'', theoretically expected for magnetospheric
voltages $\Vm = \sqrt{W_{\textrm{\tiny{}R}}/c} \approx
10^{13}\,(\dot{P}/10^{-15})^{1/2}P^{-3/2} \; \mathrm{V} \lesssim
10^{12} \equiv \Phi_{\textrm{death}}$ V (a restatement of the death
line based on curvature gamma ray emission and one photon magnetic
conversion estimated by \citet{Sturrock71}).  That this death line
describes the disappearance boundary of radio pulsars in the $P, \;
\dot{P}$ diagram reasonably well (see e.g. Fig.~15.1 in
\citet{Arons2009}) provides a strong hint that low altitude, polar cap
pair creation with high energy beam acceleration and one-photon
magnetic conversion of the curvature $\gamma$-rays has something to do
with pulsar activity.

A warm, quasi-neutral atmosphere plasma atmosphere overlying the
magnetized ocean and solid crust, with no restrictions on charged
particle motion along $B$, is the likely surface state of most
pulsars, either because of the residual heat of the neutron star, or
because of polar cap heating by precipitating particles from the
magnetosphere (from local pair discharges or from the return current
flow).  The upper atmosphere then can freely supply charge, in a
manner very similar to a space charge limited current flow from the
cathode of a classical vacuum tube.  Our simulation results on space
charge limited flow with (and without) pair creation reveal a variety
of noteworthy new features.

We have made the first determination of the polar accelerator's
behavior under conditions where the current density, not the voltage,
is held fixed.  We found that space charge limited flow is not
universally high energy.  Emission of pair creating gamma rays, does
not occur on all polar field lines, even when
$\Vm>\Phi_{\mathrm{death}}$.  The force-free model of the
magnetosphere suggests the polar current flow includes sub-GJ flow
regions with $0<j/\GJ{j}<1$, where the low energy current flow (with
the Lorentz factor $\gamma_{\mathrm{beam}}\lesssim{}3$) with no
(curvature) gamma ray emission and no pair creation (inverse Compton
gamma ray emission and $\gamma$-$\gamma$ conversion to pairs are both
negligible), with adjustment of the current density and charge density
to the force free values through formation of a trapped particle,
non-neutral hanging charge cloud.  This kind of quasi-stationary flow
occupies most of the polar flux tube for the aligned and almost
aligned rotator, but progressively disappears as the obliquity
increases.  The force free model also requires regions of distributed
return current flow, $j/\GJ{j}<0$, and, as the obliquity increases,
regions of super-GJ current flow, $j/\GJ{j}>1$, occupying most of the
polar flux tube as the obliquity approaches $90^\circ$.  Both the
return current and super-GJ regimes exhibit unsteady, high energy
current flow (an unsteady beam), driving discharges copiously creating
pairs whose character is similar to that encountered when the
atmosphere is absent and the surface is a strongly bound solid.  These
regions, as projected onto the polar cap, are shown in
Fig.~\ref{fig:bai}.  In the special case of $\jm/\GJ{j}^{\,0}$ being
slightly larger than unity, the plasma flow can sustain
quasi-stationary cascades, and a new class of such models was
described in \S\ref{sec:stationary-cascades}.  But generally speaking,
such stationary regimes represent a singular case, rarely if ever
achievable by magnetospheres described by the force-free model.

The only previous quantitative study of time-dependent cascades in
space charge limited flow is that of \citet{Levinson05}.  They used a
two-fluid model and concluded that chaotic pair formation takes place
throughout the whole zone where pair formation is possible.  Our
simulations do not support their conclusions.  Particle trapping plays
a significant role in adjusting of the plasma flow to the required
current density and, hence, the plasma can not be adequately
represented as two fluids (electrons and positrons) each with its own
unidirectional velocity.  This two fluid representations introduced
additional rigidity into the system and this, in our opinion, led to
formation of strong chaotic electric field everywhere in the domain.

The most recent analytical treatment of the problem was presented in
\citet{Beloborodov2008}.  Our results support Beloborodov's
(\citeyear{Beloborodov2008}) general conjecture about the character of
plasma flow in the force-free regime, that the sub-GJ flow is low
energy and pair formation is possible only in super-GJ or anti-GJ
current flow.  The simulations differ from his qualitative picture of
what would happen in several respects: a) the low energy flow with
trapped particles does not retain the spatial oscillations of the cold
flow, replacing those by the two-component beam/cloud structure -- the
warm beam has non-oscillatory velocity, while the trapped particles,
not included in his picture, are those with two turning points in
their orbits; b) bursts of pair cascades repeat after longer time than
$h/c$, $h$ being the gap height; and c) in most cases the gap is not
destroyed but moves as a whole, usually relativistically.

The discharges were given a novel treatment.  After the seminal
efforts of \cite{RudermanSutherland1975} but prior to our work, models
universally incorporated their assumption that when pair creation is
copious (many convertible gamma rays per primary high energy particle,
as is the case in the curvature radiation cascades typical of young,
high voltage pulsars), at and above the height where pairs start
appearing, $E_\parallel$ drops to zero and stays that way at all
greater altitudes. \cite{AronsScharlemann1979} formalized this into
the dynamics of a pair formation front (PFF), showing that when pair
creation is copious, the transition between a charged starved region
where $E_\parallel \neq 0$ and the pair dominated region where
$E_\parallel$ can be set equal to zero is thin -- that structure
necessarily involves one sided particle trapping, under pulsar
conditions, thus causing the formation of counter streaming components
in the current flow.  Our solutions exhibit this transition between
the dense plasma and the quasi-vacuum regions outside.  But, since
$E_\parallel$ was obtained from a dynamical field equation, we nowhere
assumed $E_\parallel$ had to be zero everywhere outside (above or
below) a pair discharge. Instead, the pair creation itself creates the
polarizable plasma that dynamically resets $E_\parallel$ to zero
inside the discharge, while outside the electric field was allowed to
float to whatever is dynamically required.  The resulting models thus
exhibit macroscopic intermittency -- a pair discharge shorts out
$E_\parallel$, then as the cloud of pair plasma moves out into the
magnetosphere or toward the surface, a gap reforms in which residual
particles accelerate, radiate gamma rays and form a new discharge.
The fact that pair creation opacity declines with distance from the
neutron star was modeled by setting the magnetic field equal to zero
in the upper part (typically the upper 30\%) of the simulation domain.
That \emph{Ansatz} forces the one photon pair creation opacity to be
zero in the upper part of the domain.

It must be said, however, that the resulting model of the
intermittency is very simplified -- probably over-simplified.  The
representation of the ``magnetosphere'' as a region of negligible pair
creation optical depth to gamma rays emitted from just above the polar
cap is the right general idea, but the 1D character of our simulations
almost certainly leads to an overemphasis on the coherence of the
intermittent discharges -- they form coherent slabs of pair plasma,
separated by coherent slabs of quasi-vacuum.  In reality, the low
opacity region where discharges must end appears at heights $\sim
R_*\gg\PC{r}$ = polar cap and polar flux tube transverse radius.  Then
the acceleration zone left behind a discharge created plasma cloud is
long and skinny, with the narrow ``wave-guide'' geometry (and the
physics of the polar flux tube's boundary) playing an essential role
in the nature of the quasi-vacuum field.  Under these circumstances,
it is unclear and unknown whether the discharges would preserve
anything like the coherence exhibited in the results reported here --
formation of multiple ``lightning bolts'' is perhaps more likely,
simultaneously coexisting within the polar flux tube, the electric
fields of the discharges affecting the dynamics of their neighbors,
causing the discharges to influence each other - a complex dynamical
system quite likely to be chaotic .

In addition, the extent of the gaps formed between discharges requires
consideration of all the sources of pair conversion opacity, not only
one photon conversion in the $B$ field.  Even when magnetic conversion
drops to zero, $\gamma$-$\gamma$ conversion with gamma rays colliding
with soft photons from the atmosphere (either heated polar cap or
overall warm surface, if the star is young enough) can provide
discharge initiating pairs within quasi-vacuum gaps, limiting the
extent of such gaps to less than what occurs when the opacity is set
equal to zero.  These issues require multi-dimensional modeling of the
accelerator and the photon transfer, as well extending the radiation
transfer model, improvements required before the possible direct
consequences of low altitude discharges for observations can be
addressed. Those consequences are the pair multiplicity of the
outflow, the heating of the polar caps by the discharge components
that move toward the neutron star, and possible direct collective
emission of photons by the time dependent currents in the discharges.

The multiplicity ($\kappa_\pm \equiv n_{\mathrm{pair}}/\GJ{n}$) within
individual pair creation bursts show the traditional results -- when a
discharge occurs, $\kappa_\pm \approx \gamma_{\mathrm{beam}} m_\pm
c^2/\varepsilon_{\mathrm{curvature}} \sim 10^3$
($\varepsilon_{\mathrm{curvature}}=$ energy of pair producing
curvature photons), somewhat enhanced by the conversion of the
synchrotron photons in the cascade -- thus, multiplicities within
individual bursts up to $\sim 10^4$ are observed. Intermittency
reduces the overall multiplicity of the outflow -- the magnitude of
this reduction is hard to judge without a multi-dimensional model.
The intermittency reduction is offset by continued pair creation as
the plasma clouds and their high energy leading edges drift out to
heights of order the stellar radius and more, emphasizing again the
needs for a multi-D treatment. However, the multiplicity is unlikely
to be as high as is inferred from pulsar wind nebulae
\citep{BucciantiniAronsAmato2011}, so long as the simple, star
centered dipole $B$ field model is retained. More general $B$ fields,
with smaller radii of curvature of photon orbits with respect to $B$,
can lead to enhanced pair creating opacity and pair yield -- the
offset dipole model \citep{Arons1998, Harding2011} is the most
plausible magnetic anomaly model that can be consistent with the
dipolar character of the polar flux tube revealed by radio
observations \citep{RankinIV,Kramer1998}.  Whether sufficiently large
multiplicities can be obtained while respecting the geometric
constraints obtained from the radio polarimetry is under
investigation.

Polar cap heating and consequent soft X-ray emission provides another
possible observable that can constrain the model. In common with all
discharge models, roughly half the energy in each discharge is
deposited in the crust below the atmosphere and ocean.  If
intermittency is neglected, existing observations provide strong
challenges to all polar cap discharge models, including ours.
Intermittency reduces the average energy flux.  Whether that effect
allows this polar discharge model to survive confrontation with
observations (or, better, prove useful in modeling such observations)
remains to be studied. Simulating the particle back flux from a
discharge is a particular challenge, requiring resolving the Debye
length in the pairs, which rapidly decreases as the pair density
grows.

Our results imply discharges occur on some of the polar field lines
for all inclinations.  These discharges incorporate time dependent,
quasi-coherent currents on microsecond and shorter time scales.  It
has not escaped our attention that such fluctuations might be a direct
source of radio emission from the low altitude polar flux tube, a
region strongly suggested as the site of the radio emission by the
radio astronomical phenomenology.  Although the electric fields in our
one-dimensional model are wholly electrostatic, therefore cannot leave
the plasma, in a more general multi-D setting which has substantial
spatial inhomogeneity transverse to $B$, the field components parallel
to $B$ have accompanying components $E_\perp$ perpendicular to ${B}$
which make the fluctuations fully electromagnetic.  Therefore, these
electrostatic spectra {\it may} be representative of electromagnetic
fluctuations which can leave the plasma, and the pulsar.  From the
point of view of simulations, a multidimensional, electromagnetic
treatment is required in order to investigate this
possibility\footnote{\cite{Fawley/PhDT:1978} used a linear response
  analysis to study this possibility, in analytic model of discharges
  when strong surface binding suppresses fee emission.  The work
  described in this paper is the first to study fully non-linear
  discharges in the free emission regime. Whether this idea for radio
  emission will lead to a useful model remains to be studied.}.

\begin{figure}
\begin{center}
  \includegraphics[width=\columnwidth]{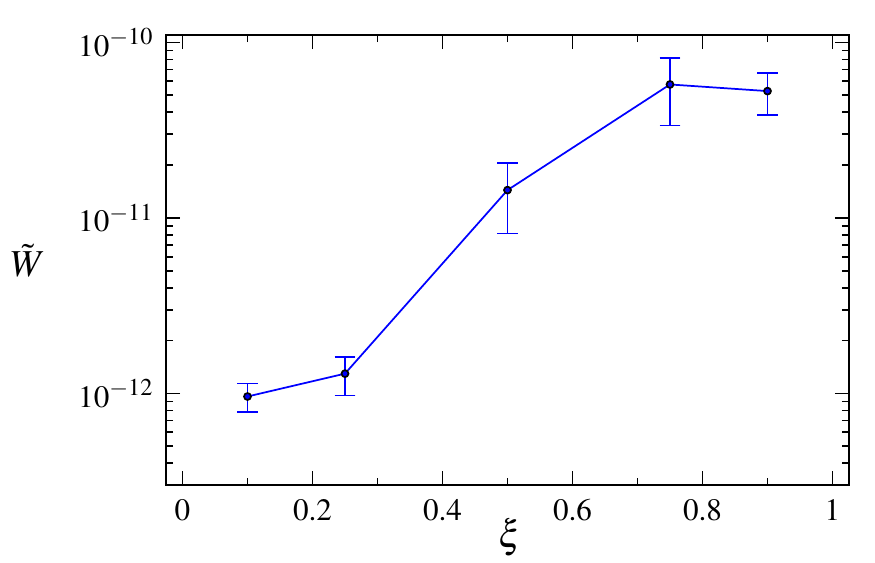}
\end{center}
\caption{Estimated energy flux $\tilde{W}$ in plasma waves for sub-GJ
  flows as a function of $\xi=\jm/\GJ{j}$. $\tilde{W}$ is normalized
  to $W_{\mathrm{md}} B_{12}^{-1} P^2 $.}
  \label{fig:w_cold_flow}
\end{figure}

In the frame of our model we can provide only estimates of the energy
available for such directly excited waves.  If the electrostatic
oscillations could form an electromagnetic wave which leaves the
plasma then the energy carried by the wave would be of the order
\begin{equation}
  \label{eq:W_oscillations}
  W_{\mathrm{r}} \sim \frac{<E^2>}{8\pi}\, c \pi \PC{r}^2\,.
\end{equation}
Both kinds of flow, with or without pair formation, cause fluctuating
electric field.  Fluctuating electric field in the sub-GJ flow turns
out to be too low to provide the energy even for the radio
emission. The electric field for the sub-GJ flow shown in
Fig.~\ref{fig:e_acc_cold} is normalized to
$E_0=|\GJ{\eta}|\pi\lambdaDGJ$; in this normalization the estimates for
$W_{\mathrm{r}}$ is given by
\begin{equation}
  \label{eq:W_R_subGJ}
  W_{\mathrm{r}} \sim  4.8\cdot10^{-9}\,<\tilde{E}^2> W_{\mathrm{md}} B_{12}^{-1} P^2  \,,
\end{equation}
where $\tilde{E}$ is the normalized electric field and
$W_{\mathrm{md}}$ is magnetodipolar energy losses
\begin{equation}
  \label{eq:W_md}
  W_{\mathrm{md}} = \frac{B_0^2 R_*^6\Omega^4}{4 c^3}\,.
\end{equation}
In Fig.~\ref{fig:w_cold_flow} we plot the estimated values of
$\tilde{W_{\mathrm{r}}}=W_{\mathrm{r}}/B_{12}^{-1}P^2W_{\mathrm{md}}$
as a function of $\xi=\jm/\GJ{j}$.  It is evident that even for the
Crab pulsar, known for its very low radio efficiency in terms of spin
down energy losses $\sim10^{-8}$, the energy in electrostatic
oscillations of the cold flow can not power the radio emission.  For
plasma flows with pair formation, screening of the electric field
proceeds similarly to cascades in the Ruderman-Sutherland model
studied in \PapI.  We observed formation of superluminal plasma waves
during discharges in space charge limited flows, similar to what was
found for strongly bound surfaces.  The range of plasma oscillation
wavelengths was rather broad, similar to what was observed in
discharges discussed in \PapI. The electric field for plasma flows
with pair formation discussed in \S\ref{sec:j_pairs} is normalized to
$E_0=|\GJ{\eta}|\pi\PC{r}$; in this normalization the estimates for
$W_{\mathrm{r}}$ is given by
\begin{equation}
  \label{eq:W_R_pairs}
  W_{\mathrm{r}} \sim \frac{1}{4} <\tilde{E}^2>  W_{\mathrm{md}}\,.
\end{equation}
The amplitude of the (normalized) oscillating electric field during
the screening phase of cascade development shown in
Fig.~\ref{fig:wave_jp05},~\ref{fig:wave_jm20} is $\sim0.1-0.01$ and
the energy in such oscillations is more than enough to power the radio
emission. So, from the energetics point of view, it seems that the
plasma flows with pair formation studied here are candidates for
powering the pulsar radio emission through direct radiation by the
discharge currents .

Finally, our work may have implications for the formation of the
particle accelerators in the outer magnetosphere required to account
for the gamma ray pulsars.  Low voltage, sub-GJ current flow may have
a \citet{Holloway1973} ``problem'', in that the non-neutral cloud and
low energy, non-neutral current-carrying beam cannot cross the null
surface where $\vec{\Omega} \cdot \vec{B} = 0$ -- in contrast, field
lines populated with dense plasma from discharges have no difficulty
with plasma flowing across the null surface, the pair plasma clouds
are quasi-neutral and easily adjust to the locally required charge
density.  Field lines passing from the low altitude trapped cloud to
the outer magnetosphere that go through the null surface might form a
physically self-consistent gap reminiscent of the earliest proposals
for ``outer gaps'', with potential for outer magnetosphere discharges
that could provide a model for the observed gamma ray emission.  This
issue also requires multidimensional modeling.  The caustic formation
exhibited by models for gamma ray pulses suggest that if such gaps can
form, they are localized (by pair creation?) to regions close to the
flux bundles where return currents flow.

Another possibility for particle accelerators in the outer
magnetosphere exists in the regions carrying the return current.  At
some time between the bursts of pair formation in the polar cap the
electric field at large latitudes can not be screened by pairs created
in the polar cap.  This field can accelerate particles and give rise
to pair creation via $\gamma$-$\gamma$ process.  As we pointed out at
the end of \S\ref{sec:j_negative}, an outer magnetosphere cascade zone
might exist along the magnetic field lines carrying the return
current. This might be a very intriguing possibility in view of the
observational evidences that the spectrum of pulsar gamma-ray emission
does not have a super-exponential cut-off, which should be present if
there were absorption of gamma-rays in the magnetic field.  Moreover,
modeling of pulsar gamma ray light curves suggest that the gamma-ray
emission originates from the regions close to the boundary between the
open and closed magnetic filed lines
\citep[e.g.][]{BaiSpitkovsky2010a}.  The return current regions for a
broad range of pulsar inclination angles are close to that boundary
or/and are within it, e.g. in Fig.~\ref{fig:bai} the return current is
flowing in and around the current sheet for pulsar inclination angles
$\alpha\lesssim30^\circ$

\section{Conclusions}
\label{sec:conclusions}

Our principal conclusion is simple -- pair creation can occur at
pulsar polar caps, but (almost) always in the form of fully time
dependent current flow (microsecond time scales for the variability).
That time dependence with pair creation allows the current to adjust
to any magnetospheric load, while simultaneously allowing the charge
density to adjust to the requirements of the force free
magnetosphere. We have also shown that a substantial fraction of the
open field lines (fraction decreasing with increasing obliquity) solve
the current flow problem with a low energy, non neutral beam carrying
the current co-existing with a non-neutral, electrically trapped
particle cloud.  This is an essentially time independent local
solution.  Exploring the consequences of these new results for global
theory and observations requires extending the calculations to
multidimensional, electromagnetic models for the accelerating electric
field, and perhaps to background magnetic field models more general
that the star centered dipole geometry used here in the choice of the
magnetic radius of curvature that enters into the pair conversion
opacity.

\section*{Acknowledgments}

We wish to thank Xuening Bai for making the plot shown in our
Fig.~\ref{fig:bai}. This work was supported by NSF grant AST-0507813;
NASA grants NNG06GJI08G, NNX09AU05G; and DOE grant DE-FC02-06ER41453.
AT was also supported by an appointment to the NASA Postdoctoral
program at NASA Goddard Space Flight center, administered by ORAU.

\bibliographystyle{mn2e} 

\bibliography{tdc_2}

\appendix

\section{One-dimensional electrodynamics of the polar cap}
\label{sec:1D-Electrodynamics}

Let us decompose the magnetic field in the acceleration zone into two
terms $\vec{B}=\vec{B}_0+\delta\vec{B}$, where $\vec{B}_0$ is the
global time-averaged field (the ``external'' field for our local
problem, $=\vec{B}_{\mathrm{magnetosphere}}$ ) -- and $\delta\vec{B}$
is the fluctuating magnetic field due to local currents in the system
caused by charge motion in the acceleration zone.

In our 1D approximation the characteristic size of the acceleration
zone in longitudinal direction, along magnetic field lines, $l$ is
much smaller than its characteristic width $w$ (the latter being of
the order of $\PC{r}$).  The characteristic timescale of
electromagnetic field variations due to relativistic charge motion is
$\tau\sim{}l/c$, as charges move along magnetic field lines. The
characteristic scale of the global magnetic field variation in
longitudinal direction $l_{\mathrm{mag}}$ is much larger than both $l$
and $w$.  The relation between these scales is
$l\ll{}w\ll{}l_{\mathrm{mag}}$.

Electromagnetic field can be determined from Ampere's and Faraday's
laws
\begin{eqnarray}
  \nabla\times\vec{B}_0 + \nabla\times\delta\vec{B} & = &
  \frac{4\pi}{c}\vec{j} +
  \frac{1}{c}\frac{\partial\vec{E}}{\partial{}t}   
  \label{eq:Ampere_Law_B_deltaB_Apendix} 
  \\ 
  \nabla\times\vec{E} & = & 
  - \frac{1}{c}\frac{\partial\,\delta\vec{B}}{\partial{}t}\,.
  \label{eq:Faraday_Law_B_deltaB_Apendix}
\end{eqnarray}
Combining Ampere's (\ref{eq:Ampere_Law_B_deltaB_Apendix}) and
Faraday's (\ref{eq:Faraday_Law_B_deltaB_Apendix}) laws by eliminating
the electric field we get an equation for the magnetic field
$\delta\vec{B}$:
\begin{equation}
  \label{eq:deltaB_Equation_General_Apendix}
  \frac{1}{c^2}\frac{\partial^2\,\delta\vec{B}}{\partial{}t^2} -
  \nabla^2\delta\vec{B} =
  \frac{4\pi}{c}\nabla\times\vec{j} + \nabla^2\vec{B}_0\,.
\end{equation}
Only the perpendicular components of the magnetic field
$\vec{B}_{0,\perp},\delta\vec{B}_\perp$ affect the accelerating
electric field $\vec{E}_\parallel$.  Now using the perpendicular
component of eq.~(\ref{eq:deltaB_Equation_General_Apendix}) we
estimate $\delta\vec{B}_{\perp}$.

The estimates of each of the terms in
eq.~(\ref{eq:deltaB_Equation_General_Apendix}) are as follows.  For
terms with $\delta\vec{B}_\perp$ we have
\begin{equation}
  \label{eq:Laplacian_OOM_deltaB_Apendix}
  \left(\nabla^2\delta{}\vec{B}\right)_\perp \sim \frac{\delta{}B_\perp}{l^2} +
  \frac{\delta{}B_\perp}{w^2} \sim
  \frac{\delta{}B_\perp}{l^2}\,
\end{equation}
and 
\begin{equation}
  \label{eq:Laplacian_OOM_d2Bdt2_Apendix}
  \left(\frac{1}{c^2}\frac{\partial^2\,\delta\vec{B}}{\partial{}t^2}\right)_\perp
  \sim \frac{\delta{}B_\perp}{(c\tau)^2} \sim
  \frac{\delta{}B_\perp}{l^2}\,.
\end{equation}
The perpendicular component of the global magnetic field changes in
the longitudinal direction on the scale $l_{\mathrm{mag}}$, in the lateral
direction it changes on the scale $w$ and, hence,
\begin{equation}
  \label{eq:Laplacian_deltaB_OOM_Apendix}
  \left(\nabla^2\vec{B}_0\right)_\perp \sim \frac{B_{0,\perp}}{l_{\mathrm{mag}}^2} +
  \frac{B_{0,\perp}}{w^2} \sim
  \frac{B_{0,\perp}}{w^2}\,.
\end{equation}
Current flows along magnetic field lines and its perpendicular
component is of the order of
$\left(B_{0,\perp}/B_0\right)j\sim{}(l/\rho_c)j$, and as the radius of
curvature of magnetic field lines $\rho_c$ is much larger than any of
the scales in our problem we have
\begin{equation}
  \label{eq:Curl_j_OOM_Apendix}
  \left(\nabla\times\vec{j}\right)_\perp \sim 
  \frac{j}{w} - \frac{j_\perp}{l} \sim
  \frac{j}{w} - \frac{l}{\rho_c}\frac{j}{l} \sim
  \frac{j}{w}\,.
\end{equation}
Combining estimates
(\ref{eq:Laplacian_OOM_deltaB_Apendix})-(\ref{eq:Curl_j_OOM_Apendix})
from eq.~(\ref{eq:deltaB_Equation_General_Apendix}) we have
\begin{equation}
  \label{eq:deltaB_OOM_Apendix}
  \frac{\delta{}B_\perp}{w} \sim \left(\frac{l}{w}\right)^2 \left( \frac{4\pi}{c}j + \frac{B_{0,\perp}}{w}\right)\,.
\end{equation}

The order-of-magnitude version of Ampere's law
(\ref{eq:Ampere_Law_B_deltaB_Apendix}) is
\begin{equation}
  \label{eq:Ampere_Law_OOM_Apendix}
  \frac{B_{0,\perp}}{w} + \frac{\delta{}B_\perp}{w} \sim \frac{4\pi}{c}j + \frac{E_\parallel}{c\tau}\,,
\end{equation}
and from eq.~(\ref{eq:deltaB_OOM_Apendix}) it follows that the
fluctuating magnetic field due to charge motion introduces only a
second order term in $l/w$ and it can be neglected.  So, the Ampere
law in 1D has the form
\begin{equation}
  \label{eq:Ampere_Law_1D_Apendix}
  \frac{\partial{}E_\parallel}{\partial{}t} = 
  - 4\pi{} 
  \left( j - \frac{c}{4\pi}(\nabla\times\vec{B}_0)_\parallel \right) \equiv 
  - 4\pi{} ( j - \jm )\,. 
\end{equation}

The same expression for the accelerating electric field through the
current density as in eq.~(\ref{eq:Ampere_Law_1D_Apendix}) can be
obtained from the Gauss law and charge conservation (see e.g. \PapI,
Appendix A), in that case $\jm$ emerges as an integration constant
corresponding to the time-average current density flowing trough the
system.  In 1D the system is essentially electrostatic, charges create
only electric field, and naturally both the Ampere's law and the
Gauss' law reduce to the same equation.  The magnetic field
perpendicular to the background stellar $B$ field is negligible, both
due to the background magnetospheric current and due to the rapidly
variable currents within the acceleration zone considered in this
study - in particular, to lowest order in $(l/\rho_c)^2$, the particle
orbits are determined by the stellar field $B_0$ and consideration of
the full electromagnetism of the discharges is not required to
characterize the discharge dynamics.

\section{Stationary one-dimensional Space Charge Limited Flow}
\label{sec:stat-sclf}

\subsection{General equations}
\label{sec:sclf_general_Appendix}

Let us consider a stationary one-dimensional beam of equal particles
with mass $m$ and charge $q$ having the same sign as the GJ charge
density $\GJ{\eta}$ flowing from the surface of the NS at $x=0$ into
the magnetosphere at $x>0$.  The current density of the beam is a
fraction $\xi$ of the GJ charge density,
\begin{equation}
  \label{eq:j_ksi_jGJ_Appendix}
  j = \xi\GJ{j} \,.
\end{equation}
For the energy of particles in the beam we have
\begin{equation}
  \label{eq:energy_eq}
  mc^2\gamma + q\phi = mc^2\gamma_0 + q\phi_0\,,
\end{equation}
where $\gamma=(1-v^2/c^2)^{-1/2}$ is particle's Lorentz factor, $v$ is
particle's velocity, $\phi$ is the electric potential; quantities with
the subscript $0$ refer to their values at NS surface, at $x=0$.  The
electric potential $\phi$ is given by the Gauss law
\begin{equation}
  \label{eq:Gauss_Law_Appendix}
  \frac{d^2\phi}{dx^2} = -4\pi(\eta-\GJ{\eta})\,.
\end{equation}
where $\eta=j/v$ is the charge density of the beam. We consider the
case when $v,\gamma,\eta$, and $\phi$ are functions of distance $x$,
while $\GJ{\eta}$ is constant.

Expressing $\phi$ through $\gamma$ from eq.~(\ref{eq:energy_eq}) and
$\eta$ through $j= v\eta$ we get an equation for the beam's Lorentz
factor
\begin{equation}
  \label{eq:d2_gamma_dx_Appendix}
  \frac{d^2\gamma}{dx^2} = \frac{4\pi\GJ{\eta}q}{mc^2}
  \left(\xi\,\frac{\gamma}{\sqrt{\gamma^2-1}} -1\right) \,.
\end{equation}
The plasma frequency for a mildly relativistic plasma with the GJ
charge density consisting of particles with charge $q$ and mass $m$ is
\begin{equation}
  \label{eq:omegaPL_Appendix}
  \omegaPGJ = \left( \frac{4\pi\GJ{\eta}q}{m}\right)^{1/2}\,,
\end{equation}
and the Debye length (also the skin depth, since the characteristic
velocity is $c$) of such a plasma is
\begin{equation}
  \label{eq:lambda_Debye_Appendix}
  \lambdaDGJ = \frac{c}{\omegaPGJ}\,.
\end{equation}
Introducing  normalized distance $s\equiv{}x/\lambdaDGJ$
eq.~(\ref{eq:d2_gamma_dx_Appendix}) become
\begin{equation}
  \label{eq:d2_gamma_ds_Appendix}
  \frac{d^2\gamma}{ds^2} = \xi\,\frac{\gamma}{\sqrt{\gamma^2-1}} -1 \,.
\end{equation}
The first integral of this equation is easily obtained by multiplying
both part by $d\gamma/ds$
\begin{equation}
  \label{eq:d_gamma_ds_full_Appendix}
  \left(\frac{d\gamma}{ds}\right)^2 = 
  2 \left[ 
    \xi\left( \sqrt{\gamma^2-1} -\sqrt{\gamma_0^2-1}\, \right) -
    (\gamma-\gamma_0)
  \right]  + \left(\frac{d\gamma}{ds}\right)_0^2 
  \,.
\end{equation}
Under the conventional assumptions the space charge limited flow
starts at the surface with zero velocity and the electric field is
completely screened, so $\gamma_0=1$ and $(d\gamma/ds)_0=0$; with
these assumptions eq.~(\ref{eq:d_gamma_ds_full_Appendix}) become
\begin{equation}
  \label{eq:d_gamma_ds_Appendix}
  \left(\frac{d\gamma}{ds}\right)^2 = 
  2 \left( \xi\sqrt{\gamma^2-1} - \gamma + 1 \right)
\end{equation}

It is more convenient to analyze the properties of the flow in terms
of the spatial component of the 4-velocity (the normalized momentum
$p\equiv\gamma{}v/c$).  In terms of $p$
eq.~(\ref{eq:d_gamma_ds_Appendix}) is
\begin{equation}
  \label{eq:d_p_ds_Appendix}
  \left(\frac{dp}{ds}\right)^2 = 
  2\, \frac{p^2+1}{p^2} 
  \left( 1+ \xi{}p - \sqrt{p^2+1}  \right)\,.
\end{equation}
The right hand site of eq.~(\ref{eq:d_p_ds_Appendix}) is positive if
either $\xi\ge1$ or $0\le\xi<1$ \emph{and} $p<p_{\max}$, where
\begin{equation}
  \label{eq:p_max_Appendix}
  p_{\max} =\frac{2\xi}{1-\xi^2}\,.
\end{equation}
So, for $\xi\ge1$ flow will accelerate monotonically, while for
$0\le\xi<1$ the momentum will oscillate in the range $[0,p_{\max}]$.

According to eq.~(\ref{eq:d_p_ds_Appendix}) $dp/ds$ does not depend
explicitly on the distance $s$ and so its absolute value is the same
for the same value of $p$; therefore for $0\le\xi<1$ the function
$p(s)$ is symmetric around its maxima and minima and is periodic.

\subsection{Ultra-relativistic flow}
\label{sec:large_p_Appendix}

For $p\gg1$ we can neglect terms of the order $O(1/p)$ and
higher. Then eq.~(\ref{eq:d_p_ds_Appendix}) takes the form
\begin{equation}
  \label{eq:dp_ds_large_p_Appendix}
  \left(\frac{dp}{ds}\right)^2 = 
  2 \left[ 1 + p\left(\xi -1\right) \right]\,,  
\end{equation}
the solution of eq.~(\ref{eq:dp_ds_large_p_Appendix}) is
(cf. eq.~B3 in \citealt[][]{FawleyAronsScharlemann1977})
\begin{equation}
  \label{eq:p_large_Appendix}
  p = \sqrt{2}s +\frac{\xi-1}{2}s^2\,. 
\end{equation}

If $\xi\ge1$ the flow is continuously accelerating and its momentum at
large $s$ will grow as
\begin{equation}
  \label{eq:p_large_s2_Appendix}
  p = \frac{\xi-1}{2}s^2\,. 
\end{equation}

For $\xi<1$ the flow is oscillatory and its momentum periodically
reaches $p_{\max}$ (see eq.~(\ref{eq:p_max_Appendix})).  If $1- \xi
\ll1$ and $p_{\mathrm{max}} \gg 1$ the flow is relativistic almost
everywhere. The wavelength of such spatial oscillations is twice the
distance between the points where $p=0$ and $p$ reaches its maximum
value.  From eq.~(\ref{eq:p_large_Appendix}) the distance where $p$
reaches it maximum is $\sqrt{2}(1-\xi)^{-1}$ and so the period of
spatial oscillation is \citep{Beloborodov2008}
\begin{equation}
  \label{eq:s0_large_p_Appendix}
  s_0 = 2\sqrt{2}(1-\xi)^{-1}\,.
\end{equation}

\subsection{Non-relativistic flow}
\label{sec:small_p_Appendix}

Stationary space charge limited flow is non-relativistic at the
beginning, close to the NS surface; it can remain non-relativistic if
$\xi\ll{1}$ (see eq.~(\ref{eq:p_max_Appendix})).

For $p\ll{}1$ we can neglect terms of the order $O(p^3)$ and higher
and eq.~(\ref{eq:d_p_ds_Appendix}) becomes the Cycloid equation
\begin{equation}
  \label{eq:dp_ds_small_p_Appendix}
  \left(\frac{dp}{ds}\right)^2 = \frac{2\xi-p}{p}\,.
\end{equation}
The solution of this equation in parametric form is
\citep{Beloborodov2008}
\begin{eqnarray}
  p &=& \xi(1-\cos\omegaPGJ{}t)
  \label{eq:p_Cycloid_Appendix}\\
  s &=& \xi(\omegaPGJ{}t-\sin\omegaPGJ{}t)\,,
  \label{eq:s_Cycloid_Appendix}
\end{eqnarray}
where the time $t$ is measured in seconds.  The period of spatial
oscillations is then
\begin{equation}
  \label{eq:s0_small_p_Appendix}
  s_0 = 2\pi\xi{}\,.
\end{equation}
For small $\xi{}$ the flow is always non-relativistic and
eqs.~(\ref{eq:p_Cycloid_Appendix},\ref{eq:s_Cycloid_Appendix})
describe it accurately everywhere.  For large $\xi{}$
eqs.~(\ref{eq:p_Cycloid_Appendix},\ref{eq:s_Cycloid_Appendix}) are
good approximation near the starting point of the flow for all $\xi$
and for $\xi<1$ also near periodically repeating stagnation points,
with small values of $p$.

For small $s$, near the flow's starting point, the time $t$ can be
eliminated from
(\ref{eq:p_Cycloid_Appendix},~\ref{eq:s_Cycloid_Appendix}) and we get
\citep[cf. eq.~(13) in][]{FawleyAronsScharlemann1977}
\begin{equation}
  \label{eq:p_small_Appendix}
  p \simeq \left(\frac{9}{2}\right)^{1/3}\xi^{1/3}s^{2/3}
\end{equation}

\section{Boundary conditions for modeling of Space Charge Limited Flow}
\label{sec:bc_sclf}

\begin{figure}
\begin{center}
  \includegraphics[width=\columnwidth]{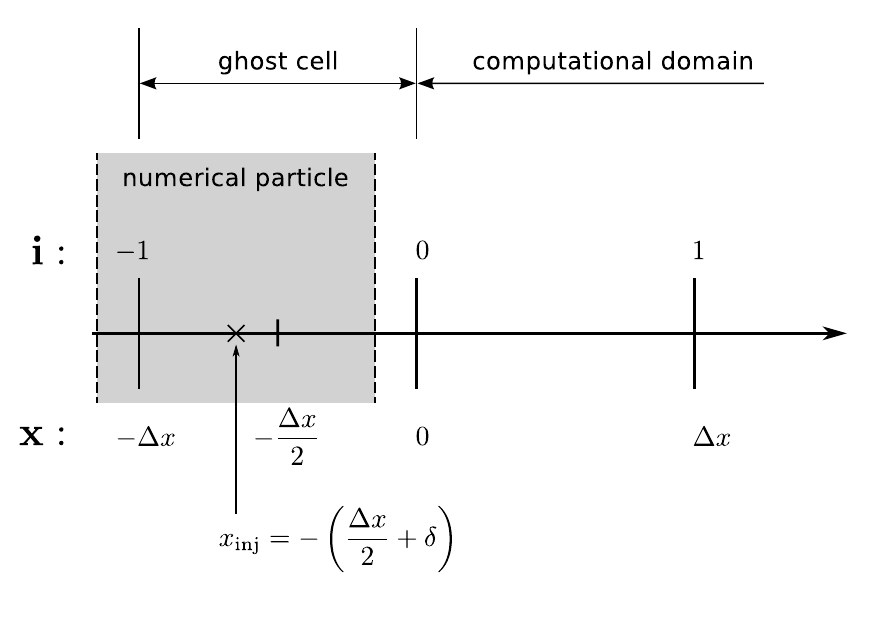}
\end{center}
\caption{Numerical implementation of boundary conditions for space
  charge limited flow. Injection of one numerical particle is shown,
  particle's center is marked by a cross.  See text for explanations.}
  \label{fig:sclf_bc}
\end{figure}

For modeling of the space charge limited flow in the polar cap one
needs to make an adequate numerical model for an infinitely large pool
of particles available at the NS surface in order to correctly
simulate the space charge limitation condition.  This is not a trivial
task. We found the following procedure works well.

The calculation domain of the length $L$ is divided in $M_x$ equal
numerical cells (a typical value of $M_x$ in our simulations is 
$\sim$ a few thousand).  The electric field $E$ and current $j$ are set
at cell boundaries $i=0,\dots,M_x$.  For calculation of the current
density we use a 1D version of the charge conservative algorithm
proposed by \citet{VillasenorBuneman92}, when charged particles are
represented by uniformly charged sheets with the width equal to the
cell size $\Delta{x}$ and the position of a particle is the position
of the sheet's center.  The fraction of the sheet passed through the
cell boundary $i$ during a time step determines the contribution of the
particle into the current $j_i$ at that point.

At each end of the calculation domain we have one ``ghost'' cell.  The
outside boundaries of the ghost shells are ``ghost'' points with
indexes $i=-1$ and $i=M_x+1$.  The equation~(\ref{eq:dE_dt}) for the
electric field is solved for points $0\dots{}M_x$.  Particles can
move into the ghost cells but when their positions are outside of the
domain $[-\Delta{x}/2; L+\Delta{x}/2]$ and they are moving outwards
they do not contribute into the current density in the domain anymore
and such particles are deleted at the end of each time step.

The electric field at the ghost points is set to zero.  We solve a 1D
\emph{initial} value problem -- the electric field in the domain is
calculated from the values of the electric field at the same points at
the previous time step and is not coupled to the electric field at the
``ghost'' points.  The electric field inside the ghost cells, at the
particles' position, are obtained by quadratic interpolation using
values $E_{-1}$, $E_{0}$, $E_{1}$ (and $E_{M_x-1}$, $E_{M_x}$,
$E_{M_x+1}$) and, therefore, setting $E$ at the ghost point to zero
($E_{-1}=E_{M_x+1}=0$) smoothly reduces the electric field inside
``ghost'' cells toward their outer ends.  Setting the electric field
at ghost points at each time step to the values obtained as
extrapolation of electric field values near the domain boundaries
(e.g. using quadratic extrapolation from points $0,1,2$ and
$M_x-2,M_x-1,M_x$) or to some non-zero values resulted only in higher
numerical noise and did not change the system behavior.

At the beginning of each time step we inject certain amount of
electrons $N_\mathrm{inj}$ and equal number of heavy positive charged
particles (``ions'') in the first ``ghost'' cell of our computation
domain at the position slightly outside the center of the fist cell
$x_\mathrm{inj}=-(\Delta{x}/2+\delta)$, see Fig.~\ref{fig:sclf_bc}.
The momentum of each injected particle is sampled from a uniform
distribution in the interval $[-p_\mathrm{inj},p_\mathrm{inj}]$.  In
this way we can model finite temperature of the particles on one hand
and populate the domain with particles more uniformly on the other
hand, as each injected particle after the first time step will have a
slightly different position.  If during this time step the electric
field inside the ``ghost'' cell cannot move particles into the
interval $[-\Delta{x}/2;L+\Delta{x}/2]$, they do not contribute to the
current density and will be deleted at the end of the time step.
Depending on the value of the electric field either positive of
negative particles will be ``sucked'' into the domain.

We experimented with different values for $N_\mathrm{inj}$ and found
that usually after $N_\mathrm{inj}$ exceeds the critical value
$N_\mathrm{inj}^\mathrm{cr}\equiv2(\jm/Qc)(c\Delta{t}/\Delta{x})$ by
$\sim20-30\%$ computational results stop depending on
$N_\mathrm{inj}$; further increase of $N_\mathrm{inj}$ results in
higher numerical noise. $N_\mathrm{inj}^\mathrm{cr}$ is twice the
number density of particles with relativistic velocities which must be
injected at every time step in order to provide the required current
density $\jm$; $Q$ is the charge of a numerical particle, $c$ is the
speed of light.  The factor 2 accounts for injected particles having
negative initial momentum -- most of them do not reach computational
domain and are deleted.  Having some injected particles with negative
momenta results in slightly lower numerical noise as well as more
realistically represents the finite temperature of the NS atmosphere
(the equivalent of the warm cathode in the analogous vacuum tube and
high current beam technologies). The computational overhead caused by
such particles is negligible, as $N_\mathrm{inj}$ is orders of
magnitude less that the total number of particles.  We also
experimented with different values for the time step and found that
values of $\Delta{t}$ such that $\Delta{x}/c\Delta{t}\sim5$ results in
relatively low level of numerical noise due to discrete events of
particle injection; smaller $\Delta{t}$ lead to larger numerical
overhead, as the stability of the leapfrog scheme requires only that
$\Delta{t}<0.5\Delta{x}/c$.

\label{lastpage}

\end{document}